\pdfoutput=1
\documentclass[
longbibliography,
prx,
twocolumn,
superscriptaddress,
amsmath,amssymb,
floatfix
]{revtex4-2}

\usepackage{graphicx}
\usepackage{dcolumn}
\usepackage{bm}
\usepackage[dvipsnames]{xcolor}
\usepackage{hyperref}
\usepackage[mathlines]{lineno}
\usepackage{mathrsfs}
\usepackage{enumitem}
\setenumerate{noitemsep,topsep=0pt,parsep=0.5em,partopsep=0pt}

\definecolor{C1}{RGB}{0,206,206}
\definecolor{C2}{RGB}{251, 77, 61}
\definecolor{C3}{HTML}{5AA9E6}
\definecolor{C4}{RGB}{202, 21, 81}

\hypersetup{colorlinks=true, linkcolor=C2, citecolor=C2, urlcolor=C2}

\usepackage{tikz}
\usetikzlibrary{fadings,decorations.pathreplacing}
\tikzset{
    partial ellipse/.style args={#1:#2:#3}{
        insert path={+ (#1:#3) arc (#1:#2:#3)}
    }
}

\usepackage{amsthm}
\newtheorem*{theorem*}{Theorem}

\newtheorem{corollary*}{Corollary}
\newtheorem*{definition*}{Definition}

\usepackage[varg]{txfonts}

\usepackage{dsfont}

\newcommand*{\spam}{\textsf{SPAM}}
\newcommand*{\rb}{\textsf{RB}}
\newcommand*{\qpt}{\textsf{QPT}}
\newcommand*{\asf}{\textsf{ASF}}
\newcommand*{\compp}{\textsf{CP}}
\newcommand*{\compptp}{\textsf{CPTP}}
\newcommand*{\povm}{\textsf{POVM}}
\newcommand*{\dof}{\textsf{d.o.f.}}
\newcommand*{\f}{\frac}
\newcommand*{\mc}{\mathcal}
\newcommand*{\mscr}{\mathscr}
\newcommand*{\dg}{\dagger}
\newcommand*{\mbb}{\mathds}
\DeclareMathOperator{\tr}{tr}
\DeclareMathOperator*{\Motimes}{\text{\raisebox{0.25ex}{\scalebox{0.6}{$\bigotimes$}}}}
\DeclareMathOperator*{\Mcirc}{\text{\raisebox{0.25ex}{\scalebox{0.7}{$\bigcirc$}}}}
\newcommand*{\dimE}{d_\mathsf{E}}
\newcommand*{\dimS}{d_\mathsf{S}}
\newcommand*{\env}{\mathsf{E}}
\newcommand*{\syst}{\mathsf{S}}
\newcommand*{\markov}{{\scriptscriptstyle{(\mathrm{M})}}}

\begin{document}

\title{Randomized Benchmarking for Non-Markovian Noise}

\author{Pedro Figueroa--Romero}
\email{pedro.romero@foxconn.com}
\affiliation{Hon Hai Quantum Computing Research Center, Taipei, Taiwan}

\author{Kavan Modi}
\email{kavan.modi@monash.edu}
\affiliation{ARC Centre for Engineered Quantum Systems \& School of Physics and Astronomy, Monash University, Victoria, Australia}

\author{Robert J. Harris}
\affiliation{ARC Centre for Engineered Quantum Systems \&  School of Mathematics and Physics, The University of Queensland, Brisbane, Australia}

\author{Thomas M. Stace}
\affiliation{ARC Centre for Engineered Quantum Systems \&  School of Mathematics and Physics, The University of Queensland, Brisbane, Australia}

\author{Min-Hsiu Hsieh}
\affiliation{Hon Hai Quantum Computing Research Center, Taipei, Taiwan}

\date{\today}

\begin{abstract}
Estimating the features of noise is the first step in a chain of protocols that will someday lead to fault-tolerant quantum computers. The randomized benchmarking (\rb) protocol is designed with this exact mindset, estimating the average strength of noise in a quantum processor with relative ease in practice. However, {\rb}, along with most other benchmarking and characterization methods, is limited in scope because it assumes that the noise is temporally uncorrelated (Markovian), which is increasingly evident not to be the case. Here, we combine the {\rb} protocol with a recent framework describing non-Markovian quantum phenomena to derive a general analytical expression of the average sequence fidelity (\asf) for non-Markovian {\rb} with the Clifford group. We show that one can identify non-Markovian features of the noise directly from the {\asf} through its deviations from the Markovian case, proposing a set of methods to collectively estimate these deviations, non-Markovian memory time scales, and diagnose (in)coherence of non-Markovian noise in an {\rb} experiment. Finally, we demonstrate the efficacy of our proposal by means of several proof-of-principle examples. Our methods are directly implementable and pave the way for a better understanding of correlated noise in quantum processors.
\end{abstract}

\maketitle

\section{\label{sec: introduction}Introduction}
The biggest challenge faced in any quantum computation can almost unequivocally be said to be the presence of errors. Among these, noise arising from interactions with the surroundings of a system represent an important class that is still far from being well understood. Given the current widespread interest in designing complex fault-tolerant quantum systems, together with the fundamental restriction that no system can ever be fully isolated from its surroundings, the need to advance our understanding of this type of noise cannot be understated.

Over the last decade, the approach known as randomized benchmarking (\rb)~\cite{Emerson_2005, Levi_2007, Knill2008, PhysRevLett.106.180504, PhysRevA.85.042311} has become the gold standard to certify the performance of gate sets and characterize the noise in computations involving these sets. \rb~generally refers to an experimental protocol allowing estimation of error rates of a gate set by quantifying their control fidelity as a function of the number of gates~\cite{helsen2020general}. Moreover, it does so in an efficient way that is robust to state preparation and measurement (\spam) errors, as opposed to approaches such as quantum process tomography (\qpt)~\cite{Chuang_1997}. It is important to point out, however, that the two approaches are rather complementary~\cite{PhysRevLett.121.170502}, as \rb~extracts less information about the noise, namely average error rates of average gates, but requires little resources for high confidence~\cite{Wallman_2014}, while \qpt~allows to fully reconstruct noise but with a higher resource cost~\cite{PhysRevA.77.032322}. Aside from \qpt~and \rb, there is a plethora of other methods lying in between, such as gate set tomography~\cite{nielsen2020gate}, compressed sensing~\cite{PhysRevLett.105.150401, Flammia_2012} or direct fidelity estimation~\cite{PhysRevLett.106.230501, PhysRevLett.107.210404, PhysRevLett.109.070504}, to name a few, to characterize quantum devices. The main reason why \rb~has become an essential tool for quantum technologies is thus its practicality and applicability to realistic experimental settings.

The most common versions of \rb~protocols are executed for sequences of Clifford gates~\cite{Knill2008}, and consider noise that is both time and gate independent, in particular Markovian and context independent. In this case, it is observed that the so-called average sequence fidelity (\asf), i.e. a figure of merit relating to the gate fidelity of the noise\textsuperscript{\footnote{ In Ref.~\cite{PhysRevLett.119.130502}, it was pointed out that this relation between the \asf~and the gate fidelity of the noise is not unique due to gauge invariance. An in-depth discussion can be seen in Ref.~\cite{Wallman_2018, merkel2018randomized} with an overview and generalisation in \cite{helsen2020general}.}}, behaves as a decaying exponential in the number of gates applied in the sequence. Nevertheless, progress for time-dependent~\cite{Wallman_2014} and gate-dependent noise~\cite{PhysRevLett.119.130502, Wallman_2018, Helsen2019}, as well as different gate sets~\cite{PhysRevA.92.060302, Hashagen_2018, Helsen2019} or other figures of merit has also been made~\cite{PhysRevLett.109.080505, Wallman_2015}. Despite this, \rb~has generally remained elusive to a characterization in the presence of temporally-correlated, so-called non-Markovian noise, and has rather been identified when the \asf~does not behave as a decaying exponential in numerical and experimental studies~\cite{Ryan_2009,Park2016,PhysRevA.89.062321, PhysRevA.92.022326, Mavadia_2018}. Hence it is not an overstatement that overcoming the Markovianity assumption in \rb~remains one of the most important hurdles to clear towards fault tolerance in quantum computers.

Correlated noise has been thoroughly examined in particular scenarios, such as that of dephasing noise. For classical correlations, e.g. in Ref.~\cite{PhysRevA.93.022303, Mavadia_2018} (and similarly in Ref.~\cite{fong2017randomized}), the noise is modeled as rotations of a qubit around the $z$-axis as determined by a classical random variable, and deviations from the uncorrelated case are found. For the quantum counterpart, in Ref.~\cite{PhysRevA.103.022607} this is generalized to correlations being mediated by a bath, modeled as a multi-mode bosonic field interacting with the qubit. Similarly, correlations arising as interaction between neighboring qubits, so-called crosstalk~\cite{PhysRevLett.109.240504} have been addressed in multi-qubit \rb~protocols, generally noticing that averaging over a single qubit generally leads to a non-exponential decay of the \asf.

The study of temporal correlations in quantum systems necessarily require in its foundations a theory of quantum stochastic processes. The development of such a theory has much older origins than \rb~but has often been contentious and faced conceptual problems still widely discussed in the community~\cite{milz2020quantum}. Nevertheless, approaches in terms of higher-order maps~\cite{PhysRevLett.101.060401, PhysRevA.80.022339} have proved successful in providing a general theory of quantum stochastic processes~\cite{milz2020quantum}, in particular unambiguously establishing a Markov condition~\cite{Costa_2016, PhysRevLett.120.040405} and providing an operational framework to characterize non-Markovian processes~\cite{PhysRevA.97.012127}.

In this manuscript, we derive an analytical expression for the \asf~of an \rb~experiment with the Clifford group under non-Markovian gate-independent noise. This allows to study the behavior of \asf~decays due to non-Markovianity, and particularly of deviations from exponential decays, given a model for the noise. We also discuss ways in which the relevant time scales, i.e. sequence lengths, for finite non-Markovian noise can be determined, and deviations from a Markovian decay can be quantified, both with or without an \emph{a priori} model of the noise. The main limitation to these methods is precisely the \rb~protocol itself, as the non-Markovian \asf ---as we show--- is not a simple function of sequence length anymore. Nevertheless, just as in the standard Markovian case, the relevance of \rb~lies in its simplicity, as it allows to analyze and quantify non-Markovian features from experiment with relative ease. Overcoming these restrictions thus can be a focus of future research towards a practical and more complete characterization of temporally-correlated noise.

The manuscript is structured as follows. In Section~\ref{sec: randomised benchmarking} we introduce the \rb~protocol and discuss the theoretical setting employed in the remainder of the paper. In Section~\ref{sec: quantum processes and non-Markovianity} we introduce the process tensor framework and elaborate on how it is a natural framework for non-Markovian \rb. In Section~\ref{sec: main average sequence fidelity} we present our main result within Eq.~\eqref{eq: average fidelity nonMarkovian} and discuss some of its properties and consequences, including containment of the Markovian case, the issue of initial correlations and the impact of \spam~errors. In Section~\ref{sec: nm RB measure} we introduce a theoretical measure for non-Markovian \rb~by means of Eq.~\eqref{eq: rb non-Markovianity}, discussing the case of classical correlations and the possibility of blindness to non-Markovian noise by \rb. In Section~\ref{sec: finite} we discuss the more realistic scenario of finite non-Markovian noise, with which we can operationally approach the problem of determining sequence lengths, i.e. time scales, at which temporal correlations in the noise are relevant, as well as quantifying deviations from an exponential decay whenever a model for the noise is unknown. In Section~\ref{sec: numerics 1} we show a proof-of-principle numerical example finding agreement with our analytical result, and discuss the effect of \spam~errors and non-Markovianity blindness. Finally, in Section~\ref{sec: numerics 2} we demonstrate numerically how the memory length of a finite non-Markovian noise process can be estimated in practice, non-Markovian deviations quantified, and how to diagnose (in)coherence of non-Markovian noise. We conclude in Section~\ref{sec: conclusions} with an overview of our results and a perspective for future work.

\section{\label{sec: randomised benchmarking}Randomized Benchmarking}
While there are many variants of \rb, and a general framework encompassing these can be established~\cite{helsen2020general}, for concreteness here we consider an \rb~protocol employing the Clifford group. This has been the most common approach in \rb~mainly because the elements on the Clifford group can be realized efficiently on a quantum processor~\cite{Gottesman1997, Gottesman1998, PhysRevA.70.052328}. The \rb~protocol is then as follows:

\begin{enumerate}[leftmargin=*]
    \item Prepare an initial state $\rho$.
    \item Sample $m$ distinct elements, $\mc{G}_1,\mc{G}_2,\ldots,\mc{G}_m$, uniformly at random from the Clifford group. Let $\mc{G}_{m+1} := \Mcirc_{i=m}^1\mc{G}_i^\dg=\mc{G}_1^\dg\circ\cdots\circ\mc{G}_m^\dg$, where $\circ$ denotes composition of maps and $\mc{G}^\dg(\cdot)=G^\dg(\cdot)G$ for any Kraus representation with unitaries $G$ of the map $\mc{G}$. We refer to $\mc{G}_{m+1}$ as an undo gate.
    \item Apply the composition $\Mcirc_{i=1}^{m+1}\mc{G}_i$ on $\rho$. In practice, this amounts to applying a noisy sequence $\mc{S}_m:=\Mcirc_{i=1}^{m+1}\hat{\mc{G}}_i$ of length $m$ on $\rho$, where $\hat{\mc{G}}_i$ are the physical noisy gates associated to $\mc{G}$.
    \item Estimate the probability $f_m=\tr\left[\mc{M}{\mc{S}}_m\left(\rho\right)\right]$ via a positive operator-valued measure (\povm)~element $\mc{M}$.
    \item Repeat $n$ times the steps 1 to 4 for the same initial state $\rho$, same \povm~element $\mc{M}$, and different sets of gates chosen uniformly at random $\{\mc{G}_i^{(1)}\}_{i=1}^m,\{\mc{G}_i^{(2)}\}_{i=1}^m,\ldots,\{\mc{G}_i^{(n)}\}_{i=1}^m$ from the Clifford group to obtain the probabilities $f_m^{(1)},f_m^{(2)},\ldots{f}_m^{(n)}$. Compute the average $\mc{F}_m=1/n\sum_{i=1}^nf_m^{(i)}$. We refer to $\mc{F}_m$ as an average sequence fidelity (\asf).
    \item Examine the behavior of the \asf~$\mc{F}_m$ over different sequence lengths $m$.
\end{enumerate}

The important insight in the \rb~protocol is that the \asf~contains the average noise rate of the applied sequences, which can be extracted efficiently by analyzing it over varying sequence lengths. Specifically, when the noise is approximated as both independent of the gates applied and the time-step at which these are applied, the \asf~is given by
\begin{equation}
    \mc{F}_m=Ap^m+B,
    \label{eq: asf markov time-gate ind}
\end{equation}
where the error rate of the noise, or so-called noise strength, is given by $p\in[0,1]$ and $A$, $B$ are constants determined by state preparation and measurement (\spam) errors~\cite{PhysRevA.85.042311}.
This implies that having performed an \rb~experiment, the data of the experimental \asf{s} can be fitted to an exponential, from which the noise-strength $p$ and the \spam~factors can be extracted. The noise strength is directly related\textsuperscript{\cite{Note1}} to the gate fidelity of the noise with respect to the identity~\cite{PhysRevA.92.022326}, and hence the labeling of $\mc{F}_m$ as a fidelity, but similarly other figures of merit can be used to learn average error rates through \rb~\cite{PhysRevLett.109.080505, Wallman_2015}.

It is important to mention that \spam~errors are implicit to steps 1 and 4, that is, in an execution of the protocol, neither the initial state preparation nor the measurement of the output state might be perfect. In the time and gate independent scenario for the Clifford group, however, as seen in Eq.~\eqref{eq: asf markov time-gate ind}, \spam~errors are constants both independent of the error rate and the sequence length.

The exponential decay in Eq.~\eqref{eq: asf markov time-gate ind} can be obtained by modeling each noisy gate as $\hat{\mc{G}}_i:=\Lambda\circ\mc{G}_i$ for some completely positive trace preserving (\compptp) map $\Lambda$; then the analytical average of the survival probabilities is given by the average over gates $\mc{G}_i$. For our purposes, we just care that the gates belong to a {unitary 2-design}, i.e.,~any distribution of gates replicating up to the second moment of the unitary group with the uniform Haar measure~\cite{PhysRevA.80.012304}, such as the Clifford group. This implies that averaging over gates can be replaced with that over the Haar measure to obtain $\mc{F}_m$, and similarly the use of higher unitary designs could serve to characterize higher-order statistical properties of noise in \rb~\cite{nakata2021quantum}. Detail about how such averaging is carried out can be seen in Appendix~\ref{appendix: average gate sequence}.

Importantly, one sees deviations from an exponential decay for more complex noise profiles, including non-Markovian noise. Nevertheless, within a Markovian assumption, \rb~generally renders a linear combination of exponential decays for the \asf~\cite{helsen2020general}, with the particular case of gate-dependence rendering a single perturbation term that decays exponentially as well with the sequence length~\cite{PhysRevLett.119.130502,Wallman_2018}. Here, together with the assumption that the gates $\mc{G}$ belong either to the multi-qubit Clifford group or to a 2-design, we make the assumption that the noise modeled by the maps $\Lambda$ is gate-independent. Other than these two assumptions, we are interested in temporal correlations in the noise described as being mediated by an external environment.

\section{\label{sec: quantum processes and non-Markovianity}Quantum processes and non-Markovianity}
The setting we consider is that of a bipartite quantum system, labeled $\syst\env$, composed of a $\dimS$-dimensional system $\syst$ and a $\dimE$-dimensional environment $\env$. An experimenter, in principle, would apply the sequence $\mc{S}_m=\Mcirc_{i=1}^{m+1}\mc{G}_i$ of Clifford gates $\mc{G}_i$ solely on $\syst$, and not have access to $\env$. We consider different scenarios for the initial state $\rho$, which is solely prepared on $\syst$, but can nevertheless get correlated with $\env$ afterwards, accounting for a new type of \spam~error. We now can model the noisy gates as $\hat{\mc{G}}_i=\Lambda_i\circ(\mc{I}_\env\otimes\mc{G}_i)$, where $\Lambda_i$ acts on the full $\syst\env$ system and $\mc{I}_\env$ is an identity map on $\env$. In particular, we require the (gate-independent) noise maps $\Lambda_i$ to be at least completely positive (\compp) trace non-increasing and allow a further time-dependence, $\Lambda_i\neq\Lambda_j$ for $i\neq{j}$; this can further be constrained to requiring trace-preservation (e.g. if the device never fails), unitarity (e.g. if the device is perfectly isolated), or time-independence (the noise does not change between time-steps).

\begin{figure}[t!]
\centering
    \includegraphics[width=\linewidth]{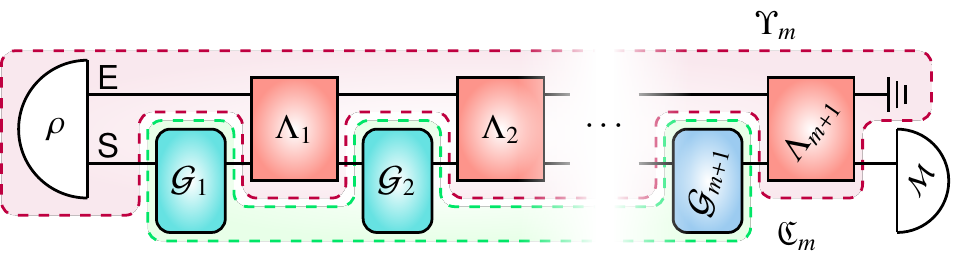}
    \caption{{\textbf{A non-Markovian noisy \rb~sequence as a process tensor contraction}}. An initial system-environment ($\syst\env$) state $\rho$ is acted on with a random Clifford element $\mc{G}_1$ on $\syst$ alone, inducing a noise map $\Lambda_1$ on $\syst\env$, followed by a random Clifford $\mc{G}_2$ inducing noise $\Lambda_2$, and so on until the undo map $\mc{G}_{m+1}$ is applied with some noise $\Lambda_{m+1}$. Finally, a \povm~element $\mc{M}$ is applied on $\syst$ and the environment is traced out (grounding symbol). This can be described by the contraction of a tensor $\Upsilon_m$ (upper box) comprising the initial state together with the noise process, with a tensor $\mathfrak{C}_m$ describing the  sequence of applied Clifford gates, all followed by applying $\mc{M}$.}
    \label{fig: non-Markov sequence}
\end{figure}

The sequence $\mc{S}_m$ can be understood as a particular example of a quantum stochastic process where the underlying dynamics are given by the noise inherent to the computation on the whole $\syst\env$. Motivated by what is done operationally in a laboratory, the process tensor framework~\cite{PhysRevA.97.012127, PhysRevLett.120.040405, milz2020quantum, Taranto_2020} provides the means by which we can treat the underlying noise source separately from what the experimenter has control over, which are the gates they apply. This effectively means that we can treat the whole noise in the sequence, together with the initial state, as a tensor $\Upsilon_m$. We may contract this tensor with the set of Clifford gates, which too can be incorporated in a tensor $\mathfrak{C}_m$. This can be depicted as in the circuit of Fig.~\ref{fig: non-Markov sequence}.

These tensors, $\Upsilon_m$ and $\mathfrak{C}_m$, just as any quantum map, can have different representations~\cite{Milz_2017}. Here we employ the Choi-state representation, which simply is a generalization of the Choi-Jamio\l{}kowski isomorphism for quantum channels~\cite{watrous2018theory}. Specifically, these can be written as
\begin{equation}
    \Upsilon_m = \tr_\env\left\{\left[\Mcirc_{i=1}^{m+1}(\Lambda_i\otimes\mc{I}_\mathsf{aux})\circ\mathscr{S}_i\right]\,\rho\otimes\psi^{\otimes m+1}\right\},
    \label{eq: def upsilon}
\end{equation}
where $\mc{I}_\mathsf{aux}$ is an identity map on an auxiliary space $\mathsf{aux} = \mathsf{A}_1 \mathsf{B}_1\cdots\mathsf{A}_{m+1}\mathsf{B}_{m+1}\simeq\syst^{\otimes2(m+1)}$ composed of $m+1$ pairs of $\syst$ systems, $\mathscr{S}_i$ is a swap gate between $\syst$ and one of these pairs in the $i$\textsuperscript{th} auxiliary space, say $\mathsf{A}_i$, and $\psi := \sum|ii\rangle\!\langle{jj}|$ is an unnormalized maximally entangled state; on the other hand, for the gate sequence,\textsuperscript{\footnote{ Strictly speaking, here we are defining $\mathfrak{C}_m=\mbb1\otimes\mathfrak{A}_m$, where $\mathfrak{A}_m$ is the Choi state of the sequence of gates $\mc{G}_i$, and $\mc{G}_i$ can be defined to act on either auxiliary space $\mathsf{A}_i$ or $\mathsf{B}_i$, the choice only depends on what auxiliary space the swap $\mscr{S}_i$ on the definition of $\Upsilon_m$ swaps with, so that the contraction $\Upsilon_m\mathfrak{C}_m^\mathrm{T}$ contracts the correct spaces.}}
\begin{equation}
    \mathfrak{C}_m = \mbb1_\syst\otimes\left[\Motimes_{i=1}^{m+1}\left(\mc{I}_{\mathsf{A}_i}\otimes\mc{G}_i\right)\right]\psi^{\otimes{m+1}}.
    \label{eq: def mathfrakC}
\end{equation}

Detail about the definitions in Eq.~\eqref{eq: def upsilon} and Eq.~\eqref{eq: def mathfrakC} can be consulted in Appendix~\ref{appendix: process tensor}; we highlight, however, that we do not require the physical construction of these tensors nor access e.g. to the space \textsf{aux} or the states $\psi$, but rather they are part of the theoretical framework that will let us study the \rb~protocol when the $\Lambda_i$ maps are temporally correlated by means of the environment $\env$. More broadly, the process tensor framework generalizes the notion of a stochastic process to the quantum domain in a fully consistent way~\cite{Milz2020kolmogorovextension, nurdin2021heisenberg}, resolving problems such as the initial correlation and not-\compp~problems, and fully accounting for memory effects~\cite{milz2020quantum}.

The notion of Markovianity is formalized in the process tensor framework through a proper operational Markov condition~\cite{PhysRevLett.120.040405} as an independence of past observations, in turn containing the classical definition of Markovianity and unifying all quantum Markov conditions that had been proposed thus far~\cite{PhysRevLett.120.040405, PhysRevLett.123.040401, milz2020quantum}. Markovianity, and hence the absolute absence of temporal correlations in a process tensor, implies that no information is passed through $\env$ between time-steps. This is mathematically manifest in the Choi-state, which takes the form of a product of individual Choi states of quantum channels joining each step, as for $\mathfrak{C}_m$ in Eq.~\eqref{eq: gate sequence PT}. That is, temporal correlations in the process tensor correspond to spatial correlations in the Choi-state representation, and more precisely then, a process tensor $\Upsilon_m^\markov$ will contain only Markovian noise if and only if there are noise maps $\Lambda_i^\markov$ acting solely on $\syst$ such that 
\begin{equation}
    \Upsilon_m^\markov=\rho_\syst\otimes\left[\Motimes_{i=1}^{m+1}\left(\Lambda_i^\markov\otimes\mc{I}_{\mathsf{B}_i}\right)\right]\psi^{\otimes{m+1}}.
\end{equation}

Non-Markovianity can then naturally be quantified by means of any operationally meaningful distinguishability measure $D$ with 
\begin{equation}
    \mc{N} := \min_{\Upsilon_m^\markov} D\left(\Upsilon_m,\Upsilon_m^\markov\right),
    \label{eq: general nonMarkovianity measure}
\end{equation}
where the choice of such distance measure is rather a matter of practicality, as the minimization over all Markovian processes will often make the computation of $\mc{N}$ unfeasible. This can be alleviated either by choosing a measure $D$ such as relative entropy, where the min argument is just a product of marginals, $\Upsilon^\markov_m=\rho_\syst\Motimes_{i=1}^{m+1}\tr_{\overline{i:i-1}}\left[\Upsilon_m\right]$, where $\tr_{\overline{j:i}}$ means trace over all except between steps $i$ to $j$, or otherwise placing relevant bounds on $\mc{N}$ for Schatten-norm measures, as done in Ref.~\cite{2019almostmarkovian, 2020markovianization} to study some statistical properties of non-Markovian processes. Here, we care about quantifying how non-Markovian an \rb~experiment is, which will boil down to quantifying how distinguishable a non-Markovian \asf~is from a sensible Markovian counterpart.

We can now write the probability with $m$ noisy gates as per step 4 of the \rb~protocol, $f_m = \tr\left[\mc{M}\mc{S}_m(\rho)\right]$, in terms of the process tensor with
\begin{align}
    \mc{S}_m(\rho) &= \,\tr_\env\left\{\left[\Mcirc_{i=1}^{m+1}\Lambda_i\circ(\mc{I}_\env\otimes\mc{G}_i)\right]\,\rho\right\} \nonumber\\
    &= \tr_{\overline{\syst}}\left(\Upsilon_m\mathfrak{C}_m^\mathrm{T}\right),
\end{align}
where $\tr_{\overline{\syst}}$ here means a partial trace over all intermediate input and output systems except the final $\syst$ and $\mathrm{T}$ denotes a transpose. Computing the \asf, $\mc{F}_m$, then amounts to computing the average of $\mathfrak{C}^\mathrm{T}$ over the applied gates, $\mc{G}_i$. This is a simplification allowing to deal with the average over gates separately from the underlying noise. Furthermore, given that here we deal with the Clifford group, as explained in Section~\ref{sec: randomised benchmarking}, we can replace averaging over Clifford gates with averaging over the unitary group with the uniform Haar measure. To finally obtain $\mc{F}_m$, we have to contract the average gate sequence tensor with the noise tensor $\Upsilon_m$, which will contain the noise inherent to the \rb~sequence, and in particular can be labeled as non-Markovian if the individual noise is correlated between time-steps or Markovian otherwise.

We now present a general expression for the \asf~$\mc{F}_m$ for \rb~of the Clifford group under non-Markovian noise and explore some of its consequences.

\section{Average Sequence Fidelity for non-Markovian noise}\label{sec: main average sequence fidelity}
Given an \rb~sequence with $m$ Clifford gates affected by non-Markovian noise, we can construct the noise and gate sequence process tensors, compute the average gate tensor and contract with the noise tensor to get the average sequence fidelity (\asf). This yields the following:

\begin{theorem*}
Let $\rho$ be the initial state of a system-environment, $\syst\env$, composite with $\dimS\dimE=\mathrm{dim}(\syst\env)$. Let $\mc{S}_m(\rho)$ describe a randomized benchmarking (\rb) sequence of length $m$ over Clifford gates with the \compp~map $\Lambda_n$ acting on $\syst\env$ being the associated noise at the $n$\textsuperscript{th} time-step. Then, the average sequence fidelity (\asf)~$\mc{F}_m$ with a \povm~element $\mc{M}$ is given by
\begin{align}
    \mc{F}_m &= \tr\{\mc{M}\,\mbb{E}[\mc{S}_m(\rho)]\}\nonumber\\
    &=  \tr\left[\mc{M}\,\tr_\env\circ\Lambda_{m+1}\circ\left(\mscr{A}_m+\mscr{B}_m\right)\rho\right],
    \label{eq: average fidelity nonMarkovian}
\end{align}
where $\mbb{E}$ denotes average over Clifford gates, $\circ$ denotes composition of maps, and
\begin{align}
    \mscr{A}_m(\rho) &:= \f{\displaystyle{\Mcirc_{n=1}^m}\left(\$_{\Lambda_n}-\Theta_{\Lambda_n}\right)\otimes\mc{I}_\syst}{\left(\dimS^2-1\right)^m}\left(\rho-\rho_\env\otimes\f{\mbb1}{\dimS}\right), \label{eq: main curlyA}\\
    \mscr{B}_m(\rho) &:= \left(\Mcirc_{n=1}^m\Theta_{\Lambda_n}\right)\rho_\env\otimes\f{\mbb1}{\dimS}, \label{eq: main curlyB}
\end{align}
with $\rho_\env:=\tr_\syst(\rho)$ being the reduced initial state in $\env$; here $\$_{\Lambda_n},\Theta_{\Lambda_n}$ are maps acting solely on $\env$ as defined by
\begin{align}
    \$_{\Lambda_n}(\varepsilon) &:=\sum_{s,s^\prime=1}^{\dimS} \langle{s}|\Lambda_n(\varepsilon\otimes|s\rangle\!\langle{s^\prime}|)|s^\prime\rangle \label{eq: dollar main} \\
    \Theta_{\Lambda_n}(\varepsilon) &:= \tr_\syst\left[\Lambda_n\left(\varepsilon\otimes\f{\mbb1}{\dimS}\right)\right] \label{eq: theta main},
 \end{align}
for any operator $\varepsilon$ acting on $\env$.
\end{theorem*}

The proof can be found in full in Appendix~\ref{appendix: nM gate indep noise}. As stated before, this amounts to writing the average sequence fidelity as the contraction of tensors $\mc{F}_m=\tr\left\{\mc{M}\,\tr_{\overline{\syst}}\left[\Upsilon_m\,\mbb{E}\left(\mathfrak{C}_m^\mathrm{T}\right)\right]\right\}$, where the average $\mbb{E}\left(\mathfrak{C}_m^\mathrm{T}\right)$ can be evaluated via the second moment of the unitary group with the Haar measure, given that the Clifford group constitutes a unitary 2-design.

We first notice that in the strict noiseless limit, $\Lambda_1=\Lambda_2=\ldots=\Lambda_n=\mc{I}$, we recover $\mc{F}_m\to\tr[\mc{M}\,\rho_\syst]$, where here $\rho_\syst=\tr_\env(\rho)$, so that indeed Eq.~\eqref{eq: average fidelity nonMarkovian} is bounded by one. For the ideal case of $\syst\env$ being a closed system, each $\Lambda_n$ is a unitary. If there is no external time-dependence on the noise and all temporal correlations are described by $\env$, then $\Lambda_n=\Lambda$ for all $n$.

The two relevant terms to gain some insight about Eq.~\eqref{eq: average fidelity nonMarkovian} are $\mscr{A}_m$ and $\mscr{B}_m$ in Eqs.~\eqref{eq: main curlyA} and \eqref{eq: main curlyB}, resp., where the depolarizing effect of the noise on $\syst$ is manifest, with $\mscr{A}_m$ being partially depolarizing in $\syst$ and $\mscr{B}_m$ completely depolarizing in $\syst$. The action of $\mscr{B}_m$, in particular, is independent of the initial state on $\syst$ and picks up noise solely over $\env$. Furthermore, if the initial state is uncorrelated, the effect of averaging a sequence of $m$ gates in $\syst$ is to totally decouple $\syst$ from $\env$, so that both $\mscr{A}_m$ and $\mscr{B}_m$ give a product state, with $\env$ carrying all the noise factors. Finally upon applying $\tr_\env\circ\Lambda_{m+1}$ on $\mscr{A}_m$, this would render a factor analogous to a product of noise-strengths $p_1p_2\cdots{p}_m$.

The notation we use for $\mscr{A}$ and $\mscr{B}$, which here are quantum maps, is suggestive in that these reduce to the corresponding $Ap^m$ and $B$, resp., in the time-independent Markovian limit. In a Markovian scenario the environment is superfluous and we would have $\Lambda_n\to\mc{I}_\env\otimes\Lambda^\markov_n$ together with $\rho\to\rho_\env\otimes\rho_\syst$, i.e. the noise at each step is a \compp~map acting on $\syst$ alone and the initial state on $\syst\env$ is completely uncorrelated. Then, if the noise is trace preserving as well, Eq.~\eqref{eq: average fidelity nonMarkovian} reduces to the Markovian time-dependent \asf~derived in Ref.~\cite{Wallman_2014}, 
\begin{align}
    \mc{F}^\markov_m=p_1\cdots{p}_mA+B,
    \label{eq: asf Markovian limit}
\end{align}
where,
\begin{gather}
    p_n=\f{\tr\left[\Lambda_n^\markov\right]-1}{\dimS^2-1},\\
    A=\tr\left[\mc{M}\Lambda_{m+1}^\markov\left(\rho_\syst-\f{\mbb1}{\dimS}\right)\right],\quad\,B=\tr\left[\mc{M}\Lambda_{m+1}^\markov\left(\f{\mbb1}{\dimS}\right)\right].
\end{gather}

That is, we get $\mscr{A}_m(\rho)\to{p}_1\cdots{p}_m\left(\rho-\mbb1/\dimS\right)$ and $\mscr{B}_m(\rho)\to\mbb1/\dimS$ in this limit, which makes it clear that $\mscr{B}$ renders only \spam~and non-Markovian noise contributions. Here $\tr\left[\Lambda_n^\markov\right]=\sum_\mu|\tr\lambda_\mu^\markov|^2$ where $\lambda_\mu^\markov$ are the Kraus operators of $\Lambda_n^\markov$. Furthermore, despite being complicated in the general case\textsuperscript{\footnote{ The action of the map $\$_{\Lambda_n}$ can alternatively be written as $\$_{\Lambda}(\cdot)=\sum_\mu\tr_\syst(\lambda_\mu)(\cdot)\tr_\syst(\lambda_\mu^\dg)$ with $\lambda_\mu$ the Kraus operators of $\Lambda$.}}, the map $\$_{\Lambda_n^\markov}$ simply picks up a noise multiplicative factor, $\$_{\Lambda_n^\markov}(\varepsilon)=\tr\left[\Lambda_n^\markov\right]\,\varepsilon$ and $\Theta_{\Lambda_n^\markov}$ becomes an identity map, $\Theta_{\Lambda_n^\markov}(\varepsilon)=\varepsilon$, in this limit. Finally, Eq.~\eqref{eq: asf Markovian limit} implies that we recover the decaying exponential in Eq.~\eqref{eq: asf markov time-gate ind} for time-independent Markovian noise. The recovery of the standard \asf~in this limit is shown in detail in Appendix~\ref{appendix: Markovian limit}.

On the other hand, a unique feature when considering non-Markovian noise is initial correlations~\cite{modiscirep, PhysRevLett.114.090402}; these could be particularly relevant in a non-Markovian \rb~experiment because the averaging over $\syst$ gates only depolarizes the noise in $\syst$ after the first gate is applied, but does nothing to correlations in the initial state. Furthermore, as pointed out before, if the initial state is uncorrelated, the \asf~reduces to a quantity of the form $\mc{F}_m\to\tr\left[\mc{M}\tr_\env\circ\Lambda_{m+1}\left(\sigma_\env^{(m)}\otimes\sigma_{\syst}^{(m)}\right)\right]$, and tracing the environment part would give a term analogous to a product of noise-strengths $p_1p_2\cdots{p}_m$. This implies that in general, when benchmarking non-Markovian errors with \rb, the impact of \spam~errors could potentially be relevant in general in the error rates if such errors are large and generate initial correlations. In principle the presence of such errors could also be diagnosed by an offset in the average sequence fidelity $\mc{F}_m$, as we exemplify numerically in Appendix~\ref{appendix: numerics}.

Finally, non-exponential decays in \rb~have often been attributed to non-Markovianity~\cite{PhysRevA.89.062321, PhysRevA.92.022326, Mavadia_2018,PhysRevA.103.022607}: by mere inspection, setting $\Lambda_n=\Lambda$ on all steps $n$, we get $\Mcirc_{n=1}^m(\$_{\Lambda_n}-\Theta_{\Lambda_n})=(\$_\Lambda-\Theta_\Lambda)^{\Mcirc{m}}$, which will generally not render an exponential decay in the \asf. It is important to point out that while non-Markovianity generally leads to non-exponential decays, there can also be other contextual factors~\cite{PhysRevX.9.021045}, such as gate-dependence or other rather arbitrary external time-dependence leading to such behavior.

\section{Quantifying non-Markovianity in Randomized Benchmarking}\label{sec: nm RB measure}
Non-Markovianity in a quantum process can encompass both classical and quantum correlations; the latter is manifest in the Choi-state of a process tensor whenever its components are entangled~\cite{Giarmatzi_2021, milz2020genuine}. As examples of classical correlations, in Appendix~\ref{appendix: classical} we reproduce the \asf~of the model in Ref.~\cite{PhysRevA.93.022303}, where classical temporal correlations are modeled via dephasing noise determined by a classical stochastic process; this effectively renders an \asf~analogous to one that is Markovian time-dependent with the noise parameter being a random variable. We also illustrate this via a shallow pocket model~\cite{PhysRevA.92.022102, Arenz_2018, taranto2019memory, milz2020quantum}, where the time-dependence in the \asf~is explicit but the treatment as a Markov \asf~decay remains the same. These examples suggest that while the general measure of non-Markovianity $\mc{N}$ for a process tensor in Eq.~\eqref{eq: general nonMarkovianity measure} is sensitive to any sort of temporal-correlation, this might not necessarily be the case for the \asf.

An \rb~experiment could be blind to non-Markovianity in the sense of producing equivalent data of some Markovian noise model. It is, of course, a possibility for there to be a subclass of time-independent non-Markovian processes leading to exponential or almost exponential behavior, although as mentioned above, in general a time-independent noise does not lead to an exponential behavior unless the environment is superfluous. In Appendix~\ref{appendix: numerics} we exemplify this numerically with a spin interaction as source of non-Markovian noise. Whether in general there exists a whole class of non-Markovian processes that can be classified as \rb~blind, together with criteria to decide \rb~blindness, is an open question that could potentially be addressed in the near future.

\begin{figure}[t!]
\centering
    \includegraphics[width=\linewidth]{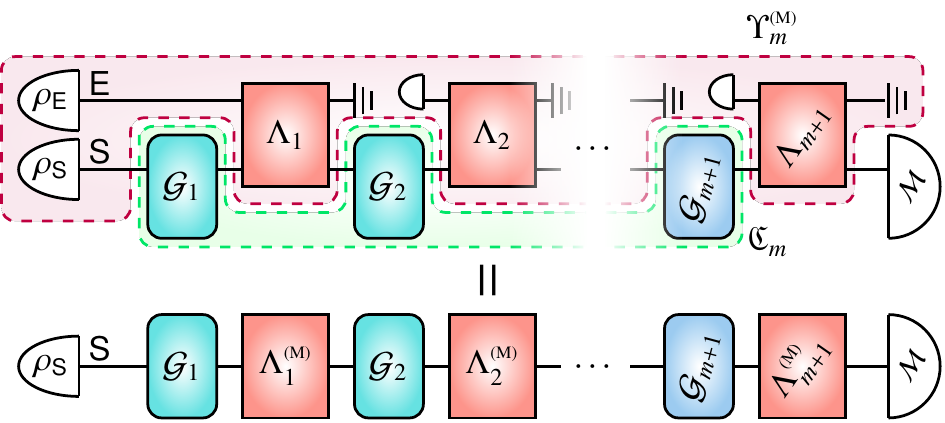}
    \caption{\textbf{{A Markovianized counterpart of a non-Markovian \rb~sequence.}} Given a non-Markovian \rb~sequence with a noise process $\Upsilon_m$, a Markovianized counterpart $\Upsilon^\markov$ can be given by one where the information carried in $\env$ is dissipated or lost between each step. The corresponding \rb~sequence has Markovian noise given by the \compp~maps $\Lambda_n^\markov$ acting as $\Lambda_n^\markov(\sigma)=\tr_\env\circ\Lambda_n(\varepsilon_n\otimes\sigma)$ for any pure state $\varepsilon_n$ on $\env$.}
    \label{fig: Markovian RB}
\end{figure}

There could be instances where having a non-Markovian noise process and being able to quantify its general non-Markovianity $\mc{N}$ with Eq.~\eqref{eq: general nonMarkovianity measure}, we really only care about how much its associated \asf~for the Clifford group deviates from a Markovian one. As in principle the set of possible Markovian processes to compare with would be restricted to a class specific to the given device to be benchmarked, ideally, we would look for a direct Markovian counterpart of the original non-Markovian noise process that we have. Thus, we propose to look at deviations from the \asf~generated by the Markovianized process $\Upsilon^\markov$ where each noise map in the original non-Markovian noise $\Lambda_n$ at time-step $n$ dissipates its $\env$ part: this amounts to taking a Markovian process with the initial state being uncorrelated $\rho\to\rho_\env\otimes\rho_\syst$, and with dynamics at each step being given by the \compp~map $\Lambda_n^\markov$ on system $\syst$ acting as $\Lambda_n^\markov(\sigma)=\tr_\env\circ\Lambda_n(\varepsilon_n\otimes\sigma)$ for an arbitrary pure state $\varepsilon_n$. This is depicted in Fig.~\ref{fig: Markovian RB}.

\begin{definition*}
Let $\mc{F}_m$ be the average sequence fidelity (\asf) of a randomized benchmarking (\rb) experiment over the Clifford group with gate-independent non-Markovian noise. We define the \rb~non-Markovianity as
\begin{gather}
\begin{split}
    \mc{N}_q^{\mc{F}_m} &:= \left\|\mc{F}_m-\mc{F}_m^\markov\right\|_q \\
    &= \left(\sum_{n=1}^m\left|\tr\left\{\mc{M}\,\tr_{\overline{\syst}}\left[\left(\Upsilon_n-\Upsilon_n^\markov\right)\mbb{E}\left(\mathfrak{C}_n^\mathrm{T}\right)\right]\right\}\right|^q\right)^{1/q},
    \label{eq: rb non-Markovianity}
\end{split}
\end{gather}
where $\mc{F}_m^\markov$ is the \asf~of the Markovian noise process associated to $\Upsilon_m$, given by $\Upsilon_m^\markov := \rho_\syst\otimes\left[\Motimes_{i=1}^{m+1}\left(\Lambda_i^\markov\otimes\mc{I}_{\mathsf{B}_i}\right)\right]\psi^{\otimes{m+1}}$, where
\begin{equation}
    \Lambda_n^\markov(\sigma) := \tr_\env\circ\Lambda_n(\varepsilon_n\otimes\sigma),
\end{equation}
for any $\sigma$ acting on $\syst$ and an arbitrary pure state $\varepsilon_n$ on $\env$.
\end{definition*}

The measure $\mc{N}_q^{\mc{F}_m}$ boils down to how well the \povm~element $\mc{M}$ can distinguish $\tr_\env\circ\Lambda\circ\mscr{A}_m(\rho)$ from $p^m\Lambda^\markov(\rho_\syst-\mbb1/\dimS)$, as well as $\tr_\env\circ\Lambda\circ\mscr{B}_m(\rho)$ from $\Lambda^\markov(\mbb1/\dimS)$ for \compptp~noise. Generic bounds can also potentially become possible with this non-Markovianity quantifier. Of course, the \rb~non-Markovianity measure in Eq.~\eqref{eq: rb non-Markovianity} also already makes it manifest that if an underlying noise process in an \rb~sequence is Markovian, then $\mc{N}_q^{\mc{F}_m}=0$. The converse, however, might not necessarily be true or deviations could be negligible in practice~\textsuperscript{\footnote{ This observation can be seen to follow e.g. because we may upper-bound the total \rb~non-Markovianity of a sequence length $m$ experiment, $\mc{N}_q^{\mc{F}_m}$ in Eq.~\eqref{eq: rb non-Markovianity}, as $\mc{N}_q^{\mc{F}_m}\leq\sum\mc{N}_q$, where $\mc{N}_q$ is general non-Markovianity for each intermediate step up to $m$ as in Eq.~\eqref{eq: general nonMarkovianity measure} with $D$ being a Schatten $q$-norm, $D=\|X\|_q:=(\tr|X|^q)^{1/q}$.}}, as we exemplify below numerically.

There could be several scenarios where Eq.~\eqref{eq: rb non-Markovianity} could be computed or estimated. One might be where either the full Markov process $\Upsilon_m^\markov$, or just an error rate is known, but once the \rb~experiment is run, deviations from $\mc{F}_m^\markov$ are observed which most plausibly could be explained by non-Markovianity. This means we could actually compute $\mc{N}_q^{\mc{F}_m}$ directly from the experimental data and e.g. analyze the observed \asf~as a time-dependent \rb~decay. On the other hand, another scenario could be that we have a plausible model for the non-Markovian noise process $\Upsilon_m$, and thus know the expression for the non-Markovian \asf~$\mc{F}_m$~in Eq.~\eqref{eq: average fidelity nonMarkovian}. Then we may construct the Markovian counterpart $\mc{F}_m^\markov=p_1\cdots{p}_mA+B$ of the \asf, compute $\mc{N}_q^{\mc{F}_m}$ in Eq.~\eqref{eq: rb non-Markovianity} and compare with the actual \rb~data.

Perhaps the most common case, however, will be that an \rb~experiment is run without a-priori knowledge of a model for the noise and a non-exponential curve for the \asf~is observed. At the same time, the observed statistics for a given physical process often depend only on a portion of their history rather than on their full past, implying that the relevant temporal correlations in the noise would likely be manifest in \rb~only over a finite sequence length. This notion of a finite memory within the noise will allow us to estimate, in practice, the amount of non-Markovian effects that are being observed in an \rb~experiment, as well as to operationally construct an analogue of a Markovianized \asf, $\mc{F}_m^\markov$, to estimate deviations from Markovianity in \rb.

\section{Models of finite non-Markovian noise}\label{sec: finite}
A possible scenario is to have an underlying noise process that is non-negligibly non-Markovian up to a given finite sequence length, with the remaining noise being effectively almost Markovian. This is related to the notion of finite quantum Markov order~\cite{PhysRevLett.122.140401, PhysRevA.99.042108, Taranto_2020}, which similar to the classical concept of finite Markov order, describes a quantum process where future statistics depend only on a finite number of the previous operations on the system and its outcomes. We have then the following.

\begin{corollary*}[Initial non-Markovian noise]\label{corollary initial}
Let $\rho$ be an initial state on a system-environment, $\syst\env$, composite and let $\mc{S}_{m}^{\ell:1}(\rho)$ describe an \rb~sequence of length $m$ with noise described by \compp~maps $\Lambda_n$ on $\syst\env$ for all $n$ up to a sequence length $\ell<m$, with  the rest of the sequence having noise \compptp~maps $\Lambda^\markov_{\tilde{n}}$ on $\syst$ and associated noise-strengths $p_{\tilde{n}}$. Then the average sequence fidelity (\asf) upon acting with a \povm~element $\mc{M}$ is given by
\begin{align}
\label{eq: asf finite main}
    \mc{F}_m &= \tr\{\mc{M}\,\mbb{E}[\mc{S}_m^{\ell:1}(\rho)]\} \\\nonumber &= p_{\ell+1}\cdots{p}_m\,\tr\left[\mc{M}\Lambda_{m+1}^\markov\circ\tr_\env\circ\mscr{A}_\ell(\rho)\right] + B\tr\left[\mscr{B}_\ell(\rho)\right],
\end{align}
where $B=\tr\left[\mc{M}\Lambda_{m+1}^\markov\left(\mbb1/\dimS\right)\right]$ and $\mscr{A}_n$, $\mscr{B}_n$ are defined in Eq.~\eqref{eq: main curlyA} and Eq.~\eqref{eq: main curlyB}, resp.
\end{corollary*}

This implies that after a sequence length $\ell$, non-Markovian noise will be manifest in an \rb~experiment as \spam~errors and not affect the subsequent decay, which for time-independent noise, would remain exponential. The assumption that the noise suddenly stops acting jointly on $\syst\env$ is at best an approximation, but one that can effectively be used whenever the non-Markovian noise effects are relevant only over some finite sequence length $\ell$.

The main reason why this is important is twofold: first, detecting non-Markovian effects with an \rb~experiment will be most likely be efficient for short sequence lengths, in the sense of requiring a small amount of fidelity samples, since there is no compounding error, so for small $\ell$ any significant non-Markovian noise effects can be resolved through \rb; and second, the time scale of the memory effects displayed by the noise, i.e. the length $\ell$ inherent in the noise process, can then potentially be determined through an \rb~experiment. This would also be related to determining the order of a finite quantum Markov order process~\cite{white2021nonmarkovian}.

\begin{figure}[t!]
\centering
    \includegraphics[width=\linewidth]{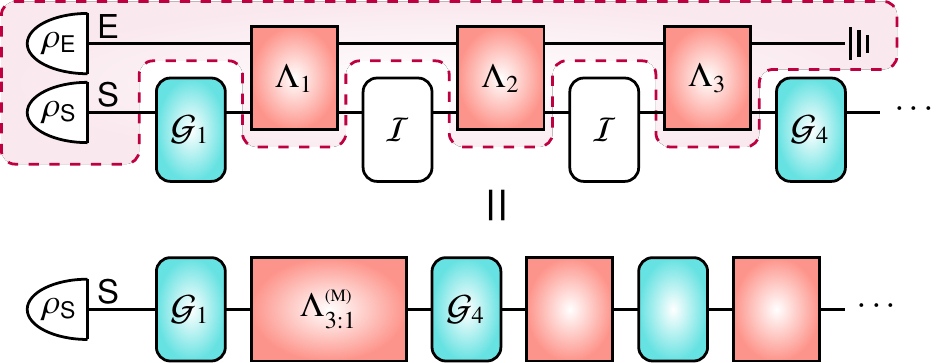} \caption{\textbf{Determining time scales of finite non-Markovian noise.} A noise process with initial finite non-Markovian noise over a sequence length $\ell=5$ will decay as described by a Markovian \asf~after such step, with the non-Markovian part contributing as \spam~error factors. By fixing to identity the gates of at least time-steps, here step 2 and step 3, the decay of the \asf~corresponding to such sequence becomes entirely Markovian; this allows to operationally determine the time scales of finite non-Markovian noise as well as to construct sensible Markovian \asf{s} to quantify \rb~non-Markovianity, as exemplified in Section~\ref{sec: numerics 2}.}
    \label{fig: finite identities}
\end{figure}

In section~\ref{sec: numerics 2} we show one such example where the sequence length $\ell$ of non-Markovian noise can be estimated from an \rb~experiment's data alone, and where a sensible time-independent Markovianized \asf, $\tilde{\mc{F}}_m^\markov$, can be constructed so as to operationally estimate non-Markovian deviations in such an experiment. This follows by noticing the following. Whenever we have finite non-Markovian noise, say over an initial sequence length $\ell$, described by \compptp~maps $\Lambda_n$, and an initial uncorrelated state, by choosing to fix $\ell-1$ Cliffords after the first one to be identities, by Eq.~\eqref{eq: asf finite main}, we get a Markovian decay with
\begin{equation}
    \mc{F}_m = {p}_{\ell+1}\cdots{p}_m(p_{\ell:1}A)+B,
    \label{eq: cor1 initial nM}
\end{equation}
where,
\begin{equation}
    p_{\ell:1} := \f{\tr\left[\Lambda_{\ell:1}^\markov\right]-1}{\dimS^2-1},\quad\,
    \Lambda_{\ell:1}^\markov(\cdot) := \tr_\env\left[\Mcirc_{n=1}^\ell\Lambda_n\left(\rho_\env\otimes\cdot\,\right)\right],
\end{equation}
that is, the initial block of finite non-Markovian noise looks like a single noise map $\Lambda_{\ell:1}^\markov$ if we randomize over a single Clifford within this block, with the remaining ones set to identities. This is more clearly seen in Fig.~\ref{fig: finite identities}. This is, again, at best an approximation, but one that serves effectively to estimate the time scales for non-Markovian noise in an \rb~experiment. Of course, presumably, in realistic cases this would be more complicated and possibly all of the noise process be time-dependent, albeit with small non-Markovianity effects.

Another scenario could be to have an almost Markovian noise initially, up to a sequence length $\ell$, after which non-Markovianity turns significant. Then we have the following.

\begin{corollary*}[Late non-Markovian noise]\label{corollary late}
Let $\rho$ be an initial state on a system-environment, $\syst\env$, composite and let $\mc{S}_{m}^{m:\ell+1}(\rho)$ describe an \rb~sequence of length $m$ with noise described by \compptp~maps $\Lambda_n^\markov$ on $\syst$ and noise strengths $p_n$ for all $n$ up to a sequence length $\ell<m$, with  the rest of the sequence having noise \compp~maps $\Lambda_{\tilde{n}}$ on $\syst\env$. Then the average sequence fidelity (\asf) upon acting with a \povm~element $\mc{M}$ is given by
\begin{align}
    \mc{F}_m &= \tr\{\mc{M}\,\mbb{E}[\mc{S}_m^{m:\ell+1}(\rho)]\}\nonumber\\ &=p_1\cdots{p}_\ell\tr[\mc{M}\,\tr_\env\circ\Lambda_{m+1}\circ\mscr{A}_{m:\ell+1}(\rho)]\nonumber\\
    &\qquad\qquad\qquad+\tr[\mc{M}\,\tr_\env\circ\Lambda_{m+1}\circ\mscr{B}_{m:\ell+1}(\rho)],
\end{align}
 where
\begin{align}
    \mscr{A}_{m:k}(\rho) &:= \f{\displaystyle{\Mcirc_{n=k}^m}\left(\$_{\Lambda_n}-\Theta_{\Lambda_n}\right)\otimes\mc{I}_\syst}{\left(\dimS^2-1\right)^{m-k+1}}\left(\rho-\rho_\env\otimes\f{\mbb1}{\dimS}\right),\\
    \mscr{B}_{m:k}(\rho)&:=\Mcirc_{n=k}^m\Theta_{\Lambda_n}(\rho_\env)\otimes\f{\mbb1}{\dimS},
\end{align}
with $\$_{\Lambda_n}$ and $\Theta_{\Lambda_n}$ defined in Eq.~\eqref{eq: dollar main} and Eq.~\eqref{eq: theta main}, resp.
\end{corollary*}

This case might be relevant in practice whenever the sequence length $\ell$ is relatively small, both because non-Markovian noise would affect relevant computations and because the onset of such non-Markovian deviations could be resolved by an \rb~experiment.

Furthermore, in the middle of this two cases, we have the possibility of noise being intermittently non-Markovian, i.e. being displayed significantly over blocks of some finite sequence length. We have then
\begin{corollary*}[Blocks of finite non-Markovian noise]\label{corollary mixed}
Let $\rho$ be an initial state on a system-environment, $\syst\env$, composite and let $\mc{S}_{m}^{\{m:\ell+1,\ell:1\}}(\rho)$ describe an \rb~sequence of length $m$ with noise given by \compp~maps $\Lambda_n$ on $\syst\env$ for all $n$ up to a sequence length $\ell<m$, then at the $\ell$\textsuperscript{th} step by $\Lambda_\ell(\cdot)\to\varepsilon\otimes\tr_\env\circ\Lambda_\ell(\cdot)$ for some $\env$ state $\varepsilon$, and with the rest of the sequence having noise \compp~maps $\Lambda_{\tilde{n}}$ on $\syst\env$. Then the average sequence fidelity (\asf) upon acting with a \povm~element $\mc{M}$ is given by
\begin{align}
    \mc{F}_m =& \tr\left\{\mc{M}\,\mbb{E}\left[\mc{S}_m^{\{m:\ell+1,\ell:1\}}(\rho)\right]\right\} \notag\\
    =&\tr\left\{\mc{M}\,\tr_\env\circ\Lambda_{m+1}\circ\mscr{A}_{m:\ell+1}\left[\varepsilon\otimes\tr_\env\circ\mscr{A}_\ell(\rho)\right]\right\} \\
    &\quad+ \tr\left[\mscr{B}_\ell(\rho)\right] \tr\left\{\mc{M}\,\tr_\env\circ\Lambda_{m+1}\circ\mscr{B}_{m:\ell+1}\left(\varepsilon\otimes\f{\mbb1}{\dimS}\right)\right\},\notag
\end{align}
with $\mscr{A}_{\ell:k}$ and $\mscr{B}_{\ell:k}$ defined as in Corollary~\ref{corollary late}. 
\end{corollary*}

This turns into a much more complicated \asf, but in essence any other combination considering finite non-Markovian noise can be considered. Of course, experimentally, there would be other challenges involved to study these more complicated finite non-Markovian noise processes, such as being restricted to short sequence lengths and/or requiring a larger amount of observations.

All cases in Corollary~\ref{corollary initial},~\ref{corollary late},~\ref{corollary mixed}, are derived in detail in Appendix~\ref{appendix: finite non-Markovian noise}.  We now turn to study two numerical examples of non-Markovian \rb.

\section{Numerical Model: Two-qubit fully non-Markovian spin noise}\label{sec: numerics 1}
As a proof of principle, we now test Eq.~\eqref{eq: average fidelity nonMarkovian} with a qubit in $\syst$ subject to time-independent unitary noise $\Lambda(\cdot)=\lambda(\cdot)\lambda^\dg$, where $\lambda=\exp(-i\delta H)$, due to interaction with another qubit, identified as $\env$, where $H$ given by the two-spin interaction
\begin{align}
    H &= J\,X_1X_2 + h_x (X_1 + X_2) + h_y (Y_1 + Y_2),
    \label{eq: Ising numerical model}
\end{align}
with $X_i,Y_i$ being Pauli matrices acting on the $i$\textsuperscript{th} site. Even though we use this as a simple theoretical construction and illustration, similar noise dynamics, albeit with many more considerations, come upon in real spin qubit quantum computers, e.g. as undesired crosstalk~\cite{heinz2021crosstalk}.
\begin{figure}[t!]
    \centering
    \includegraphics[width=1.05\linewidth]{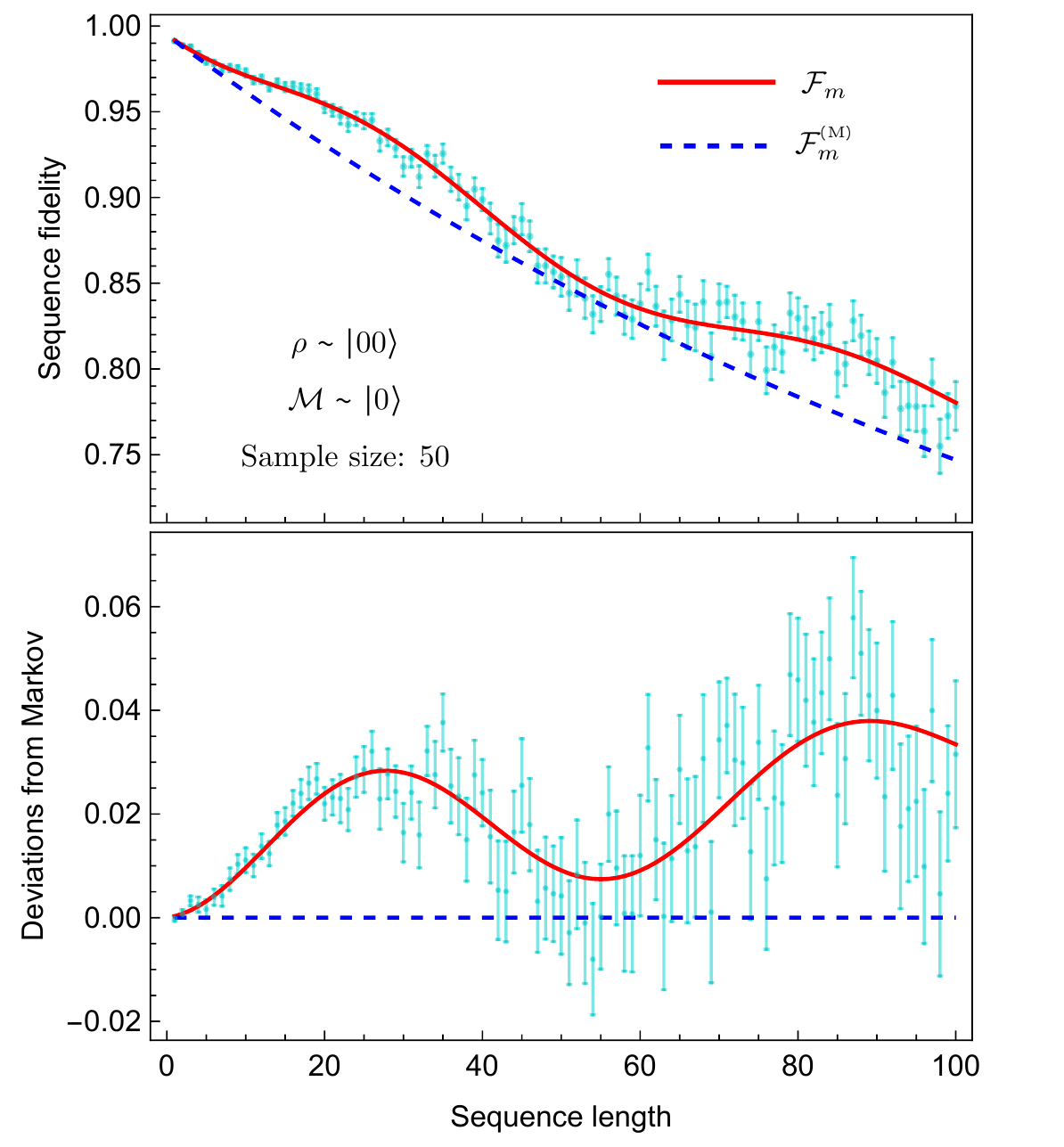}
    \caption{\textbf{{Average sequence fidelity (\asf) for time-independent unitary non-Markovian noise and deviations from its Markovianized counterpart.}} We consider the noise model described by the two-spin interaction of Eq.~\eqref{eq: Ising numerical model} with $\delta\approx0.03$, $J=1.7$, $h_x=1.47$ and $h_y=-1.05$, for a single qubit as system $\syst$. We take $\rho=|00\rangle\!\langle00|$ and  $\mc{M}=|0\rangle\!\langle0|$. \emph{Top}: The continuous (red) line denotes the analytical \asf~given by Eq.~\eqref{eq: average fidelity nonMarkovian}, with each point joined for clarity, the dots denote the numerical average of the \asf~over 50 samples, with bars being the standard deviation of the mean (uncertainty of the numerical mean from the true mean), and the dashed (blue) line denotes the analytical \asf~of the Markovianized process with time-independent noise $\Lambda^\markov(\cdot)=\tr_\env\circ\Lambda(\varepsilon\otimes\,\cdot\,)$, here with $\varepsilon=|0\rangle\!\langle0|$. \emph{Bottom}: Deviations of the \asf~by both the analytical data (continuous red line) and numerical data produced by Eq.~\eqref{eq: average fidelity nonMarkovian}, from the Markovianized \asf, $\mc{F}_m^\markov$ (dashed blue line).
    }
    \label{fig: plots main}
\end{figure}

We take $J=1.7$, $h_x=1.47$ and $h_y=-1.05$ arbitrarily, for which we compute the \asf~$\mc{F}_m$ as a function of $m$, both by numerical averaging and employing Eq.~\eqref{eq: average fidelity nonMarkovian} with $\delta=0.029475$. We take $\rho=|00\rangle\!\langle00|$ and $\mc{M}=|0\rangle\!\langle0|$ and ignore \spam~errors. We display the results in Fig.~\ref{fig: plots main} together with its Markovianized \asf, $\mc{F}_m^\markov$, whereby the time-independent noise is modeled as a \compp~map given by $\Lambda^\markov(\cdot)=\tr_\env\circ\Lambda(\varepsilon\otimes\,\cdot\,)$; specifically there we perform the numerical average over 50 samples of numerical sequence fidelities computed by sampling Haar random one-qubit unitaries, with the bars denoting the standard deviation of the mean.
\begin{figure}[t!]
\centering
    \includegraphics[width=1.05\linewidth]{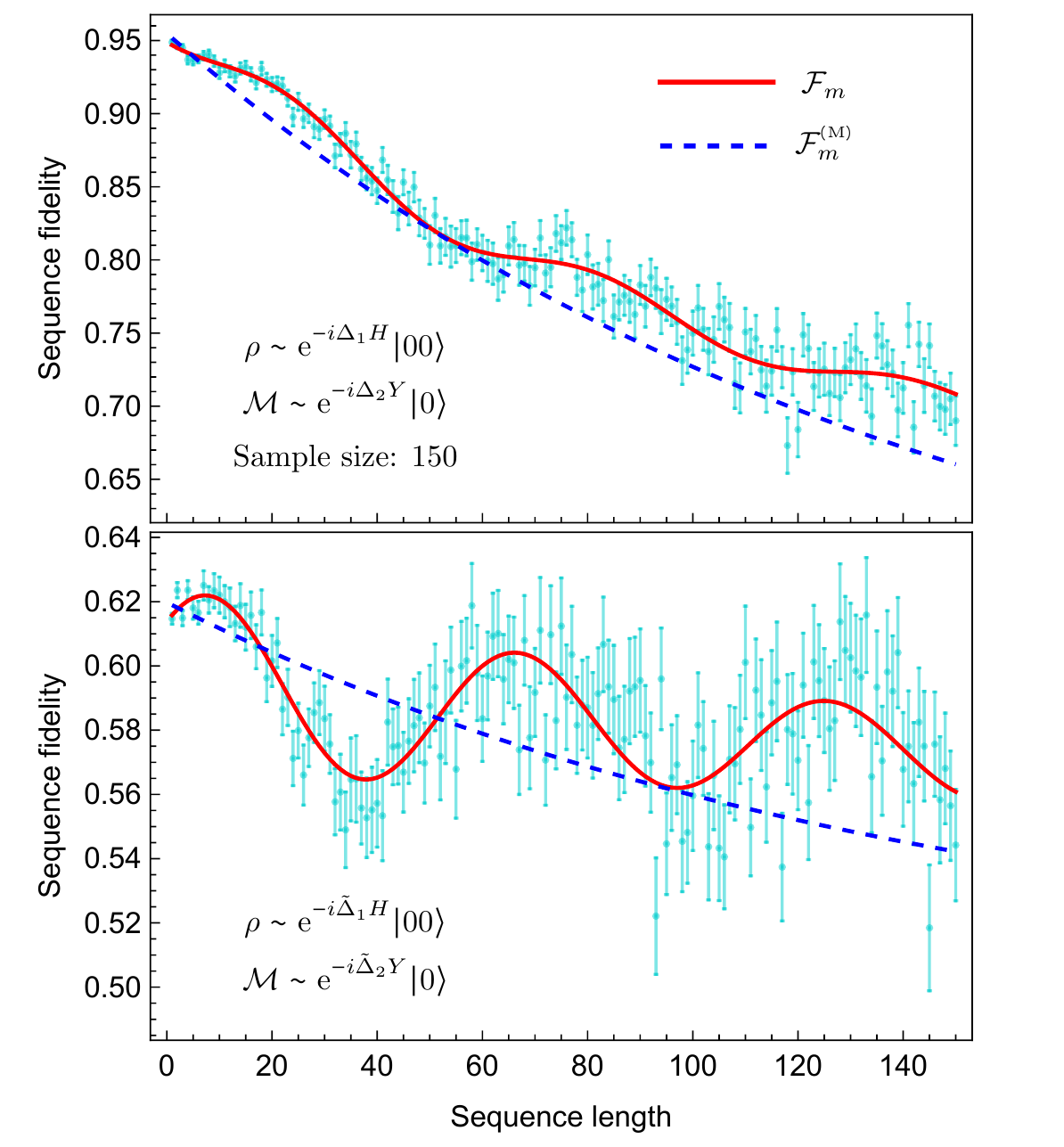}
    \caption{{\textbf{Effect of \spam~errors in the two-qubit spin noise in Eq.\eqref{eq: Ising numerical model}.}} We consider the noise model described by the two-spin interaction of Eq.~\eqref{eq: Ising numerical model} with $\delta\approx0.03$, $J=1.7$, $h_x=1.47$ and $h_y=-1.05$, for a single qubit as system $\syst$. On both plots, the continuous (red) line denotes analytical \asf~in Eq.~\eqref{eq: average fidelity nonMarkovian}, with each point joined for clarity, dots denote numerical average of the \asf~over 150 samples, with bars being the standard deviation of the mean (uncertainty of the numerical mean from the true mean), and the dashed (blue) line denotes the analytical \asf~of the Markovianized process. \emph{Top}: the initial state $\rho=|00\rangle\!\langle00|$ is affected by the sequence noise $\Lambda\sim\exp(-i\Delta_1{H})$ for a small $\Delta_1\approx0.04$ and $\mc{M}=|0\rangle\!\langle0|$ is slightly rotated via $\Lambda\sim\exp(-i\Delta_2{Y})$ with a small $\Delta_2\approx0.09$. \emph{Bottom}: $\tilde{\Delta}_1\approx0.29$ and $\Delta_2\approx0.10$ are increased considerably, amounting to large \spam~errors. In all cases the sample size is 100.}
    \label{fig: main spam}
\end{figure}

We can verify that Eq.~\eqref{eq: average fidelity nonMarkovian} effectively predicts the correct \asf, which is a rather complicated decaying function of $m$, clearly non-exponential. The numerical data remains reasonably well around the analytical prediction, with deviations becoming apparent for larger sequence lengths, which can be understood as compounded error. Despite these deviations being relatively small, they are significant enough that they can be probed numerically with a reasonable sample size for small sequence lengths, say for at least $m\lesssim50$. This also makes manifest that for larger sequence lengths, many more sample runs would be needed to reveal non-Markovianity deviations. The \rb~non-Markovianity, $\mc{N}_q^{\mc{F}_m}$ with respect to the Markovian counterpart can also be swiftly computed through the sum of absolute values of the differences between $\mc{F}_m$ and $\mc{F}_m^\markov$; in particular in Fig.~\ref{fig: plots main} the \rb~non-Markovianity is not particularly high (between $\mc{N}_1^{\mc{F}_{100}}\approx2.1$ and $\mc{N}_\infty^{\mc{F}_{100}}\approx0.04$) but it is enough to be distinguished numerically for small sequence lengths.

Let us now consider the effect of \spam~errors. Suppose the initial state $\rho$ is affected by the same $\Lambda$ error for some small $\delta=\Delta_1$, and that $\mc{M}$ is slightly rotated via $\exp(-i\Delta_2{Y})$ for a small $\Delta_2$. In Fig.~\ref{fig: main spam} we show examples for both mild, $\Delta_1=0.04232$ and $\Delta_2 = 0.09321$, and much stronger noise with, $\tilde{\Delta}_1 = 0.2932$ and $\tilde{\Delta}_2 = 0.10321$.

In Appendix~\ref{appendix: numerics}, we also show the case where the preparation affects only $\syst$ by some rotation $\exp(-i\gamma{X})$ with a small $\gamma$, but somehow does not generate correlations with $\env$. Similar to Markovian noise models, add an offset to the average fidelities. In the non-Markovian case, however, the error rates do seem to be affected, presumably mainly because of the initial correlations induced by the preparation errors, as argued before in Section~\ref{sec: main average sequence fidelity}. This is still an aspect that would need to be examined closely, as when \spam~errors are significant, the offset also appears larger in the non-Markovian case, making it more difficult to distinguish non-Markovian errors from Markovian ones numerically.

We also notice in Appendix~\ref{appendix: numerics} that the non-Markovian effect of deviating from an exponential seems to fade in increasing $\env$-qubits; this is expected but this too would need to be thoroughly studied in realistic scenarios where the dimension of the environment is effectively finite~\cite{PhysRevLett.120.030402, PhysRevLett.122.160401}. On the other hand, we notice as well that an $XX$-spin chain displays practically no deviations from an exponential \asf~decay presumably because of the absence of the external field, i.e. while the noise is non-Markovian, $\mc{N}\neq0$, \rb~displays only minimal deviations, $\mc{N}_q^{\mc{F}_m}\approx0$, and the behavior is almost exponential for all sequence lengths.

While this is mainly a numerical test of our main result, we now show an example and propose how to analyze a plausible realistic scenario for an \rb~experiment displaying finite non-Markovian noise, and having no prior knowledge of a model for such noise.

\section{Numerical example: noise memory time scales, Markovianized average sequence fidelity and coherent noise}\label{sec: numerics 2}
Consider now again a pair of qubits which up to some sequence length $\ell$ display an \asf~that is mostly non-Markovian and subsequently turns almost Markovian. Here we model the underlying noise with
\begin{equation}
    \Lambda_n^{(\ell)} = q_{n-\ell} \Lambda + (1-q_{n-\ell}) \Lambda^\markov,
    \label{eq: lambda finite}
\end{equation}
where $q_{n-\ell} = [1+\exp(n-\ell)]^{-1}$ and both $\Lambda$, $\Lambda^\markov$ are determined as in the previous example with Eq.~\eqref{eq: Ising numerical model} with the same constants, $J=1.7$, $h_x=1.47$ and $h_y=-1.05$, but we now fix $\delta\approx0.03$ for $\Lambda$ and $\delta^\markov=2.5\delta$ for $\Lambda^\markov$. In particular, we notice that $q_k$ converges rapidly to $1$ for $k<0$, i.e. for a sequence lengths $m<\ell$, similarly converges rapidly to $0$ for the remaining $k>0$, meaning sequence lengths $m>\ell$, and finally $q_k=0.5$ at $k=0$, i.e. for a sequence length $m=\ell$.
\begin{figure}[t!]
\centering
\includegraphics[width=1.05\linewidth]{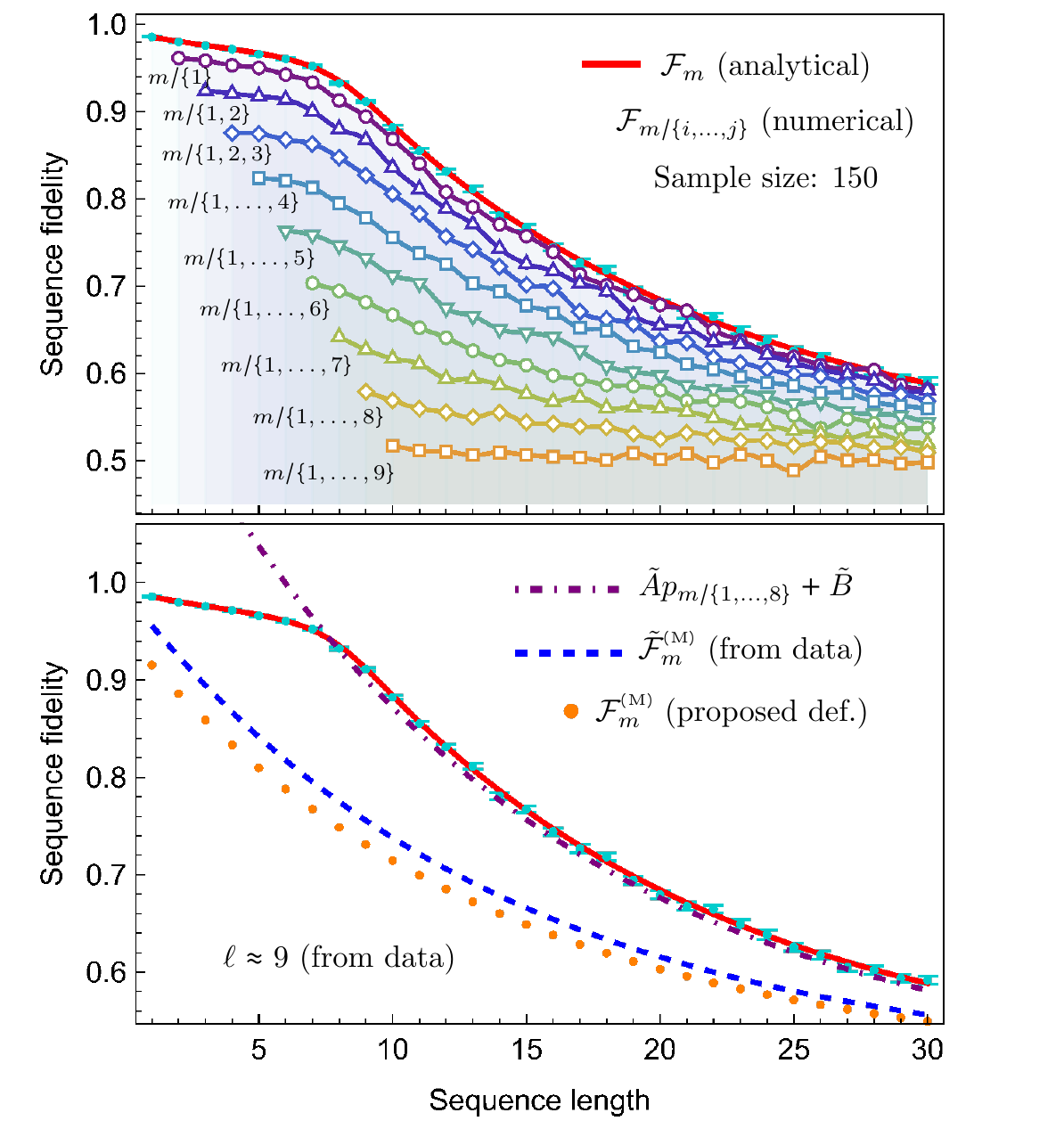}
    \caption{\textbf{{Determining the sequence length of finite non-Markovian noise.}} \emph{Top}: An \rb~experiment might display deviations from an exponential over a finite sequence length, as shown by the first \asf~from the top, with the continuous red line denoting the underlying analytical \asf, $\mc{F}_m$. Such non-Markovian noise sequence length can be determined experimentally by fixing Cliffords to identities, running corresponding \rb~protocols to obtain \asf{s} $\mc{F}_{m/\{i,\ldots,j\}}$, where $\{i,\ldots,j\}$ denotes steps taken to identity (shown joined in the plot for clarity), fitting exponentials to these, and approximately matching their decay rates $p_{m/\{i,\ldots,j\}}$ with the one in the manifestly exponential part in the original data. \emph{Bottom}: The non-Markovian noise sequence length was determined to be $\ell\approx9$; the dot-dash purple line denotes the curve with the decay rate $p_{m/\{1,\ldots,8\}}$ and constants $A$, $B$ of the fitted exponential of the original data starting at $m=9$. Once $\ell$ is determined, a sensible Markovianized \asf, shown as a dashed blue line, can be taken with $p=p_{m/\{1,\ldots,8\}}$ and reasonable criteria for fixing $A$, $B$; here we choose $A\approx{B}$ assuming low \spam~errors. The analytical Markovianized time-independent \asf~of the form we proposed in Section~\ref{sec: nm RB measure} is shown with orange dots just as a comparison.}
    \label{fig: plots finite MO}
\end{figure}

Henceforth we assume that an experimenter would not know both what the noise maps $\Lambda_n^{(\ell)}$ are, nor what the non-Markovian finite sequence length $\ell$ is. Given Corollary~\ref{corollary initial}, however, we know that whenever we have finite time-independent non-Markovian noise, within the Markovian part the decay will be practically exponential with the non-Markovian part acting as \spam~errors. Specifically, here we would get an \asf~of the form of Eq.~\eqref{eq: cor1 initial nM} for almost time-independent noise (i.e. with almost equal noise-strengths $p_{\ell+1}\approx\ldots\approx{p}_m$) after such sequence length $\ell$. Our expression assumes that the transition to Markovian noise occurs from step to step, however, even if dissipation occurs smoothly and non-Markovianity never entirely fades, we can still estimate at which sequence length the memory of the noise stops being relevant by identifying exponential decays. This also allows to identify a Markovianized time-independent \asf~with which the experimenter can estimate the impact of non-Markovian errors.

A way to achieve this in practice is by fixing Clifford gates to identity wherever the decay appears non-exponential; this will give an exponential decay of the \asf~whenever there is at most one random Clifford within the non-Markovian sequence. In Fig.~\ref{fig: plots finite MO} we display the \asf, $\mc{F}_m$, both analytical and numerical, for a finite noise memory process with noise modeled by Eq.~\eqref{eq: lambda finite}, again taking $\rho=|00\rangle\!\langle00|$ and $\mc{M}=|0\rangle\!\langle0|$. We also display numerical \asf{s}, denoted $\mc{F}_{m/\{i,\ldots,j\}}$, with fixed identities at sequence lengths $i,\ldots,j$. The corresponding \asf{s}~$\mc{F}_{m/\{i,\ldots,j\}}$ will normally be decreasing as $\mc{F}_m>\mc{F}_{m/\{1\}}>\mc{F}_{m/\{1,2\}}>\cdots>\mc{F}_{m/\{1,2,\ldots,10\}}$ given that fixing identities at subsequent steps is equivalent to set compounding error over such steps, which can be thought of simply as leaving the noise as a dynamical process to accumulate in time.

The non-Markovian sequence length can be identified by matching approximately the decay rate $p_{m/\{i,\ldots,j\}}$ of one of these sequences $\mc{F}_{m/\{i,\ldots,j\}}$ with the corresponding one of the manifestly Markovian part in the full sequence. Once the decay rate has been determined, a sensible time-independent Markovianized \asf, $\tilde{\mc{F}}_m^\markov$, can be constructed by making reasonable assumptions for the \spam~factors $A$ and $B$. Detail of this process is shown in Appendix~\ref{appendix: numerics}. For the case of the \rb~experiments in Fig.~\ref{fig: plots finite MO}, the non-Markovian noise sequence length was determined to be $\ell\approx9$ by approximately matching $p_{m/{1,\ldots,8}}$ with the corresponding one for the exponential fit between sequence lengths $15\leq{m}\leq{30}$ of the original data. Notice that in our model in Eq.\eqref{eq: lambda finite}, at sequence length $m=9$ the noise still has half probability of acting jointly on $\syst\env$; the found $\ell\approx9$ just says that after such sequence length the decay turns mostly exponential. We then finally constructed a Markovianized \asf~with $\tilde{\mc{F}}_m^\markov={A}p_{m/\{1,\ldots,8\}}+B$ with $A\approx{B}$ supposing \spam~errors to be small; we compared this with a Markovianized construction as proposed in Section~\ref{sec: nm RB measure}, with time-independent noise given throughout by $\Lambda^\markov$.

This practical approach can work reasonably well, as we show in this example, and allow both to determine the amount of memory within the noise, i.e, for how long the noise is being meaningfully non-Markovian, as well as to operationally construct a time-independent Markovianized \asf~with which the impact of non-Markovianity in the noise can be quantified. The approach is consistent as well, in the sense that applying it to an exponential decay yields $\ell=1$ and at most a numerical error due to fixing an identity on the first step.

\begin{figure}[t!]
\includegraphics[width=1.05\linewidth]{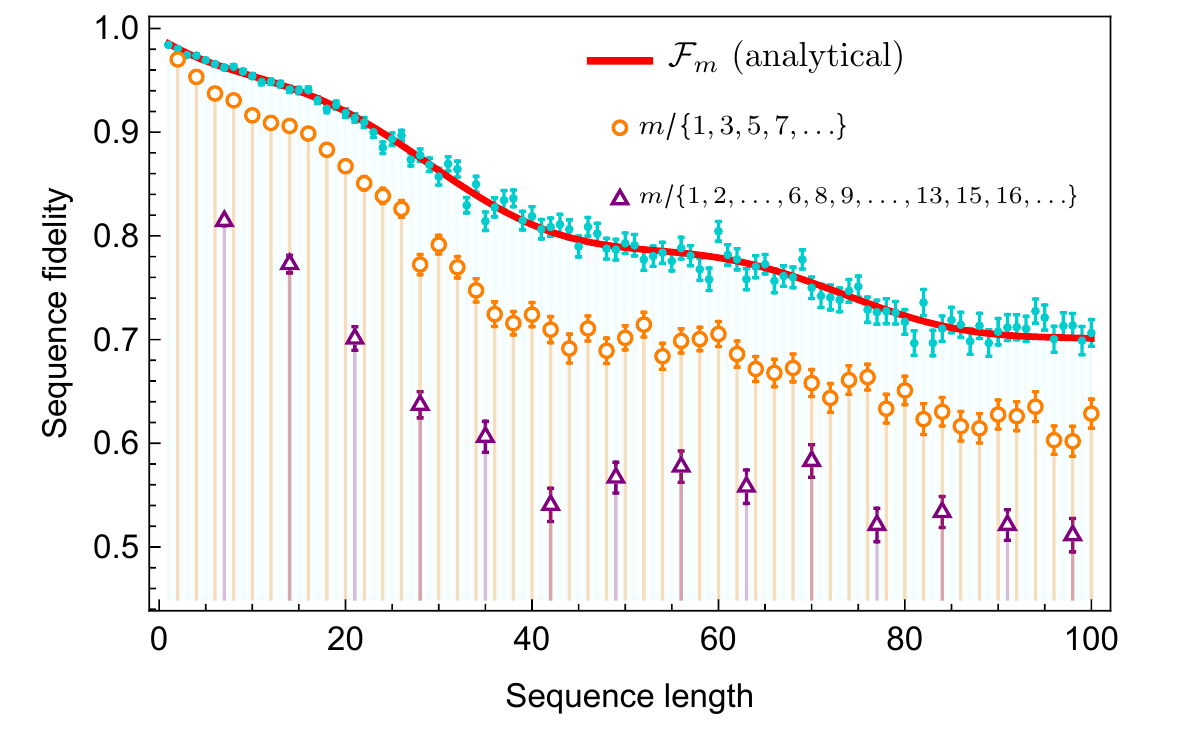}
    \caption{\textbf{Diagnosing coherent non-Markovian noise.} Interleaving identities can allow to determine whether non-Markovian noise is coherent: in a continuous red line we show the analytical \asf~with the same model of Eq.~\eqref{eq: Ising numerical model}, with $\rho\sim|00\rangle$, $\mc{M}\sim|0\rangle$, and same parameters as displayed in Fig.~\ref{fig: plots main}, with numerical data displayed by teal points with bars denoting uncertainty around the mean; in circles and triangles are shown numerical \asf{s} $\mc{F}_{m/\{i,\ldots,j\}}$ with Cliffords at time-steps $\{1,3,5,\ldots\}$ and $\{1,2,\ldots,6,8,9,\ldots,13,15,16,\ldots\}$ set to identity, resp. Here we show only two examples of $\mc{F}_{m/\{i,\ldots,j\}}$, not describing an exponential decay, for clarity, but similar behavior occurs interleaving identities at any set of steps. Numerical averages were done with 200 sequence fidelity samples.}
    \label{fig: plot inteleaved ids}
\end{figure}

There are, however, two apparent downsides to this approach, one is having to run another set of experiments requiring a higher amount of samples, given that the noise accumulates and makes it harder to get reliable data, and the second is that the \asf{s} with fixed identities $\mc{F}_{m/\{i,\ldots,j\}}$ can eventually get too low if the noise memory is too high and not provide useful information. These are issues that could be resolved easily or otherwise depending on the particular case at hand.

Finally, while this approach cannot be used generally on fully non-Markovian noise, i.e. one over all sequence lengths, to determine operationally a sensible Markovianized \asf~$\tilde{\mc{F}}_m^\markov$, it can nevertheless tell us whether the non-Markovian noise we are dealing with is coherent. This is important because whenever coherent noise can be diagnosed and characterized, e.g. with via unitarity measures~\cite{PhysRevLett.121.170502, Wallman_2015, PhysRevA.99.012315, girling2021estimation} or otherwise, in principle it could be addressed and calibrated if we have access to the $\env$ qubits. Precisely then, we may tell if the noise is unitary over the whole $\syst\env$ if we get a general non-exponential behavior described by Eq.~\eqref{eq: average fidelity nonMarkovian} no matter how many identities we fix, or if some dissipation is occurring and we rather have a scenario closer to that of Corollary~\ref{corollary mixed} of finite non-Markovian blocks of noise. We use the model of the previous section in Eq.~\eqref{eq: Ising numerical model} to exemplify this, as shown in Fig.~\ref{fig: plot inteleaved ids}. The way we can proceed is to run \rb~experiments with a given number of identities interleaved; if the deviations from an exponential disappear, or fade considerably, this might point out to some dissipation, otherwise we would be able to identify the noise as highly coherent. Here once again the challenge is rather with numerical precision and compounded error, as interleaving identities highly degrades the \asf.

\section{\label{sec: conclusions}Conclusions and Discussion}
We have, \emph{i}) derived a general analytical expression for the average sequence fidelity (\asf) of a randomized benchmarking (\rb) experiment with the Clifford group subject to gate-independent non-Markovian noise, \emph{ii}) proposed a theoretical measure to quantify non-Markovian deviations in an \asf, \emph{iii}) derived the \asf~for the case of finite non-Markovian noise, allowing to operationally estimate both non-Markovian noise time scales and the measure of deviations from Markovianity, and \emph{iv}) exemplified all these with two proof of principle numerical examples. Along the manuscript we also discuss the effect of state preparation and measurement (\spam) errors in \rb~with non-Markovian noise, as well as the case of classical correlations, which we argue can be treated as a Markovian time-dependent problem, and more generally the idea of \rb~blindness to a subclass of non-Markovian noise processes.

\begin{figure}[t!]
\centering
	\includegraphics[width=\linewidth]{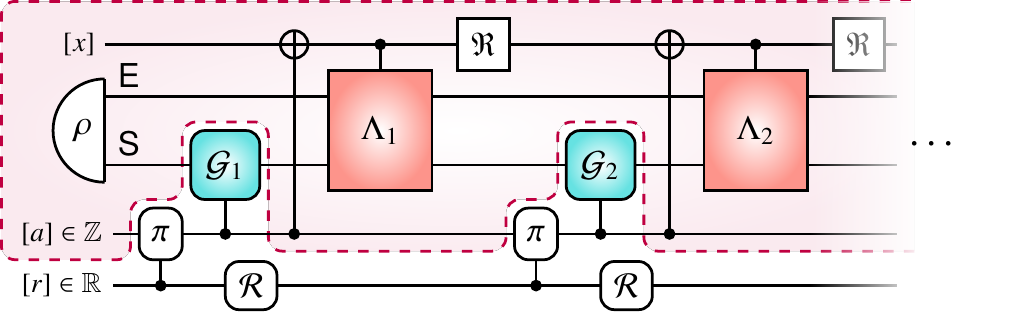}
	\caption{\textbf{{A circuit describing both gate-dependence and non-Markovianity.}} The full system consists of a $\syst\env$ system in state $\rho$ and classical registers $[x]$, $[a]$ and $[r]$. The $\pi$ operations are permutations, $\mc{R}$ are randomizing operations and $\mathfrak{R}$ are reset operations. Vertical lines joining with $\bullet$ denote control with the corresponding classical register: essentially the \textsf{\textsc{cnot}} gates (control with $\boldsymbol{\oplus}$ in the $[x]$ extreme) will carry the dependence from the applied gates at every step. Finally in \rb~the inverse operation (with an associated error) would be applied and a measurement in $\syst$ would be performed.}
	\label{fig: gate dependence}
\end{figure}

The \asf~in our main result makes the depolarizing effect of averaging over Clifford gates on the system of interest manifest, while taking all of the noise in the sequence to the environment. The reduction of our main result to the standard Markovian gate-independent scenario is then straightforward with the trace over the environment giving rise to the noise strength and the \spam~error constants, and similarly one may consider cases where non-Markovian noise is finite over a subset of sequence lengths. Our main result also makes it clear that in general, non-Markovian noise will display non-exponential behavior, although we point out that there could be a subclass of non-Markovian models that do display an almost exponential decay that in practice would be almost impossible to resolve.  We exemplified numerically how for small sequence lengths, deviations from Markovianity can be observed efficiently, as well as how the relevant time-frames for finite non-Markovianity can be operationally determined and non-Markovian deviations in the \asf~quantified.

We highlight as well that the methods to quantify non-Markovian effects, determine memory time scales, and diagnose coherence of non-Markovian noise, could be implemented beyond the randomized benchmarking framework on other noise benchmarking, characterization or mitigation approaches whenever temporal correlations should be taken into account, e.g. for cross-talk or leakage errors~\cite{PhysRevA.100.032325, arm2020learningbased, ParradoRodriguez2021crosstalk}, similar to how it is done within the framework of resource theories in Ref.~\cite{berk2021extracting} with the notion of temporal coarse-graining.

Needless to say, there are countless ways to move forward in the study of time-correlated errors in quantum computing. Arguably, the clearest ones arising from our manuscript within the \rb~procedure, would be to have a model-independent \asf, similar to the Markovian case, to benchmark other experimentally relevant groups or more generally arbitrary gate sets, to study context-dependent errors, with one possible way being the one we propose in Fig.~\ref{fig: gate dependence}, or to incorporate non-Markovianity in the general \rb~framework carefully constructed in Ref.~\cite{helsen2020general}. All of these extensions have already been studied in quite some depth for Markovian errors and doing the same for the non-Markovian case would be a natural step forward. Other than this, there are questions that still would need to be understood such as the impact of non-Markovianity in decay rates as a function of sequence length, or explicitly how a higher or lower amount of non-Markovianity affects the average gate fidelity. More generally, there is still a need to deeply understand errors arising from temporal correlations, and our result represents a step in this direction.

\begin{acknowledgments}
We thank Felix A. Pollock for conversations. KM is supported through Australian Research Council Future Fellowship FT160100073 and Discovery Project grant DP210100597. KM was a recipient of the International Quantum U Tech Accelerator award by the US Air Force Research Laboratory. RH is supported by the Australian Research Council Centre of Excellence for Engineered Quantum Systems (Grant No. CE 170100009).
\end{acknowledgments}

\onecolumngrid
\appendix

\section{\label{appendix: process tensor}The process tensor of the noise and gate sequences}
The process tensor is a multi-linear map taking \compp~maps as input and giving a single quantum state as output. The operational scenario is the following: an initial quantum state $\rho$ on the joint $\syst\env$ composite is acted on with an operation $\mc{G}_1$ solely on system $\syst$, which in general is given by a \compp~map; subsequently the whole composite evolves unitarily through a unitary map $\mc{U}_1$, after which an operation $\mc{G}_2$ is performed on $\syst$, then the whole evolves unitarily under a unitary map $\mc{U}_2$, and so on, until an intervention $\mc{G}_k$, followed finally by a unitary map $\mc{U}_k$. This means the final state in system $\syst$ will be given by
\begin{equation}
    \rho_\mathsf{S}^{(k)}=\tr_\mathsf{E}\left[\left(\Mcirc_{i=1}^{k}\mc{U}_i\circ\mc{G}_i\right)\,\rho\right],
    \label{eq: process tensor open}
\end{equation}
where here we implicitly write $\mc{G}_\ell$ for $\mc{I}_\mathsf{E}\otimes\mc{G}_\ell$. The process tensor is thus a map $\mc{T}_{k:1}:\mscr{B}(\mscr{H}_\mathsf{S})^{\otimes{2k}}\to\mscr{B}(\mscr{H}_\mathsf{S})$, where $\mscr{B}(\mscr{H})$ means space of bounded linear operators over the Hilbert space $\mscr{H}$, taking $k$ \compp~maps as arguments and giving a quantum state as output at time-step $k$, i.e.
\begin{equation}
    \mc{T}_{k:1}\left[\vec{\mc{G}}_{k:1}\right]=\rho_\mathsf{S}^{(k)},
\end{equation}
where $\vec{\mc{G}}_{k:1}=(\mc{G}_1,\mc{G}_2,\ldots,\mc{G}_k)$. Such operations $\mc{G}_i$ are said to form an intervention and belong to an instrument, which can be understood as a generalization of a \povm, and the particular outcomes of each intervention yield a joint probability distribution describing a stochastic process.

\begin{figure}[ht!]
    \centering
    \includegraphics[width=0.5\textwidth]{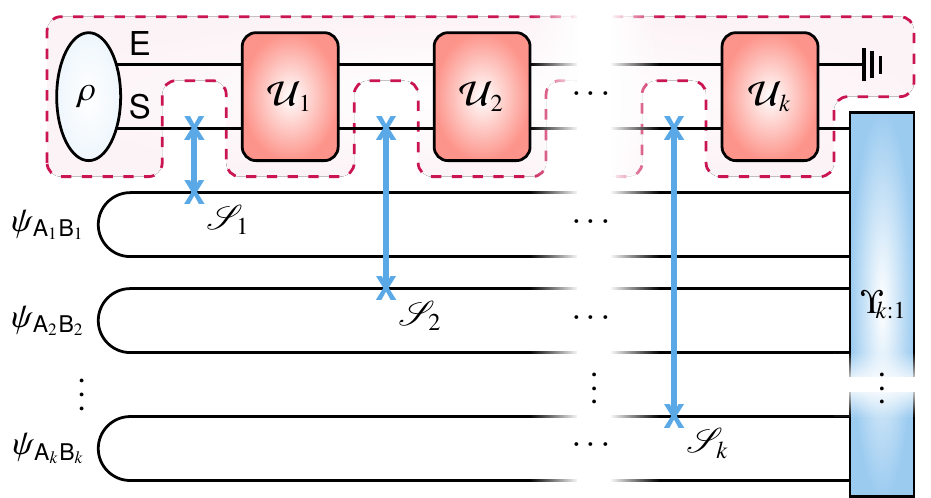}
    \caption[Choi state representation of the process tensor]{{\textbf{The Choi-state representation of a $k$-step process tensor}}, denoted $\Upsilon_{k:1}$, can be obtained by swapping out the system, $\mathscr{S}_i$ with half a maximally entangled state, $\psi_{\mathsf{A}_i\mathsf{B}_i}$, at each step $i$. The final state is an unnormalized many-body state acting on a $d_\mathsf{S}^{2k+1}$ dimensional system.}
    \label{fig: Choi process tensor}
\end{figure}

The generalization of a Choi-state, as given by the Choi-Jamio\l{}kowski isomorphism~\cite{watrous2018theory, bengtsson2017geometry}, for a $k$-step process tensor follows by introducing $k$ maximally entangled states $\psi_{\mathsf{A}_i\mathsf{B}_i}\in\mscr{B}(\mscr{H}_{\mathsf{A}_i}\otimes\mscr{H}_{\mathsf{B}_i})$, where $\mscr{H}_{\mathsf{A}_i}\cong\mscr{H}_\mathsf{S}$ and similarly for $\mathsf{B}$, and letting half of each (that is, the part of either subspace $\mathsf{A}_i$ or $\mathsf{B}_i$) act as an input at every step by swapping the input spaces with the corresponding auxiliary space. This is more clearly illustrated in Fig.~\ref{fig: Choi process tensor}.

Specifically, the Choi-state of the process tensor takes the form
\begin{equation}
    \Upsilon_{k:1}=\tr_\env\left[\left(\Mcirc_{i=1}^k\mc{U}_i\circ\mscr{S}_i\right)\rho\otimes\psi^{\otimes{k}}\right],
    \label{eq: process tensor Choi state}
\end{equation}
where here we are implicitly writing $\mc{U}_i$ for $\mc{U}_i\otimes\mc{I}_{\mathsf{A}_1\mathsf{B}_1\cdots\mathsf{A}_k\mathsf{B}_k}$ and $\psi^{\otimes{k}}=\psi_{\mathsf{A}_1\mathsf{B}_1}\otimes\cdots\otimes\psi_{\mathsf{A}_k\mathsf{B}_k}$. The generalized swap $\mscr{S}_i$ between system $\syst$ and an auxiliary space $\mathsf{A}_i$ at time-step $i$ is defined by $\mscr{S}_i(\cdot) := \varsigma_i(\cdot)\varsigma_i$, where here
\begin{equation}
    \varsigma_i\!:=\!\!\sum_{\ell,j=1}^{d_\mathsf{S}}\mc{I}_\mathsf{E}\otimes|\ell\rangle\!\langle{j}|\otimes\mc{I}_{\mathsf{A}_1\mathsf{B}_1\cdots\mathsf{A}_{i-1}\mathsf{B}_{i-1}}\otimes|j\rangle\!\langle\ell|\otimes\mc{I}_{\mathsf{B}_i\mathsf{A}_{i+1}\mathsf{B}_{i+1}\cdots\mathsf{A}_k\mathsf{B}_k}.
\end{equation}

Similar to the case of quantum channels, the isomorphism between the action of the process tensor and its Choi representation is manifest~\cite{figueroaromero2021equilibration} through the relationship
\begin{equation}
    \mc{T}_{k:1}\left[\vec{\mc{G}}_{k:1}\right] = \tr_{\overline{\syst}}\left[\Upsilon_{k:1}\left(\mbb1_\syst\otimes\mathfrak{Y}_{k:1}^\mathrm{T}\right)\right],
    \label{eq: process tensor Choi}
\end{equation}
where here
\begin{equation}
    \mathfrak{Y}_{k:1}=\left(\Motimes_{i=1}^k\mbb1_{\mathsf{A}_i}\otimes\mc{G}_i\right)\psi^{\otimes{k}},
\end{equation}
is the Choi-state for the operations $\vec{\mc{G}}_{k:1}$, the notation $\tr_{\overline{\syst}}$ stands for partial trace over all except output in $\syst$, and $\mathrm{T}$ denotes a transpose.

\begin{figure}[ht!]
    \centering
    \includegraphics[width=0.5\textwidth]{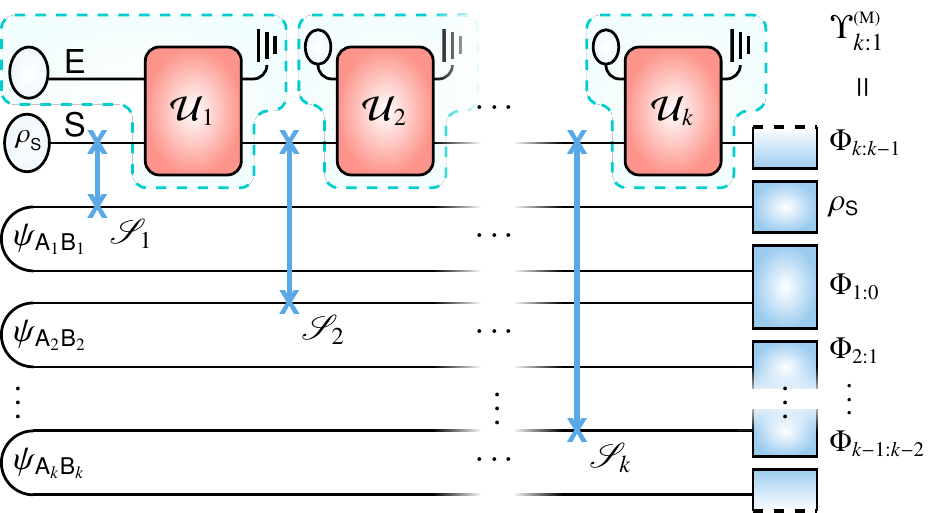}
    \caption{{\textbf{The process tensor for a Markov process $\Upsilon^\markov_{k:1}$}}, having no temporal correlations, can be described by the Choi-state $\Upsilon^\markov_{k:1}$ taking a product of Choi-states $\Phi_{j:i}$ connecting adjacent time-steps $i$ to $j$. Each $\Phi_{j:i}$ corresponds to a \compp~map describing the (open) system evolution between time-step $i$ and $j$.}
    \label{fig: Markovian processes}
\end{figure}

The tensor $\mathfrak{Y}_{k:1}$ is an example of a Markovian process tensor in the sense that it does not have any temporal correlations and thus assumes a product form. For a Markovian dynamical $k$-step process, $\Upsilon^\markov_{k:1}=\Motimes_{i=1}^k\Phi_{i:i-1}\otimes\rho_\syst$, the Choi-states $\Phi_{j:i}$ can either correspond to a closed system dynamics between steps $i$ and $j$, or in general to a \compp~dynamics, e.g. that of an open system, where the environment is discarded between each step and no information passes on to the next step, as shown in Figure~\ref{fig: Markovian processes}. The order of the spaces will be relevant whenever two tensors are contracted and can be written generally through swaps with auxiliary spaces and half maximally entangled states.

In this manuscript we describe a noise \rb~sequence as a process tensor with the dynamics being described by the noise at each step, $\mc{U}_i\to\Lambda_i$. Explicitly, in the main text we denote the tensor Choi-state for a $m+1$-step process for the noise as
\begin{equation}
    \Upsilon_m = \tr_\env\left\{\left[\Mcirc_{i=1}^{m+1}(\Lambda_i\otimes\mc{I}_\mathsf{aux})\circ\mathscr{S}_i\right]\,\rho\otimes\psi^{\otimes m+1}\right\},
    \label{eq: noise process tensor}
\end{equation}
where $\mc{I}_\mathsf{aux}$ is an identity map on an auxiliary space $\mathsf{aux}=\mathsf{A}_1\mathsf{B}_1\cdots\mathsf{A}_{m+1}\mathsf{B}_{m+1}\simeq\syst^{\otimes2(m+1)}$ composed of $m+1$ pairs of $\syst$ systems, $\mathscr{S}_i$ is a swap gate between $\syst$ and one of these pairs in the $i$\textsuperscript{th} auxiliary space, say $\mathsf{A}_i$, and $\psi := \sum|ii\rangle\!\langle{jj}|$ is an unnormalized maximally entangled state. The tensor state for the gate sequence, on the other hand, can be defined simply
\begin{equation}
    \mathfrak{C}_m = \mbb1_\syst\otimes\left[\Motimes_{i=1}^{m+1}\left(\mc{I}_{\mathsf{A}_i}\otimes\mc{G}_i\right)\right]\psi^{\otimes{m+1}},
    \label{eq: gate sequence PT}
\end{equation}
where strictly speaking, here we are defining $\mathfrak{C}_m=\mbb1\otimes\mathfrak{A}_m$, where $\mathfrak{A}_m$ is the Choi state of the sequence of gates $\mc{G}_i$, and $\mc{G}_i$ can be defined to act on either auxiliary space $\mathsf{A}_i$ or $\mathsf{B}_i$, the choice only depends on what auxiliary space the swap $\mscr{S}_i$ on the definition of $\Upsilon_m$ swaps with, so that the contraction $\Upsilon_m\mathfrak{C}_m^\mathrm{T}$ contracts the correct spaces.

\section{\label{appendix: average gate sequence}Average gate sequence}
Representing the noise and gate sequences as process tensors implies that computing the \asf~just requires computing the average gate sequence, $\mbb{E}\left(\mathfrak{C}_m^\mathrm{T}\right)$.

Whenever the gates entering this sequence belong to at least a unitary 2-design, we can simply replace the average over gates by that over unitaries distributed uniformly, i.e. according to the Haar measure, say $\mu$, over the $\dimS$-dimensional unitary group, which we denote $\mbb{U}(\dimS)$. The Haar measure is the unique measure on $\mbb{U}(d)$ satisfying invariance under left and right multiplication, i.e. it is invariant under arbitrary rotations. Specifically, given a subset $\mbb{V}\subseteq\mbb{U}(d)$, we have $\mu(\mbb{W})=\int_\mbb{W}d\mu(U)$ for the Haar measure, $\mu$, with the (left-right invariance) property
\begin{align}
   \mu(\mbb{W}) &= \mu(V\mbb{W}) = \int_\mbb{W}d\mu(VU) = \int_\mbb{W}d\mu(UV) =\mu(\mbb{W}V),
\end{align}
for any fixed $V\in\mbb{U}(d)$. For any quantity $f$ depending on a unitary $U\in\mbb{U}(d)$, we denote integration over such unitary by $\mbb{E}[f(U)]$ and refer to it as the Haar or uniform average of $f$.

Let the action of the unitary maps $\mc{G}_i$ be given by $\mc{G}_i(\cdot)=G_i(\cdot)G_i^\dg$, then let us compute
\begin{align}
    &\mbb{E}\left[\mathfrak{C}_m^\dg\right] = \mbb1_\syst\Motimes_{i=1}^{m+1}\sum_{\ell_i,k_i=1}^{\dimS}\int\limits_{\mbb{U}(\dimS)}\left(|\ell_i\rangle\!\langle{k_i}|\otimes{G}_i^\dg|\ell_i\rangle\!\langle{k}_i|G_i\right)\,d\mu(G_1)\cdots{d}\mu(G_m),
\end{align}
where crucially, $G_{m+1} = G_1^\dg{G}_2^\dg\cdots{G}_m^\dg$. This means that we need to be able to compute integrals with two pairs of $G_i$ and $G_i^\dg$ terms. One way to do this is by employing the 2-moment of $\mbb{U}(d)$, given by~\cite{Collins_2003, Collins_2006}
\begin{align}
    \int\limits_{\mathbb{U}(d)}&U_{v_1u_1}U^*_{v^\prime_1u^\prime_1}U_{v_2u_2}U^*_{v^\prime_2u^\prime_2}\,d\mu(U) = \sum_{\sigma,\tau\,\in\,\mbb{S}_2}\delta_{u_1u^\prime_{\sigma(1)}}\delta_{u_2u^\prime_{\sigma(2)}}\delta_{v_1v^\prime_{\tau(1)}}\delta_{v_2v^\prime_{\tau(2)}}\mathrm{Wg}(\tau\sigma^{-1},d),
\end{align}
where here the subindices $v_1u_1,v_1^\prime u_1^\prime,v_2u_2,v_2^\prime u_2^\prime$ refer to components of the same matrix $U$ (with $U^*$ denoting complex conjugate), with the extra subindices $1,2$ being mere labels, and where $\mbb{S}_2$ is the symmetric group on 2 elements. The symbol $\mathrm{Wg}$ is the Weingarten function, which here takes the values
\begin{equation}
    \mathrm{Wg}[(1)(2),d] = \f{1}{d^2-1},\qquad\mathrm{Wg}[(1,2),d]=-\f{1}{d(d^2-1)},
\end{equation}
on the two possible permutations $\tau\sigma^{-1}\in\mbb{S}_2$.

Then we can let $G_\ell=\sum{G}_{v^{(\ell)}{u}^{(\ell)}}|v^{(\ell)}\rangle\!\langle{u}^{(\ell)}|$ for each $G_\ell$ and employ the 2-moment above; let us take the integral over $G_1$ first,
\begin{align}
	&\mbb{E}\left[\mathfrak{C}_m^\dg\right] = \mbb1_\syst\otimes\sum_{u,v=1}^{\dimS}\sum_{\sigma,\tau\in\mbb{S}_2}\mathrm{Wg}(\tau\sigma^{-1},\dimS)|v^{\prime(1)}_1{u}^{\prime\,(1)}_1\rangle\!\langle{v}^{\prime\,(1)}_{\tau(1)}u_{\sigma(1)}^{\prime\,(1)}|\nonumber\\
	&\qquad\Motimes_{\ell=2}^m\sum_{i_\ell,j_\ell}\,\int\limits_{\mbb{U}(d)}\left(|i_\ell\rangle\!\langle{j}_\ell|\otimes{G}_\ell^\dg|i_\ell\rangle\!\langle{j}_\ell|G_\ell\right) \otimes\,|v_{\tau(2)}^{\prime\,(1)}\rangle\!\langle{v}_{2}^{\prime\,(1)}|\otimes{G}_{m}\cdots{G}_2|{u}_{\sigma(2)}^{\prime\,(1)}\rangle\!\langle{u}_{2}^{\prime\,(1)}|G^\dg_2\cdots{G}^\dg_m\,d\mu(G_2)\cdots{d}\mu(G_m),
\end{align}
then we can do similarly with all remaining unitaries by also labeling each permutation mapping $\sigma_\ell$ and $\tau_\ell$ for the corresponding integral over each $G_\ell$, i.e.
\begin{align}
    \mbb{E}\left[\mathfrak{C}_{m}^\dg\right] &= \mbb1_\syst\otimes\sum\mathrm{Wg}_1\cdots\mathrm{Wg}_m|v^{(1)}_1{u}^{(1)}_1\cdots{v}^{(m)}_1{u}^{(m)}_1{u}_{\sigma_m(2)}^{(m)}v_{\tau_1(2)}^{(1)}\rangle\!\langle{v}^{(1)}_{\tau_1(1)}u_{\sigma_1(1)}^{(1)}\cdots{v}^{(m)}_{\tau_m(1)}{u}_{\sigma_m(1)}^{(m)}{u}_{2}^{(m)}{v}_{2}^{(1)}|\nonumber\\
    &\hspace{1in}\delta(u_2^{(1)},v_2^{(2)})\cdots\delta(u_2^{(m-1)},v_2^{(m)})\,\delta(u_{\sigma_1(2)}^{(1)},v_{\tau_2(2)}^{(2)})\cdots\delta(u_{\sigma_{m-1}(2)}^{(m-1)},v_{\tau_m(2)}^{(m)}),
    \label{eq: average gate sequence}
\end{align}
where for easiness of notation we dropped the primes and denoted $\mathrm{Wg}_\ell:=\mathrm{Wg}(\tau_\ell\sigma_\ell^{-1},\dimS)$, where a sum is implicit over basis vectors and permutations on $\mbb{S}_2$, and where $\delta(a,b)$ stands for the usual Kronecker $\delta_{ab}$. We finally notice that $\mbb{E}\left[\mathfrak{C}_m^\dg\right]=\mbb{E}\left[\mathfrak{C}_m^\mathrm{T}\right]$.

\section{Average sequence fidelity}\label{appendix: nM gate indep noise}
\textbf{{The Markovian case.}} As a first case let us verify that the average gate sequence given by Eq.~\ref{eq: average gate sequence} reproduces an \asf~described by a decaying exponential in the number of gates when the noise is Markovian.

Consider first a single gate, $m=1$. We have
\begin{equation}
    \mbb{E}\left[\mathfrak{C}_m^{\mathrm{T}}\right]=\mbb1_\syst\otimes\sum\mathrm{Wg}|v_1{u}_1u_{\sigma(2)}v_{\tau(2)}\rangle\!\langle{v}_{\tau(1)}u_{\sigma(1)}u_2v_2|,
\end{equation}
where $\mathrm{Wg}$ implicitly depends on $\tau\sigma^{-1}$, with each $\tau$ and $\sigma$ being summed over the symmetric group $\mathbb{S}_2$, and with an implicit sum over each $u$ and $v$.

For the Markovian process tensor, we consider noise described by some $\dimS$-dimensional \compp~map $\Lambda_n^\markov$ at time-step $n$ with Kraus representation $\Lambda_n^ \markov(\cdot)=\sum_i\lambda^ \markov_{i_n}(\cdot)\lambda^ \markov_{i_n}$, so that
\begin{align}
    &\Upsilon_1^\markov = (\Lambda_2^ \markov\otimes\mc{I})\circ\mscr{S}_2\circ(\Lambda_1^ \markov\otimes\mc{I})\circ\mscr{S}_1\left(\rho\otimes\psi^{\otimes2}\right)\nonumber\\
    &= \sum(\lambda_{i_2}^ \markov\otimes\mbb1)\varsigma_2(\lambda_{i_1}^ \markov\otimes\mbb1)\varsigma_1(\rho\otimes\psi^{\otimes2})\varsigma_1^\dg(\lambda_{i_1}^{\markov\,\dg}\otimes\mbb1)\varsigma_2^\dg(\lambda_{i_2}^{\markov\,\dg}\otimes\mbb1)\nonumber\\
    &= \sum\lambda_{i_2}^ \markov|\alpha_2\rangle\!\langle\beta_2|\lambda_{i_1}^ \markov|\alpha_1\rangle\!\langle\beta_1|\rho|\delta_1\rangle\!\langle\gamma_1|\lambda_{i_1}^{\markov\,\dg}|\delta_2\rangle\!\langle\gamma_2|\lambda_{i_2}^{\markov\,\dg}\otimes|\beta_1\alpha_1\beta_2\alpha_2\rangle\!\langle\delta_1\gamma_1\delta_2\gamma_2|.
\end{align}

Let us simply denote $\dimS$ as $d$, as there is no environment to care about. Then we obtain the average sequence
\begin{align}
	&\mbb{E}[\mc{S}_1(\rho)] = \tr_{\overline{S}}\left\{\Upsilon_1^\markov\,\mbb{E}\left[\mathfrak{C}_m^{\mathrm{T}}\right]\right\}\nonumber\\
	&= \Lambda_2^ \markov\left[\sum\mathrm{Wg}|v_2\rangle\!\langle{u}_2|\lambda_{i_1}^ \markov|u_{\sigma(1)}\rangle\!\langle{v}_{\tau(1)}|\rho|v_1\rangle\!\langle{u}_1|\lambda_{i_1}^{\markov\,\dg}|{u}_{\sigma(2)}\rangle\!\langle{v}_{\tau(2)}|\right]\nonumber\\
	&=\Lambda_2\Bigg[\f{1}{d^2-1}\left(\,\underbrace{\tr\left(\sum_i\lambda_{i_1}^{\dg\markov}\lambda_{i_1}^{\markov}\right)\,\mbb1}_{\sigma=\tau=\boldsymbol{1}}+\sum_i\underbrace{\tr(\lambda_{i_1}^ \markov)\tr(\lambda_{i_1}^{\markov\,\dg})\,\rho}_{\sigma=\tau=(1,2)}\right) - \f{1}{d(d^2-1)}\left(\,\underbrace{\tr\left(\sum_i\lambda_{i_1}^{\dg\markov}\lambda_{i_1}^{\markov}\right)\,\rho}_{\sigma=\boldsymbol{1},\,\tau=(1,2)}\!\!+\sum_i\underbrace{\tr(\lambda_{i_1}^ \markov)\tr(\lambda_{i_1}^{\markov\,\dg})\,\mbb1}_{\sigma(1,2),\,\tau=(1)(2)}\right)\Bigg]\nonumber\\
    &= \Lambda_2\left[\f{d\tr\left(\sum_i\lambda_{i_1}^{\dg\markov}\lambda_{i_1}^{\markov}\right)-\sum_i|\tr(\lambda_{i_1}^ \markov)|^2}{d^2-1}\,\left(\f{\mbb1}{d}\right)+\f{d\sum_i|\tr(\lambda_{i_1}^ \markov)|^2-\tr\left(\sum_i\lambda_{i_1}^{\dg\markov}\lambda_{i_1}^{\markov}\right)}{d(d^2-1)}\,\rho\right].
\label{eq: length-one-usual}
\end{align}

Now, if the noise is trace-preserving as well, we have $\tr\left(\sum_i\lambda_{i_1}^{\dg\markov}\lambda_{i_1}^{\markov}\right)=\tr(\mbb1)=d$. Then we get
\begin{align}
	\mbb{E}[\mc{S}_1(\rho)] = \Lambda_2^ \markov\circ\mscr{D}_p(\rho),
\end{align}
where we define $\mscr{D}_p(X) := pX+(1-p)\f{\mbb1}{d}$ as a depolarising map with the so-called noise-strength,
\begin{equation}
    p := \f{\sum_i|\tr(\lambda_{i_1}^ \markov)|^2-1}{d^2-1}\in[0,1],
\end{equation}
which has to be constrained to $[0,1]$. If we denote the noise map $\Lambda_n^\markov=\sum_i\lambda_{i_n}^\markov\otimes\lambda_{i_n}^{\markov\,\dg}$, with the Kraus operators acting on the respective system $\syst$ Hilbert space and conjugate space, resp., we can simply write
\begin{equation}
    p = \f{\tr\left[\Lambda_1^ \markov\right]-1}{d^2-1}.
\end{equation}

The noise-strength can be shown to be related to the gate fidelity of $\Lambda_1$ with respect to the identity~\cite{PhysRevA.92.022326}, i.e. $\mathsf{f}_{\Lambda_1,\mc{I}}=\int{d}\psi\langle\psi|\Lambda_1^ \markov(|\psi\rangle\!\langle\psi|)|\psi\rangle$, as $p=\f{df_{\Lambda_1,\mc{I}}-1}{d-1}$. This is the relevant parameter which in practice can be recovered by running several sequences for different lengths and averaging the resulting probabilities.

To generalize to an arbitrary number of time-steps, we now use the fact that the action of the depolarizing channel can be written as
\begin{equation}
    \mscr{D}_p(X) = \sum\mathrm{Wg}|v_2\rangle\!\langle{u}_2|\lambda_i^ \markov|u_{\sigma(1)}\rangle\!\langle{v}_{\tau(1)}|X|v_1\rangle\!\langle{u}_1|\lambda_i^{\markov\,\dg}|{u}_{\sigma(2)}\rangle\!\langle{v}_{\tau(2)}|,
\label{eq: depolarising Wg}
\end{equation}
for any $X$, which follows from Eq.~\eqref{eq: length-one-usual}. This then implies that for an arbitrary sequence length,
\begin{align}
	\mbb{E}\left[\mc{S}_m(\rho)\right] &= \tr_{\overline{\syst}}\left\{\Upsilon_m^\markov\mbb{E}\left[\mathfrak{C}_m^{\mathrm{T}}\right]\right\}\nonumber\\
	&= \Lambda_{m+1}\circ\mscr{D}_{p_m}\circ\cdots\circ\mscr{D}_{p_1}(\rho)\nonumber\\
	&=p_1p_2\cdots{p}_m\,\Lambda_{m+1}^ \markov\left(\rho-\f{\mbb1}{d}\right)+\Lambda_{m+1}^ \markov\left(\f{\mbb1}{d}\right),
	\label{eq: Markovian RB avg time-dep}
\end{align}
where here now $p_n:=\f{\tr[\Lambda_n^\markov]-1}{\dimS^2-1}$, as expected. The case $p_1\neq{p_2}\neq\cdots\neq{p_m}$ corresponds to the Markovian time-dependent noise case as in Ref.~\cite{Wallman_2014}. When the noise-strengths are the same this gives the usual fitting model for the average probabilities
\begin{align}
    \mc{F}_m
    &= p^m\tr\left[\mc{M}\Lambda_{m+1}^ \markov\left(\rho-\f{\mbb1}{d}\right)\right]+\tr\left[\mc{M}\Lambda_{m+1}^ \markov\left(\f{\mbb1}{d}\right)\right]:= Ap^m+B,
    \label{eq: average fidelity Markov time indep}
\end{align}
with $A:=\tr\left[\mc{M}\Lambda^ \markov\left(\rho-\f{\mbb1}{d}\right)\right]$ and $B:=\tr\left[\mc{M}\Lambda^ \markov\left(\f{\mbb1}{d}\right)\right]$, which relate to state preparation and measurement errors.

\textbf{{General non-Markovian gate-independent noise.}} We now consider the general situation where the noise is correlated across each step through an external environment as depicted in Fig.~\ref{fig: non-Markov sequence}. Let us take first the simplest case $m=1$; the process tensor for the noise sequence is
\begin{align}
    &\Upsilon_1 = \tr_\env\left[\Lambda_2\otimes\mc{I})\circ\mscr{S}_2\circ(\Lambda_1\otimes\mc{I})\circ\mscr{S}_1\left(\rho\otimes\psi^{\otimes2}\right)\right]\nonumber\\
    &= \tr_\env\left[(\lambda_2\otimes\mbb1)\varsigma_2(\lambda_1\otimes\mbb1)\varsigma_1(\rho\otimes\psi^{\otimes2})\varsigma_1^\dg(\lambda_1^\dg\otimes\mbb1)\varsigma_2^\dg(\lambda_2^\dg\otimes\mbb1)\right]\nonumber\\
    &=\tr_\env\left[\lambda_2\zeta_{\alpha_2\beta_2}\lambda_1\zeta_{\alpha_1\beta_1}\rho\,\zeta_{\delta_1\gamma_1}\lambda_1^\dg\zeta_{\delta_2\gamma_2}\lambda_2^\dg\right]\otimes|\beta_1\alpha_1\beta_2\alpha_2\rangle\!\langle\delta_1\gamma_1\delta_2\gamma_2|
\end{align}
where here $\zeta_{ab}:=\mbb1_\env\otimes|a\rangle\!\langle{b}|$, hence we get
\begin{align}
\mbb{E}\left[\mc{S}_1(\rho)\right]&=\tr_{\overline{\syst}}\left\{\Upsilon_1\mbb{E}\left(\mathfrak{C}_1^{\mathrm{T}}\right)\right\}\nonumber\\
&=\tr_\env\left\{\Lambda_2\left[\sum\mathrm{Wg}\,\zeta_{v_2u_2}\lambda_{i_1}\zeta_{u_{\sigma(1)}v_{\tau(1)}}\rho\,
\zeta_{v_1u_1}\lambda_{i_1}^\dg\zeta_{u_{\sigma(2)}v_{\tau(2)}}\right]\right\}.
\label{avg_seq_nM_length2}
\end{align}

Let us write the Kraus operators of the $n$\textsuperscript{th} noise map, $\Lambda_n$, as
\begin{equation}
    \lambda_{i_n}:=\sum_{e_n^{(\prime)}=1}^{d_\env}\sum_{s_n^{(\prime)}=1}^{d_\syst}\mathrm{L}_{i_n}^{\substack{e_ne_n^\prime\\s_ns_n^\prime}}|e_ns_n\rangle\!\langle{e}_n^\prime s_n^\prime|,
\end{equation}
where the $e$ and $s$ indices refer to systems $\env$ and $\syst$, resp.; the subindex $n$ is simply a label for the $n$\textsuperscript{th} Kraus operator. Then
\begin{align}
    &\mbb{E}\left[\mc{S}_1(\rho)\right]=\sum\mathrm{L}_{i_1}^{\substack{e_1e_1^\prime\\s_1s_1^\prime}}\mathrm{L}_{i_1}^{*\,\substack{\epsilon_1\epsilon_1^\prime\\\zeta_1\zeta_1^\prime}}\,\mathrm{Wg}\tr_\env\bigg\{\Lambda_2\bigg[\left(|e_1\rangle\!\langle{e}_1^\prime|\otimes|v_2\rangle\!\langle{u}_2|s_1\rangle\!\langle{s}_1^\prime|u_{\sigma(1)}\rangle\!\langle{v}_{\tau(1)}|\right)\rho\left(|\epsilon_1^\prime\rangle\!\langle\epsilon_1|\otimes|v_1\rangle\!\langle{u}_1|\zeta_1^\prime\rangle\!\langle\zeta_1|u_{\sigma(2)}\rangle\!\langle{v}_{\tau(2)}|\right)\bigg]\bigg\},
\label{S2int}
\end{align}
and let us now similarly write the initial state as
\begin{equation}
    \rho=\sum_{e,e^\prime=1}^{d_\env}\sum_{s,s^\prime=1}^{d_\syst}\chi^{\substack{ee^\prime\\ss^\prime}}|es\rangle\!\langle{e^\prime s^\prime}|,\qquad\text{where}\qquad\sum\chi^{\substack{ee\\ss}}=1,
\end{equation}
then also
\begin{align}
    &\mbb{E}\left[\mc{S}_1(\rho)\right]= \sum\mathrm{L}_{i_1}^{\substack{e_1e_1^\prime\\s_1s_1^\prime}}\chi^{\substack{ee^\prime\\ss^\prime}}\mathrm{L}_{i_1}^{*\,\substack{\epsilon_1\epsilon_1^\prime\\\zeta_1\zeta_1^\prime}}\tr_\env\left\{\Lambda_2\left[|e_1\rangle\!\langle{e}_1^\prime|e\rangle\!\langle{e}^\prime|\epsilon_1^\prime\rangle\!\langle\epsilon_1|\otimes\sum\mathrm{Wg}|v_2\rangle\!\langle{u}_2|s_1\rangle\!\langle{s}_1^\prime|u_{\sigma(1)}\rangle\!\langle{v}_{\tau(1)}|s\rangle\!\langle{s}^\prime |v_1\rangle\!\langle{u}_1|\zeta_1^\prime\rangle\!\langle\zeta_1|u_{\sigma(2)}\rangle\!\langle{v}_{\tau(2)}|\right]\right\}\nonumber\\
    &:= \sum\mathrm{L}_{i_1}^{\substack{e_1e\\s_1s_1^\prime}}\chi^{\substack{ee^\prime\\ss^\prime}}\mathrm{L}_{i_1}^{*\,\substack{\epsilon_1e^\prime\\\zeta_1\zeta_1^\prime}}\tr_\env\left\{\Lambda_2\left[|e_1\rangle\!\langle\epsilon_1|\otimes\Phi_{s_1\zeta_1}(|s\rangle\!\langle{s}^\prime|)\right]\right\}\nonumber\\
    &=\sum\mathrm{L}_{i_2}^{\substack{e_2e_2^\prime\\s_2s_2^\prime}}\mathrm{L}_{i_1}^{\substack{e_1e\\s_1s_1^\prime}}\chi^{\substack{ee^\prime\\ss^\prime}}\mathrm{L}_{i_1}^{*\,\substack{\epsilon_1e^\prime\\\zeta_1\zeta_1^\prime}}\mathrm{L}_{i_2}^{*\,\substack{\epsilon_2\epsilon_2^\prime\\\zeta_2\zeta_2^\prime}}\tr\left[|e_2\rangle\!\langle{e}_2^\prime|e_1\rangle\!\langle\epsilon_1|\epsilon_2^\prime\rangle\!\langle\epsilon_2|\right]\,|s_2\rangle\!\langle{s}_2^\prime|\Phi_{s_1\zeta_1}(|s\rangle\!\langle{s}^\prime|)|\zeta_2^\prime\rangle\!\langle\zeta_2|\nonumber\\
    &=\sum\mathrm{L}_{i_2}^{\substack{e_2e_1\\s_2s_2^\prime}}\mathrm{L}_{i_1}^{\substack{e_1e\\s_1s_1^\prime}}\chi^{\substack{ee^\prime\\ss^\prime}}\mathrm{L}_{i_1}^{*\,\substack{\epsilon_1e^\prime\\\zeta_1\zeta_1^\prime}}\mathrm{L}_{i_2}^{*\,\substack{e_2\epsilon_1\\\zeta_2\zeta_2^\prime}}\,\langle{s}_2^\prime|\Phi_{s_1\zeta_1}(|s\rangle\!\langle{s}^\prime|)|\zeta_2^\prime\rangle\,|s_2\rangle\!\langle\zeta_2|,
    \label{eq: nonMarkov exp2}
\end{align}
where the second line follows by Eq.~\eqref{eq: depolarising Wg} and by defining
\begin{equation}
    \Phi_{s_n\zeta_n}(X) := \f{\dimS\,\delta_{s_ns^\prime_n}\delta_{\zeta_n\zeta^\prime_n}-\delta_{s_n\zeta_n}\delta_{s^\prime_n\zeta^\prime_n}}{\dimS(\dimS^2-1)}\,X + \f{\dimS\,\delta_{s_n\zeta_n}\delta_{s^\prime_n\zeta^\prime_n}-\delta_{s_ns^\prime_n}\delta_{\zeta_n\zeta^\prime_n}}{\dimS^2-1}\,\left(\f{\mbb1_\syst}{\dimS}\right).
\end{equation}

Now let
\begin{equation}
    \alpha_{s_n^{(\prime)}\zeta_n^{(\prime)}} := \f{\dimS\delta_{s_ns^\prime_n}\delta_{\zeta_n\zeta^\prime_n}-\delta_{s_n\zeta_n}\delta_{s^\prime_n\zeta^\prime_n}}{\dimS(\dimS^2-1)}, \qquad \beta_{s_n^{(\prime)}\zeta_n^{(\prime)}} := \f{\delta_{s_n\zeta_n}\delta_{s^\prime_n\zeta^\prime_n}}{\dimS}-\alpha_{s_n\zeta_n}
\end{equation}
so that
\begin{align}
    \Phi_{s_n\zeta_n}(X) &= \alpha_{s_n^{(\prime)}\zeta_n^{(\prime)}}\,X + \beta_{s_n^{(\prime)}\zeta_n^{(\prime)}}\left(\f{\mbb1_\syst}{\dimS}\right) = \alpha_{s_n^{(\prime)}\zeta_n^{(\prime)}}\,\left(X-\f{\mbb1_\syst}{\dimS}\right)+\f{\delta_{s_n\zeta_n}\delta_{s^\prime_n\zeta^\prime_n}}{\dimS}\left(\f{\mbb1_\syst}{\dimS}\right).
\end{align}

Now, we can also define
\begin{align}
    \mc{E}_{\boldsymbol{s}^{(\prime)}\boldsymbol{\zeta}^{(\prime)}}^{(2)} &:= \sum_{i=1}^{d_\lambda}\sum_{e=1}^{\dimE}\mathrm{L}_{i_2}^{\substack{e_4e_2\\s_2s_2^\prime}}\mathrm{L}_{i_1}^{\substack{e_2e_0\\s_1s_1^\prime}}\chi^{\substack{e_0e_1\\ss^\prime}}\mathrm{L}_{i_1}^{*\,\substack{e_3e_1\\\zeta_1\zeta_1^\prime}}\mathrm{L}_{i_2}^{*\,\substack{e_4e_3\\\zeta_2\zeta_2^\prime}}\nonumber\\
    &= \sum\langle{e_4s_2}|\lambda_{i_2}|e_2s_2^\prime\rangle\!\langle{e_2s_1}|\lambda_{i_1}|e_0s_1^\prime\rangle\!\langle{e_0s}|\rho|e_1s^\prime\rangle\!\langle{e_1\zeta_1^\prime}|\lambda_{i_1}^\dg|e_3\zeta_1\rangle\!\langle{e_3\zeta_2^\prime}|\lambda_{i_2}^\dg|e_4\zeta_2^\prime\rangle \nonumber\\
    &= \tr\left[(\mbb1_\env\otimes\langle{s_2}|)\lambda_{i_2}(\mbb1_\env\otimes|s_2^\prime\rangle\!\langle{s_1}|)\lambda_{i_1}(\mbb1_\env\otimes|s_1^\prime\rangle\!\langle{s}|)\rho(\mbb1_\env\otimes|s^\prime\rangle\!\langle{\zeta_1^\prime}|)\lambda_{i_1}^\dg(\mbb1_\env\otimes|\zeta_1\rangle\!\langle{\zeta_2^\prime}|)\lambda_{i_2}^\dg(\mbb1_\env\otimes|\zeta_2^\prime\rangle)\right],
\end{align}
where summation is over all $i_1,i_2,$\ldots and $e_0,e_1,$\ldots, and which contains all information about the noise within the whole $\syst\env$ and the correlations in between the two. We can simply write this as $\mc{E}_{\boldsymbol{s}^{(\prime)}\boldsymbol{\zeta}^{(\prime)}}^{(2)} = \tr\left[\langle{s_2}|\lambda_{i_2}|s_2^\prime\rangle\!\langle{s_1}|\lambda_{i_1}|s_1^\prime\rangle\!\langle{s}|\rho|s^\prime\rangle\!\langle{\zeta_1^\prime}|\lambda_{i_1}^\dg|\zeta_1\rangle\!\langle{\zeta_2^\prime}|\lambda_{i_2}^\dg|\zeta_2^\prime\rangle\right]$ as in the main text, where an identity on $\env$ is implicit. With this we can write Eq.~\eqref{eq: nonMarkov exp2} as
\begin{align}
    \mbb{E}\left[\mc{S}_1(\rho)\right] &= \sum\mc{E}_{\boldsymbol{s}^{(\prime)}\boldsymbol{\zeta}^{(\prime)}}^{(2)}\,\left[\alpha_{s_1\zeta_1}\left(\delta_{ss_2^\prime}\delta_{s^\prime\zeta_2^\prime}-\f{\delta_{ss^\prime}\delta_{s_2^\prime\zeta_2^\prime}}{\dimS}\right) + \f{\delta_{s_1\zeta_1}\delta_{s_1^\prime\zeta_1^\prime}}{\dimS}\left(\f{\delta_{ss^\prime}\delta_{s_2^\prime\zeta_2^\prime}}{\dimS}\right)\right]\,|s_2\rangle\!\langle\zeta_2|.
    \label{eq: avg sequence s2}
\end{align}

In general, for an arbitrary sequence length, we have
\begin{align}
    \mc{E}_{\boldsymbol{s}^{(\prime)}\boldsymbol{\zeta}^{(\prime)}}^{(m)} &:= \sum_{\boldsymbol{e},\boldsymbol{\epsilon}}\chi^{\substack{ee^\prime\\ss^\prime}}\tr\left[\left(\prod_{\eta=m}^1\mathrm{L}_{i_\eta}^{\substack{e_\eta e^\prime_\eta\\s_\eta s^\prime_\eta}}|e_\eta\rangle\!\langle{e}_\eta^\prime|\right)|e\rangle\!\langle{e}^\prime|\left(\prod_{n=1}^m\mathrm{L}_{i_n}^{*\,\substack{\epsilon_n\epsilon^\prime_n\\\zeta_n\zeta^\prime_n}}|\epsilon_n^\prime\rangle\!\langle\epsilon_n|\right)\right]\nonumber\\
    &= \sum_{i=1}^{d_\lambda} \tr\left[\left(\prod_{\eta=m}^1\langle{s}_\eta|\lambda_{i_\eta}|{s}_\eta^\prime\rangle\right)\langle{s}|\,\rho\,|s^\prime\rangle\left(\prod_{n=1}^m\langle\zeta^\prime_n|\lambda_{i_n}^\dg|\zeta_n\rangle\right)\right],
\end{align}
so that,
\begin{align}
    \mbb{E}\left[\mc{S}_m(\rho)\right] &= \sum\mc{E}_{\boldsymbol{s}^{(\prime)}\boldsymbol{\zeta}^{(\prime)}}^{(m+1)}\,\langle{s}^\prime_{m+1}|\Phi_{s_m\zeta_m}\circ\cdots\circ\Phi_{s_1\zeta_1}(|s\rangle\!\langle{s}^\prime|)|\zeta^\prime_{m+1}\rangle\,|s_{m+1}\rangle\!\langle\zeta_{m+1}|.
    \label{eq: arbitrary length nonM avg sequence}
\end{align}

The sequential application of $\Phi$ maps is given by
\begin{equation}
    \Phi_{s_m\zeta_m}\circ\cdots\circ\Phi_{s_1\zeta_1}(X) := \boldsymbol{\alpha}^{(m)}_{\boldsymbol{s},\boldsymbol{\zeta}}\left(X-\f{\mbb{1}}{\dimS}\right) + \boldsymbol{\Delta}^{(m)}_{\boldsymbol{s},\boldsymbol{\zeta}}\,\f{\mbb{1}}{\dimS},
\end{equation}
where
\begin{equation}
    \boldsymbol{\alpha}^{(m)}_{\boldsymbol{s},\boldsymbol{\zeta}} := \prod_{n=1}^m\alpha_{s_n^{(\prime)}\zeta_n^{(\prime)}},
\end{equation}
and $\boldsymbol{\Delta}^{(m)}_{\boldsymbol{s},\boldsymbol{\zeta}}$ is a sum of all $m$-term product combinations of $\alpha_{s_1^{(\prime)}\zeta_1^{(\prime)}},\cdots,\alpha_{s_m^{(\prime)}\zeta_m^{(\prime)}}$ and $\beta_{s_1^{(\prime)}\zeta_1^{(\prime)}},\cdots,\beta_{s_m^{(\prime)}\zeta_m^{(\prime)}}$, that is,
\begin{align}
    \boldsymbol{\Delta}^{(1)}_{\boldsymbol{s},\boldsymbol{\zeta}} &\sim \alpha_1 + \beta_1 \\
    \boldsymbol{\Delta}^{(2)}_{\boldsymbol{s},\boldsymbol{\zeta}} &\sim \alpha_1\alpha_2 + \alpha_1\beta_2 + \beta_1\alpha_2 + \beta_1\beta_2\\
    \boldsymbol{\Delta}^{(3)}_{\boldsymbol{s},\boldsymbol{\zeta}} &\sim \alpha_1\alpha_2\alpha_3+\alpha_1\alpha_2\beta_3 + \alpha_1\beta_2\alpha_3 + \alpha_1\beta_2\beta_3 + \beta_1\alpha_2\alpha_3 + \beta_1\alpha_2\beta_3 + \beta_1\beta_2\alpha_3 + \beta_1\beta_2\beta_3\\[-0.1in]
    &\,\,\vdots\nonumber
\end{align}
where $\alpha_i=\alpha_{s_i^{(\prime)}\zeta_i^{(\prime)}}$ and similarly for $\beta_i$; in general there are $2^m$ of these summands on $\boldsymbol{\Delta}^{(m)}_{\boldsymbol{s},\boldsymbol{\zeta}}$. However, notice that as $\beta_i\sim\f{1}{\dimS}\delta_{s_i\zeta_i}\delta_{s_i^\prime\zeta_i^\prime}-\alpha_i$, every term simplifies to products of deltas, i.e.
\begin{align}
    \boldsymbol{\Delta}^{(1)}_{\boldsymbol{s},\boldsymbol{\zeta}} \sim \f{\delta_{s_1\zeta_1}\delta_{s_1^\prime\zeta_1^\prime}}{\dimS},\quad
    \boldsymbol{\Delta}^{(2)}_{\boldsymbol{s},\boldsymbol{\zeta}} \sim \f{\delta_{s_1\zeta_1}\delta_{s_1^\prime\zeta_1^\prime}\delta_{s_2\zeta_2}\delta_{s_2^\prime\zeta_2^\prime}}{\dimS^2},\quad\cdots\quad,
    \boldsymbol{\Delta}^{(m)}_{\boldsymbol{s},\boldsymbol{\zeta}} \sim \f{\prod_{i=1}^m\delta_{s_i\zeta_i}\delta_{s_i^\prime\zeta_i^\prime}}{\dimS^m}.
\end{align}

Thus we can rewrite Eq.~\eqref{eq: arbitrary length nonM avg sequence} as
\begin{align}
    \mbb{E}&\left[\mc{S}_m(\rho)\right] = \sum\mc{E}_{\boldsymbol{s}^{(\prime)}\boldsymbol{\zeta}^{(\prime)}}^{(m+1)}\,\left[\boldsymbol{\alpha}^{(m)}_{\boldsymbol{s},\boldsymbol{\zeta}}\left(\delta_{ss_{m+1}^\prime}\delta_{s^\prime\zeta_{m+1}^\prime}-\f{\delta_{ss^\prime}\delta_{s_{m+1}^\prime\zeta_{m+1}^\prime}}{\dimS}\right) + \boldsymbol{\Delta}^{(m)}_{\boldsymbol{s},\boldsymbol{\zeta}}\left(\f{\delta_{ss^\prime}\delta_{s_{m+1}^\prime\zeta_{m+1}^\prime}}{\dimS}\right)\right]\,|s_{m+1}\rangle\!\langle\zeta_{m+1}|,
\end{align}
and so for a measurement $\mc{M}$, on average,
\begin{align}
    \mc{F}_m = \sum_{s,\zeta,s^\prime,\zeta^\prime=1}^{\dimS}\mc{E}_{\boldsymbol{s}^{(\prime)}\boldsymbol{\zeta}^{(\prime)}}^{(m+1)}\,\left( \mc{A}^{(m+1)}_{\boldsymbol{s}^{(\prime)}\boldsymbol{\zeta}^{(\prime)}} + \mc{B}^{(m+1)}_{\boldsymbol{s}^{(\prime)}\boldsymbol{\zeta}^{(\prime)}}\right),
    \label{eq: appendix main expr sum}
\end{align}
where
\begin{align}
    \mc{A}^{(m+1)}_{\boldsymbol{s}^{(\prime)}\boldsymbol{\zeta}^{(\prime)}} := \boldsymbol{\alpha}^{(m)}_{\boldsymbol{s},\boldsymbol{\zeta}}\left(\delta_{s_{m+1}^\prime s}\delta_{s^\prime\zeta_{m+1}^\prime}\!-\f{\delta_{ss^\prime}\delta_{s_{m+1}^\prime\zeta_{m+1}^\prime}}{\dimS}\right)\langle\zeta_{m+1}|\mc{M}|s_{m+1}\rangle,\qquad
    \mc{B}^{(m+1)}_{\boldsymbol{s}^{(\prime)}\boldsymbol{\zeta}^{(\prime)}} := \boldsymbol{\Delta}^{(m)}_{\boldsymbol{s},\boldsymbol{\zeta}}\, \left(\f{\delta_{ss^\prime}\delta_{s_{m+1}^\prime\zeta_{m+1}^\prime}}{\dimS}\right)\langle\zeta_{m+1}|\mc{M}|s_{m+1}\rangle.
\end{align}

This expression contains $4\dimS(m+1)$ terms, and could potentially be useful whenever the underlying noise model is not known, as all this information will be contained solely on the factors $\mc{E}_{\boldsymbol{s}^{(\prime)}\boldsymbol{\zeta}^{(\prime)}}^{(m+1)}$.

We can, however, write this expression in a more succinct way. We have
\begin{align}
     &\sum_{s,\zeta,s^\prime,\zeta^\prime=1}^{\dimS}\mc{E}_{\boldsymbol{s}^{(\prime)}\boldsymbol{\zeta}^{(\prime)}}^{(m+1)}\mc{A}^{(m+1)}_{\boldsymbol{s}^{(\prime)}\boldsymbol{\zeta}^{(\prime)}} \nonumber\\
    &= \sum \tr\left[\left(\prod_{\eta=m+1}^1\langle{s}_\eta|\lambda_{i_\eta}|{s}_\eta^\prime\rangle\right)\langle{s}|\,\rho\,|s^\prime\rangle\left(\prod_{n=1}^{m+1}\langle\zeta^\prime_n|\lambda_{i_n}^\dg|\zeta_n\rangle\right)\right]\left(\f{\prod_{N=1}^m\left(\dimS\delta_{s_Ns^\prime_N}\delta_{\zeta_N\zeta^\prime_N}-\delta_{s_N\zeta_N}\delta_{s^\prime_N\zeta^\prime_N}\right)}{\dimS^m(\dimS^2-1)^m}\right)\left(\delta_{s_{m+1}^\prime s}\delta_{s^\prime\zeta_{m+1}^\prime}\!-\f{\delta_{ss^\prime}\delta_{s_{m+1}^\prime\zeta_{m+1}^\prime}}{\dimS}\right)\nonumber\\
    &\qquad\qquad\qquad\langle\zeta_{m+1}|\mc{M}|s_{m+1}\rangle,
\end{align}
so now let us define the following. Let
\begin{align}
    \$_{\Lambda_n}(\varepsilon) &:=\sum_{i=1}^{d_\lambda} \tr_\syst(\lambda_{i_n})\,\varepsilon\,\tr_\syst(\lambda_{i_n}^\dg),\label{eq: def dollar appendix} \\
    \Theta_{\Lambda_n}(\varepsilon) &:= \tr_\syst\left[\Lambda_n\left(\varepsilon\otimes\f{\mbb1}{\dimS}\right)\right],
    \label{eq: def Theta appendix}
 \end{align}
for any operator $\varepsilon$ acting on $\env$. Then we notice that
\begin{align}
    &\f{1}{\dimS^m}\sum\left(\prod_{\eta=m}^1\langle{s}_\eta|\lambda_{i_\eta}|{s}_\eta^\prime\rangle\right)\langle{s}|\rho|{s}^\prime\rangle\left(\prod_{n=1}^{m}\langle\zeta^\prime_n|\lambda_{i_n}^\dg|\zeta_n\rangle\right)\prod_{N=1}^m\left(\dimS\delta_{s_Ns^\prime_N}\delta_{\zeta_N\zeta^\prime_N}-\delta_{s_N\zeta_N}\delta_{s^\prime_N\zeta^\prime_N}\right)\nonumber\\
    &= \f{1}{\dimS^m}\sum\left(\prod_{\eta=m}^2\langle{s}_\eta|\lambda_{i_\eta}|{s}_\eta^\prime\rangle\right)\left(\dimS\tr_\syst(\lambda_{i_1}){\langle{s}|\rho|{s}^\prime\rangle}\tr_\syst(\lambda_{i_1}^\dg)-\tr_\syst[\Lambda_1(\langle{s}|\rho|{s}^\prime\rangle\otimes\mbb1)]\right)\left(\prod_{n=2}^{m}\langle\zeta^\prime_n|\lambda_{i_n}^\dg|\zeta_n\rangle\right)\prod_{N=2}^m\left(\dimS\delta_{s_Ns^\prime_N}\delta_{\zeta_N\zeta^\prime_N}-\delta_{s_N\zeta_N}\delta_{s^\prime_N\zeta^\prime_N}\right)\nonumber\\
    &= \f{1}{\dimS^{m-1}}\sum\left(\prod_{\eta=m}^2\langle{s}_\eta|\lambda_{i_\eta}|{s}_\eta^\prime\rangle\right)\left(\$_{\Lambda_1}-\Theta_{\Lambda_1}\right)\langle{s}|\rho|{s}^\prime\rangle\left(\prod_{n=2}^{m}\langle\zeta^\prime_n|\lambda_{i_n}^\dg|\zeta_n\rangle\right)\prod_{N=2}^m\left(\dimS\delta_{s_Ns^\prime_N}\delta_{\zeta_N\zeta^\prime_N}-\delta_{s_N\zeta_N}\delta_{s^\prime_N\zeta^\prime_N}\right)\nonumber\\
    &\quad\vdots\nonumber\\
    &= \left[\Mcirc_{n=1}^m\left(\$_{\Lambda_n}-\Theta_{\Lambda_n}\right)\right]\langle{s}|\rho|{s}^\prime\rangle,
\end{align}
where as before there are implicit identities which should be clear by context, i.e. for example $\tr_\syst(\lambda_i)$ means $\tr_\syst(\lambda_i)\otimes\mbb1_\syst$. Then this means that
\begin{align}
     &\sum_{s,\zeta,s^\prime,\zeta^\prime=1}^{\dimS}\mc{E}_{\boldsymbol{s}^{(\prime)}\boldsymbol{\zeta}^{(\prime)}}^{(m+1)}\mc{A}^{(m+1)}_{\boldsymbol{s}^{(\prime)}\boldsymbol{\zeta}^{(\prime)}} = \f{\tr\left\{\mc{M}\,\tr_\env\circ\Lambda_{m+1}\left[\Mcirc_{n=1}^m\left(\$_{\Lambda_n}-\Theta_{\Lambda_n}\right)\otimes\mc{I}_\syst\right]\left(\rho-\rho_\env\otimes\f{\mbb1}{\dimS}\right)\right\}}{(\dimS^2-1)^m},
\end{align}
where $\rho_\env := \tr_\syst(\rho)$. Now for the second term, similarly (again we omit implicit identity operators),
\begin{align}
    \sum_{s,\zeta,s^\prime,\zeta^\prime=1}^{\dimS}&\mc{E}_{\boldsymbol{s}^{(\prime)}\boldsymbol{\zeta}^{(\prime)}}^{(m+1)}\mc{B}^{(m+1)}_{\boldsymbol{s}^{(\prime)}\boldsymbol{\zeta}^{(\prime)}} \nonumber\\
    &= \f{1}{\dimS^{m+1}}\sum_{s,s^\prime=1}^{\dimS}\sum_{i=1}^{d_\lambda} \tr\left[\langle{s}_{m+1}|\lambda_{i_{m+1}}|s^\prime_{m+1}\rangle\left(\prod_{\eta=m}^1\langle{s}_\eta|\lambda_{i_\eta}|{s}_\eta^\prime\rangle\right)\langle{s}|\,\rho\,|s\rangle\left(\prod_{n=1}^{m}\langle{s}^\prime_n|\lambda_{i_n}^\dg|s_n\rangle\right)\langle{s_{m+1}^\prime}|\lambda_{i_{m+1}}^\dg|\zeta_{m+1}\rangle\right]\langle\zeta_{m+1}|\mc{M}|s_{m+1}\rangle \nonumber\\
    &= \tr\left[\left(\mbb1_\env\otimes\mc{M}\right)\circ\Lambda_{m+1}\circ\left(\Mcirc_{n=1}^m\Theta_{\Lambda_n}\otimes\mc{I}_\syst\right)\left(\rho_\env\otimes\f{\mbb1}{\dimS}\right)\right].
\end{align}

Thus we can finally write
\begin{align}
    \mc{F}_m = \tr\left[\mc{M}\,\tr_\env\circ\Lambda_{m+1}\circ\left(\mscr{A}_m+\mscr{B}_m\right)\rho\right],
\end{align}
where
\begin{align}
    \mscr{A}_m(\rho) &:= \f{\displaystyle{\Mcirc_{n=1}^m}\left(\$_{\Lambda_n}-\Theta_{\Lambda_n}\right)\otimes\mc{I}_\syst}{\left(\dimS^2-1\right)^m}\left(\rho-\rho_\env\otimes\f{\mbb1}{\dimS}\right)\\
    \mscr{B}_m(\rho) &:= \Mcirc_{n=1}^m\Theta_{\Lambda_n}\left(\rho_\env\right)\otimes\f{\mbb1}{\dimS},
\end{align}
with $\$_{\Lambda_n}$ and $\Theta_{\Lambda_n}$ defined in Eq.\eqref{eq: def dollar appendix} and Eq.~\eqref{eq: def Theta appendix}, resp.

\section{Markovian limit}\label{appendix: Markovian limit}
For the Markovian limit we take $\Lambda_n\to\mc{I}_\env\otimes\Lambda^\markov_n$ and $\rho=\rho_\env\otimes\rho_\syst$. First, let us notice that, assuming $\Lambda_n^\markov$ are \compptp,
\begin{align}
    \$_{\Lambda_n^\markov}(\varepsilon) = \tr\left[\Lambda_n^\markov\right]\,\varepsilon, \qquad
    \Theta_{\Lambda_n^\markov}(\varepsilon) = \tr\left[\Lambda_n^\markov\left(\f{\mbb1}{\dimS}\right)\right]\varepsilon=\varepsilon,
\end{align}
for any operators $\varepsilon$ acting on $\env$ and $\sigma$ on $\syst$. Then this implies that
\begin{align}
    \tr_\env\circ\mscr{A}_m(\rho_\env\otimes\rho_\syst)&\to\tr_\env\circ\f{\displaystyle{\Mcirc_{n=1}^m}\left(\$_{\Lambda_n^\markov}-\Theta_{\Lambda_n^\markov}\right)}{\left(\dimS^2-1\right)^m}(\rho_\env)\otimes\left(\rho_\syst-\f{\mbb1}{\dimS}\right) \nonumber\\
    &= \f{\tr\left[\Lambda_1^\markov\right]-1}{\left(\dimS^2-1\right)^m}\tr_\env\circ\left[\Mcirc_{n=2}^m\left(\$_{\Lambda_n^\markov}-\Theta_{\Lambda_n^\markov}\right)\right](\rho_\env)\otimes\left(\rho_\syst-\f{\mbb1}{\dimS}\right) \nonumber\\
    &\,\vdots \nonumber\\
    &= \f{\prod_{n=1}^m \left(\tr\left[\Lambda_n^\markov\right]-1\right)}{\left(\dimS^2-1\right)^m}\left(\rho_\syst-\f{\mbb1}{\dimS}\right) \nonumber\\
    &=p_1\cdots{p}_m \left(\rho_\syst-\f{\mbb1}{\dimS}\right),
\end{align}
where here as well $p_n:=\f{\tr\left[\Lambda_n^\markov\right]-1}{\dimS^2-1}$ is the noise-strength of $\Lambda_n^\markov$, and
\begin{align}
    \tr_\env\circ\mscr{B}_m(\rho_\env\otimes\rho_\syst) &\to \tr_\env\Mcirc_{n=1}^m\Theta_{\Lambda_n^\markov}\left(\rho_\env\right)\otimes\f{\mbb1}{\dimS} = \f{\mbb1}{\dimS},
\end{align}
which implies that $\mc{F}_m\to{p}_1\cdots{p}_m\tr[\mc{M}\,\Lambda_{m+1}(\rho-\mbb1/\dimS)]+\tr[\mc{M}\,\Lambda_{m+1}(\mbb1/\dimS)]$ under Markovian noise.

\section{Finite non-Markovian noise}\label{appendix: finite non-Markovian noise}
{\textbf{Initial non-Markovian noise.}} Suppose a quantum noise process $\hat{\Upsilon}_m$ is non-Markovian up to some time-step $\ell<m$ and almost Markovian in the remaining steps, i.e. $\hat{\Upsilon}_m\simeq\Upsilon_\ell\otimes\Upsilon_{m:\ell+1}^\markov$, where $\Upsilon_{m:\ell}^\markov$ is a Markov process from time-step $\ell+1$ to time-step $m$. This effectively would mean that $\env$ is traced at the $\ell$\textsuperscript{th} step and the remaining noise maps act only on $\syst$. We can describe this by replacing the action of the noise map at the $\ell$\textsuperscript{th}-step as $\Lambda_\ell(X)\to\varepsilon\otimes\tr_\env[\Lambda_\ell(X)]$, where $X$ is the joint $\syst\env$ state at such step, and where $\varepsilon$ is some fiducial state of $\env$. The remaining noise maps will be given by $\Lambda_n\to\mc{I}_\env\otimes\Lambda_n^\markov$ for $\ell<n\leq{m+1}$ with some \compptp~maps $\Lambda^\markov_n$. This implies that
\begin{align}
    \tr_\env\circ\mscr{A}_m(\rho)&\to\tr_\env\circ\f{\displaystyle{\Mcirc_{n=1}^m}\left(\$_{\Lambda_n^\markov}-\Theta_{\Lambda_n^\markov}\right)\otimes\mc{I}_\syst}{\left(\dimS^2-1\right)^m}\left(\rho-\rho_\env\otimes\f{\mbb1}{\dimS}\right) = p_{\ell+1}\cdots{p}_m\tr_\env\circ\mscr{A}_\ell(\rho),
\end{align}
and also $\tr_\env\circ\mscr{B}_m(\rho)=\tr_\env\circ\mscr{B}_\ell(\rho)=\tr\left[\mscr{B}_\ell(\rho)\right]\mbb1/\dimS$. In particular if the final noise were trace-preserving, we would have $\tr\left[\mscr{B}_\ell(\rho)\right]=1$. In general, however, this implies
\begin{align}
    \mc{F}_m \to p_{\ell+1}\cdots{p}_m\,\tr\left[\mc{M}\Lambda_{m+1}^\markov\circ\tr_\env\circ\mscr{A}_\ell(\rho)\right] + \tr\left[\mscr{B}_\ell(\rho)\right]\, \tr\left[\mc{M}\Lambda_{m+1}^\markov\left(\f{\mbb1}{\dimS}\right)\right]\qquad\text{with}\qquad\ell<m,
    \label{eq: appendix finite mo}
\end{align}
where here again $p_n=\f{\tr\left[\Lambda_n^\markov\right]-1}{\dimS^2-1}$ is the noise-strength corresponding to $\Lambda_n^\markov$.

This means, as one would expect, that in such a case if non-Markovian noise cannot be resolved with an \rb~sequence length $\ell$, it would amount to \spam~errors, with any subsequent \asf~decay being Markovian. Notice however, that for short sequence lengths, non-Markovian noise could be resolved on average with a few runs of the \rb~protocol; as explained in the main text, this would allow to estimate the degree of non-Markovianity in the underlying process.

{\textbf{Late non-Markovian noise.}} Now consider the opposite, where the noise process is initially Markovian but somehow $\env$ stops being superfluous after some time-step $\ell<m$, i.e. $\hat{\Upsilon}\simeq\Upsilon_\ell^\markov\otimes\Upsilon_{m:\ell+1}$. Now we have
\begin{align}
    \mscr{A}_m(\rho_\env\otimes\rho_\syst)&\to\f{\displaystyle{\Mcirc_{n=1}^m}\left(\$_{\Lambda_n}-\Theta_{\Lambda_n}\right)\otimes\mc{I}_\syst}{\left(\dimS^2-1\right)^m}\left[\rho_\env\otimes\left(\rho_\syst-\f{\mbb1}{\dimS}\right)\right] = p_1\cdots{p}_\ell\,\mscr{A}_{m:\ell+1}(\rho_\env\otimes\rho_\syst),
\end{align}
where here we defined
\begin{equation}
    \mscr{A}_{m:k}(\rho) := \f{\displaystyle{\Mcirc_{n=k}^m}\left(\$_{\Lambda_n}-\Theta_{\Lambda_n}\right)\otimes\mc{I}_\syst}{\left(\dimS^2-1\right)^{m-k+1}}\left[\rho-\rho_\env\otimes\f{\mbb1}{\dimS}\right],
\end{equation}
whilst now $\mscr{B}_m(\rho_\env\otimes\rho_\syst)=\mscr{B}_{m:\ell+1}(\rho_\env\otimes\rho_\syst)$, where similarly, $\mscr{B}_{m:k}(\rho):=\Mcirc_{n=k}^m\Theta_{\Lambda_n}(\rho_\env)\otimes\mbb1/\dimS$. Thus
\begin{equation}
    \mc{F}_m \to p_1\cdots{p}_\ell\tr[\mc{M}\,\tr_\env\circ\Lambda_{m+1}\circ\mscr{A}_{m:\ell+1}(\rho)] + \tr[\mc{M}\,\tr_\env\circ\Lambda_{m+1}\circ\mscr{B}_{m:\ell+1}(\rho)]\qquad\text{with}\qquad\ell<m,
\end{equation}
so we get a similar behavior, but in this case, as we have seen, it would generally be harder to resolve non-Markovian effects in \rb~if these occur at longer sequences.

{\textbf{Blocks of finite non-Markovian noise.}} Now we may consider the case when the noise process is split in two non-Markovian processes, i.e. the first noise process somehow approximately resets the environment at step $\ell$ and the remaining noise process is also non-Markovian until step $m$, i.e. $\hat{\Upsilon}_m\simeq\Upsilon_{\ell}\otimes\Upsilon_{m:\ell+1}.$ Now the only difference from a standard non-Markovian \asf~is that at the $\ell$\textsuperscript{th} step we have $\Lambda_\ell(X)\to\varepsilon\otimes\tr_\env\circ\Lambda_\ell(X)$, where again $\varepsilon$ is some fiducial state of $\env$ and $X$ is the state of $\syst\env$ at the $\ell$\textsuperscript{th} step. This means we can write
\begin{align}
    \mscr{A}_m(\rho)&\to\mscr{A}_{m:\ell+1}\left[\varepsilon\otimes\tr_\env\circ\mscr{A}_\ell(\rho)\right] \nonumber\\
    & = \f{\displaystyle{\Mcirc_{n=\ell+1}^m}\left(\$_{\Lambda_n}-\Theta_{\Lambda_n}\right)(\varepsilon)}{\left(\dimS^2-1\right)^{m-\ell}}\otimes\tr_\env\circ\mscr{A}_\ell(\rho),
\end{align}
whilst now,
\begin{align}
    \mscr{B}_m(\rho) &= \left(\Mcirc_{n=\ell+1}^m\Theta_{\Lambda_n}\right)\left(\Mcirc_{n=1}^\ell\Theta_{\Lambda_n}\right)\rho_\env\otimes\f{\mbb1}{\dimS}\nonumber \\[0.5em]
    &= \left(\Mcirc_{n=\ell+1}^m\Theta_{\Lambda_n}\right)\tr_\syst\left[\Lambda_n\left\{\left(\Mcirc_{n=1}^{\ell-1}\Theta_{\Lambda_n}\right)\rho_\env\otimes\f{\mbb1}{\dimS}\right\}\right]\rho_\env\otimes\f{\mbb1}{\dimS}\nonumber \\[0.5em]
    &= \tr\left[\Lambda_n\left\{\left(\Mcirc_{n=1}^{\ell-1}\Theta_{\Lambda_n}\right)\rho_\env\otimes\f{\mbb1}{\dimS}\right\}\right]\left(\Mcirc_{n=\ell+1}^m\Theta_{\Lambda_n}\right)\varepsilon\otimes\f{\mbb1}{\dimS}\nonumber \\[0.5em]
    &=\tr\left[\mscr{B}_\ell(\rho)\right]\,\mscr{B}_{m:\ell+1}(\varepsilon\otimes\mbb1/\dimS),
\end{align}
so we may write
\begin{align}
    \mc{F}_m\to\tr\left\{\mc{M}\,\tr_\env\circ\Lambda_{m+1}\circ\mscr{A}_{m:\ell+1}\left[\varepsilon\otimes\tr_\env\circ\mscr{A}_\ell(\rho)\right]\right\} + \tr\left[\mscr{B}_\ell(\rho)\right] \tr\left\{\mc{M}\,\tr_\env\circ\Lambda_{m+1}\circ\mscr{B}_{m:\ell+1}(\varepsilon\otimes\mbb1/\dimS)\right\}\qquad\text{with}\qquad\ell<m.
    \label{eq: appendix F nM prod}
\end{align}

This is a much more complicated behavior, but notice that similarly now after a sequence length $\ell$, the first block of non-Markovian noise will be manifest only as \spam~errors. Also, now in essence any other possible mixture of Markovian and non-Markovian noise can be considered, e.g. if there is Markovian noise in-between this would give rise to $p$ factors within the first summand of Eq.~\eqref{eq: appendix F nM prod} containing $\mscr{A}$, and $\tr[\mscr{B}(\rho)]$ factors in the second summand.

In particular, suppose we have two blocks of finite non-Markovian noise, first one of length $k<\ell$, and then a second block of length $\ell<m$. Then we get a recursive expression for the \asf~of the form
\begin{align}
    \mc{F}_m&\to\tr\left\{\mc{M}\,\tr_\env\circ\Lambda_{m+1}\circ\mscr{A}_{m:\ell+k+1}\left[\varepsilon_\ell\otimes\tr_\env\circ\mscr{A}_{\ell:k+1}\left[\varepsilon_k\otimes\tr_\env\circ\mscr{A}_k(\rho)\right]\right]\right\}\nonumber\\ 
    &\qquad + \tr\left[\mscr{B}_k(\rho)\right]\tr\left[\mscr{B}_{\ell:k+1}(\varepsilon_k\otimes\mbb1/\dimS)\right] \tr\left\{\mc{M}\,\tr_\env\circ\Lambda_{m+1}\circ\mscr{B}_{m:\ell+k+1}(\varepsilon_\ell\otimes\mbb1/\dimS)\right\}\qquad\text{with}\qquad{k}<\ell<m.
\end{align}

If moreover the initial state is uncorrelated, $\rho=\rho_\env\otimes\rho_\syst$, we get
\begin{align}
    \mc{F}_m&\to\f{\tr\left[\displaystyle{\Mcirc_{n=1}^k}\left(\$_{\Lambda_n}-\Theta_{\Lambda_n}\right)(\rho_\env)\right]\tr\left[\displaystyle{\Mcirc_{n=k+1}^\ell}\left(\$_{\Lambda_n}-\Theta_{\Lambda_n}\right)(\varepsilon_k)\right]}{(\dimS^2-1)^\ell}\tr\left\{\mc{M}\,\tr_\env\circ\Lambda_{m+1}\circ\mscr{A}_{m:\ell+k+1}\left[\varepsilon_\ell\otimes\rho_\syst\right]\right\}\nonumber\\ 
    &\qquad + \tr\left[\mscr{B}_k(\rho_\env\otimes\rho_\syst)\right]\tr\left[\mscr{B}_{\ell:k+1}(\varepsilon_k\otimes\mbb1/\dimS)\right] \tr\left\{\mc{M}\,\tr_\env\circ\Lambda_{m+1}\circ\mscr{B}_{m:\ell+k+1}(\varepsilon_\ell\otimes\mbb1/\dimS)\right\}\qquad\text{with}\qquad{k}<\ell<m.
\end{align}

This then generalizes to blocks with finite non-Markovianity $\Delta\ell_{n}=\ell_n-(\ell_{n-1}+\ell_{n-2}-\cdots-\ell_1)$, where $\ell_1<\ell_2<\cdots<\ell_n<m$ are sequence lengths.

\section{Classical non-Markovian noise}\label{appendix: classical}
\begin{figure}[ht!]
\centering
\begin{minipage}{0.375\textwidth}
\begin{tikzpicture}
        \node[anchor=south west, inner sep=0] (image) at (0,0) {\includegraphics[width=\textwidth]{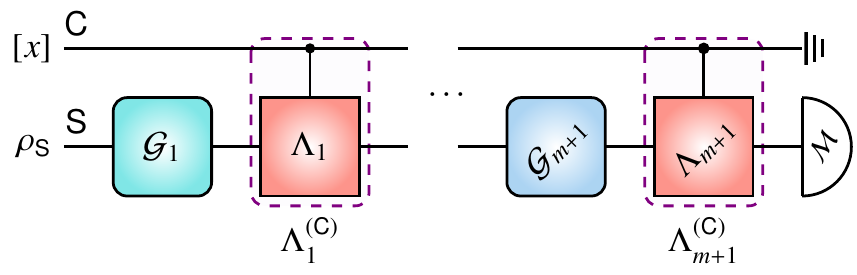}};
        \begin{scope}[x={(image.south east)},y={(image.north west)}]
        \node[below] at (0.5,-0.15) {\textsf{(a)}};
        \end{scope}
\end{tikzpicture}
\end{minipage}
	 \quad
    \begin{minipage}{0.6\textwidth}
 \begin{tikzpicture}
        \node[anchor=south west, inner sep=0] (image) at (0,0) {\includegraphics[width=\linewidth]{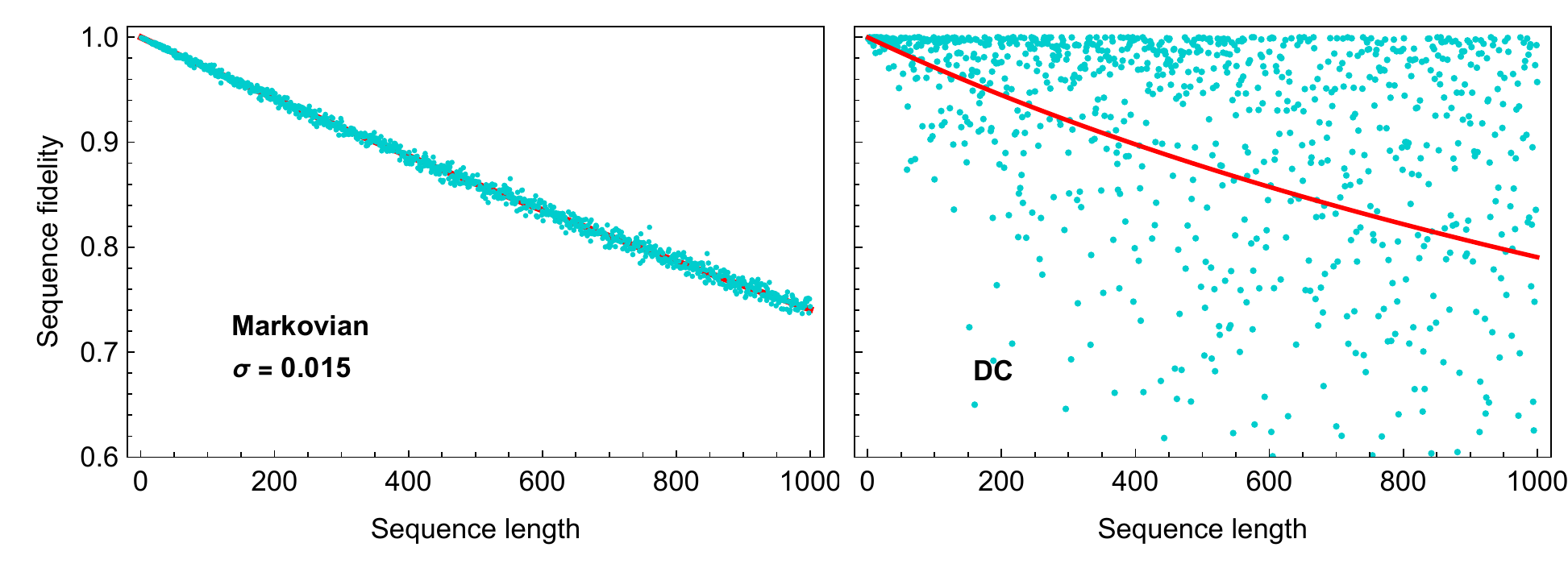}};
        \begin{scope}[x={(image.south east)},y={(image.north west)}]
        \node[below] at (0.325,0) {\textsf{(b)}};
        \node[below] at (0.785,0) {\textsf{(c)}};
      \end{scope}
    \end{tikzpicture}
    \end{minipage}
	\caption{{\textbf{Classical non-Markovian noise}}. \textsf{(a)} An \rb~sequence with classical non-Markovian noise for an initial state $\rho_\syst$, gates $\{\mc{G}_i\}$ and final measurement $\mc{M}$. Correlations are determined by a classical stochastic process $[x]$, whose output at each step $i$ controls the corresponding error map $\Lambda_i$ on $\syst$ giving some other map $\Lambda_i^{(\mathsf{C})}$. We reproduce the \asf~in~\cite{PhysRevA.93.022303} for classical dephasing noise in the cases of \textsf{(b)} classical Markovian noise and \textsf{(c)} classical DC noise with a standard deviation $\sigma=0.015$; in both cases (teal) dots represent numerical averages over 100 samples and the solid (red) curve is the analytical \asf.}
	\label{fig: classical memory}
\end{figure}
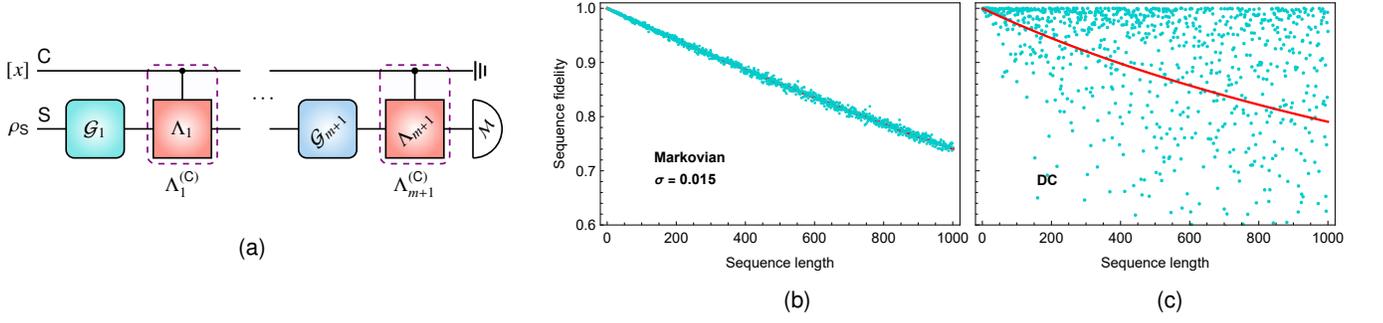

{\textbf{Classical dephasing noise.}} For the case of classical correlations we now exemplify how we may describe these through a classical memory specified by an external classical stochastic process whose outputs control the noise $\Lambda_i$ at every step $i$. We can depict a circuit for the \rb~sequence as in Fig.~\ref{fig: classical memory}\textsf{(a)}. Here we focus on the model by Ref.~\cite{PhysRevA.93.022303} and verify that we obtain the same behavior for the \asf.

In particular, such model takes a qubit system with errors $\Lambda_i^{(\mathsf{C})}(\cdot)=\lambda_i(\cdot)\lambda_i^\dg$ where $\lambda_i=\exp\left(-i\delta_i\otimes{Z}\right)=\exp\left(-i\delta_iZ\right)$ where $Z=\mathsf{diag}(1,-1)$ and with $\delta_i$ is a random variable determined by the classical external control; such variables are randomly sampled but then fixed for the whole \rb~experiment. Thus a sequence of length $m$ can be treated as a Markovian time-dependent decay $\mbb{E}\left[\mc{S}_m(\rho)\right]=p_1\cdots{p}_m\,\Lambda_{m+1}^{\textsf{(C)}}(\rho)$, where knowing the Kraus operators $\lambda$, we can compute for small $\delta$
\begin{align}
	p_i = \f{|\tr\left(\mathrm{e}^{-i\delta_i\hat{Z}}\right)|^2-1}{\dimS^2-1}=\f{4\cos^2\delta_i-1}{3}.
\end{align}

Letting the classical memory be a normally distributed discrete stochastic process $X_i\sim\mscr{N}(\mu=0,\sigma^2)$ with mean $\mu=0$ and variance $\sigma^2$, the so-called Markovian scenario considers the control operations at step $i$ giving a realization $\mathrm{X}_i=x_i$ and setting $\delta_i=x_i$. That is, all errors being independent of each other. Ignoring \spam~errors, suppose $\rho=\mc{M}=|0\rangle\!\langle0|$, so that
\begin{equation}
	\mc{F}_m^{\mathsf{C-Mark}} = \tr\left\{\mc{M}\mbb{E}\left[\mc{S}_m(\rho)\right]\right\}=\prod_{i=1}^m\left(\f{4\cos^2\delta_i-1}{3}\right).
\end{equation}

The other extreme scenario is when all noise random variables are identical so that $\delta_i=\delta$, so-called DC-noise; here the control should measure a realization with probability $p$ and update the memory with a PDF of the form $\sum_i\mbb{P}(X=x_i)\Delta(\delta-x_i)$, where here $\Delta$ is a Dirac delta distribution. Then $\mbb{E}\left[\mc{S}_m(\rho)\right]=p^m\,\mathrm{e}^{-i\delta\hat{Z}}\rho\,\mathrm{e}^{i\delta\hat{Z}}$, which similarly for $\rho=\mc{M}=|0\rangle\!\langle0|$ becomes simply
\begin{equation}
	\mc{F}_m^{\mathsf{C-DC}} = \tr\left\{\mc{M}\mbb{E}\left[\mc{S}_m(\rho)\right]\right\}=\left(\f{4\cos^2\delta-1}{3}\right)^m.
\end{equation}

For both extreme cases we see that we effectively reproduce the behavior described in~\cite{PhysRevA.93.022303}, in particular for the average sequence fidelity. Here we still have to average over the classical random variable: 
\begin{equation}
    \left\langle\mc{F}_m^{\mathsf{C-Mark}}\right\rangle = \left[\int_{-\infty}^\infty \left(\f{\exp\left(\delta^2/2\sigma^2\right)}{\sigma\sqrt{2\pi}}\right)\left(\f{4\cos^2\delta-1}{3}\right)\,d\delta\right]^m,\qquad
    \left\langle\mc{F}_m^{\mathsf{C-DC}}\right\rangle = \int_{-\infty}^\infty \left(\f{\exp\left(\delta^2/2\sigma^2\right)}{\sigma\sqrt{2\pi}}\right)\left(\f{4\cos^2\delta-1}{3}\right)^m\,d\delta.
\end{equation}

For the Markovian case, the average can be carried out to obtain a decay $\left\langle\mc{F}_m^{\mathsf{C-Mark}}\right\rangle = P^m$, where here $P$ is the true error rate together with the classical noise. For a standard deviation of $\sigma=0.015$, this gives $\left\langle\mc{F}_m^{\mathsf{C-Mark}}\right\rangle \approx (0.9997)^m$. The DC-case, as expected is more complicated, and one possibility is to expand the cosine function around $\delta=0$ to analyze the average fidelity, similar to how it is done in~\cite{PhysRevA.93.022303} with contributions up to $\delta^2$. The final behavior of $\left\langle\mc{F}_m^{\mathsf{C-DC}}\right\rangle$ differs both from an exponential and a simple product of noise-strengths. We show plots for the average fidelities in both cases with a standard deviation of $\sigma=0.015$ in Fig.~\ref{fig: classical memory}\textsf{(b)},\ref{fig: classical memory}\textsf{(c)}.

{\textbf{The shallow pocket model.}} We now consider a similar model for a qubit $\syst$ coupled to degree of freedom (\dof) on a real line, which acts as an environment. This is labeled a shallow pocket model because such \dof~cannot store energy internally. This is an interesting model for several reasons, but here mainly because it leads to completely positive and divisible dynamics of $\syst$ but it is nevertheless non-Markovian~\cite{taranto2019memory, milz2020quantum}. For \rb, however, the nature of classical correlations is what leads to a treatment of the \asf~as a time-dependent Markovian one.

The shallow pocket model now considers $\Lambda_n^{\mathsf{(C)}}(\cdot)=\lambda_n(\cdot)\lambda_n^\dg$ with $\lambda_n=\exp(-i\tau_n\,\hat{x}_n\otimes{Z})=\exp(-i\tau_n\,x_n\,Z)$, where $\hat{x}_n$ is a position operator at time-step $n$ and $\tau_n$ are time-intervals representing evolution time of the $n$\textsuperscript{th} step. This immediately implies that the average sequence is of the form $\mbb{E}\left[\mc{S}_m(\rho)\right]=p_1\cdots{p}_m\,\Lambda_{m+1}^{\textsf{(C)}}(\rho)$, where $\rho=\rho_\syst\otimes|\psi\rangle\!\langle\psi|$. The initial state of the environment \dof~is taken as $|\psi\rangle$ such that $\langle\psi|x_1\rangle=\sqrt{\gamma/\pi}\,/(x_1+i\gamma)$. Now tracing out the environment at the end of the process is equivalent to integrating $x$ over the reals with a factor $\langle\psi|x\rangle\!\langle{x}_m|\psi\rangle\delta_{xx_2}\delta_{x_2x_3}\cdots\delta_{x_m{x}_{m+1}}$. Thus we can think of the external \dof~as a classical DC noise distributed with a probability density function $|\langle\psi|x\rangle|^2$.

That is, now we have
\begin{equation}
    p_{\tau_n}(x_n) = \f{|\tr\left(\mathrm{e}^{-i\tau_n\,x_n\,\hat{Z}}\right)|^2-1}{\dimS^2-1}=\f{4\cos^2(\tau_n\,x_n)-1}{3}.
\end{equation}

Notice that all $p$'s have to be constrained to $[0,1]$, so to have a meaningful \asf~the equivalent of our distribution, namely $|\langle\psi|x\rangle|^2$, has to contain a low enough equivalent of a variance, which amounts to choosing an appropriate value for $\gamma$. Hence, now taking $\rho=|0\rangle\!\langle0|\otimes|\psi\rangle\!\langle\psi|$ and $\mc{M}=|0\rangle\!\langle0|$, we get
\begin{align}
    \mc{F}_m^{\mathsf{shallow}} = \tr\{\mc{M}\,\mbb{E}[\mc{S}_m(\rho)]\} =  \f{\gamma}{\pi}\int_{-\infty}^\infty \f{p_{\tau_1,x}\cdots{p}_{\tau_m,x}}{x^2+\gamma^2} \,dx,
\end{align}
which is somewhat harder to evaluate given that expanding around small $x$ is not a viable option. Regardless, the point we make here is that classical correlations such as the one before of dephasing noise or the shallow pocket model can be treated on \rb~with a standard Markovian time-dependent approach.

\section{Numerical calculations}\label{appendix: numerics}
\textbf{{\spam~errors.}}
As in the main text, here we consider a qubit subject to time-independent unitary noise $\Lambda(\cdot)=\lambda(\cdot)\lambda^\dg$ on a full $N$-qubit system, where $\lambda=\exp(-i\delta H)$ with $H$ given by the $N$-site Ising spin chain
\begin{align}
    H &= \sum_{i=1}^N\left(\f{J}{2}X_iX_{i+1} + h_xX_i + h_yY_i\right) = \begin{pmatrix}
 0 & h_x - i h_y & h_x - i h_y & J \\
 h_x + i h_y & 0 & J & h_x-i h_y \\
 h_x + i h_y & J & 0 & h_x-i h_y \\
 J & h_x + i h_y & h_x + i h_y & 0 \\
\end{pmatrix},
    \label{eq: appendix Ising}
\end{align}
with $X_i,Y_i$ being Pauli matrices acting on the $i$\textsuperscript{th} site. We take a closed chain so that $X_{N+1}:=X_1$. In particular, in the main text we take only $N=2$ qubits, with site $i=1$ being system $\syst$.

\begin{figure}[ht!]
\begin{tikzpicture}
    \node[anchor=south west, inner sep=0] (image) at (0,0) {\includegraphics[width=\linewidth]{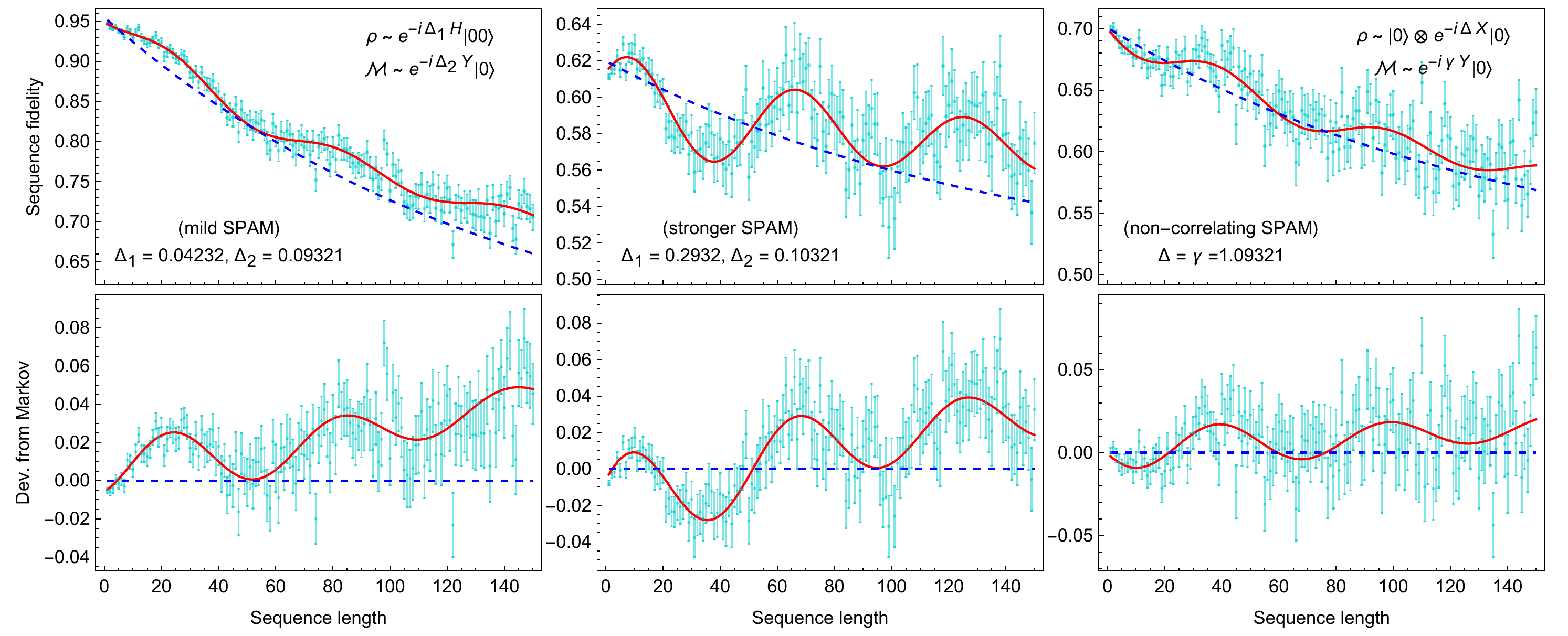}};
    \begin{scope}[x={(image.south east)},y={(image.north west)}]
    \node[below] at (0.215,0) {\textsf{(a)}};
    \node[below] at (0.535,0) {\textsf{(b)}};
    \node[below] at (0.855,0) {\textsf{(c)}};
\end{scope}
\end{tikzpicture}
    \caption{{\textbf{Effect of \spam~errors in the two-qubit spin noise in Eq.\eqref{eq: Ising numerical model}.}} In the non-Markovian case, \spam~errors result in an offset but also appear to affect the error rates. In \textsf{(a)} and \textsf{(b)} the initial state is affected by the same sequence noise with $\Lambda~\exp(-i\Delta_1{H})$ for different values of $\Delta_1$ and $\mc{M}$ is slightly rotated via $\exp(-i\Delta_2{Y})$ with a small $\Delta_2$. In \textsf{(c)} the initial state is only affected on system $\syst$ with a rotation $\exp(-i\gamma{X})$ and a relatively large $\gamma$. In all cases the sample size remained fixed at 100 samples.}
    \label{fig: appendix spam}
\end{figure}
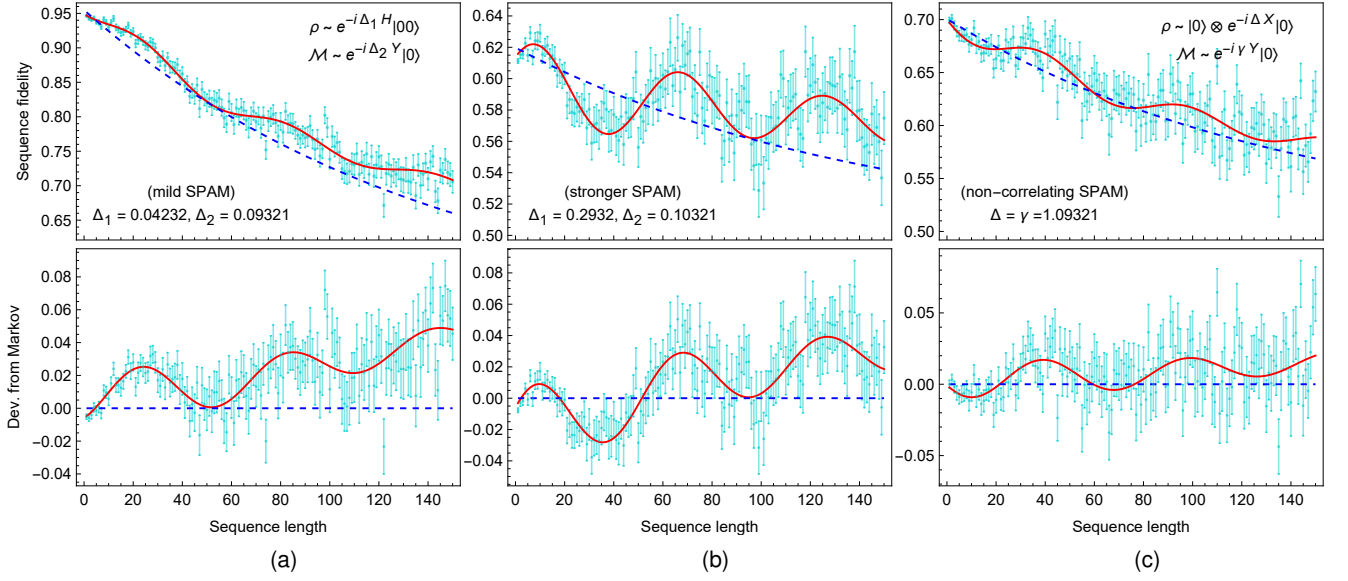

Here we pick the values $J=1.7$, $h_x=1.47$ and $h_y=-1.05$, fixing $\delta=0.029475$. To take into account \spam~errors numerically, suppose the initial state $\rho$ is previously affected by the same $\Lambda$ error for some small $\delta=\Delta_1$, and that $\mc{M}$ is slightly rotated via $\exp(-i\Delta_2{Y})$ for a small $\Delta_2$. In Fig.~\ref{fig: appendix spam} we show examples for both mild, $\Delta_1=0.04232$ and $\Delta_2 = 0.09321$, and much worse, $\Delta_1 = 0.2932$ and $\Delta_2 = 0.10321$. We also consider the case where the preparation affects only $\syst$ by some rotation $\exp(-i\gamma{X})$ with a small $\gamma$, but does not generate correlations with $\env$.

In all cases \spam~makes it harder to numerically resolve non-Markovian effects. Similar to the Markovian case, \spam~errors generate an offset of the \asf, but in general they also affect the decay rate of the errors. This can be argued to be mainly due to the correlating effect of errors but changes in the decay rates can also be seen when the preparation does not generate correlations with $\env$. The impact of \spam~in the characterization of non-Markovian noise with \rb~is thus an issue that still has to be studied in greater detail.

\begin{figure}[ht!]
    \centering
    \includegraphics[width=0.8\linewidth]{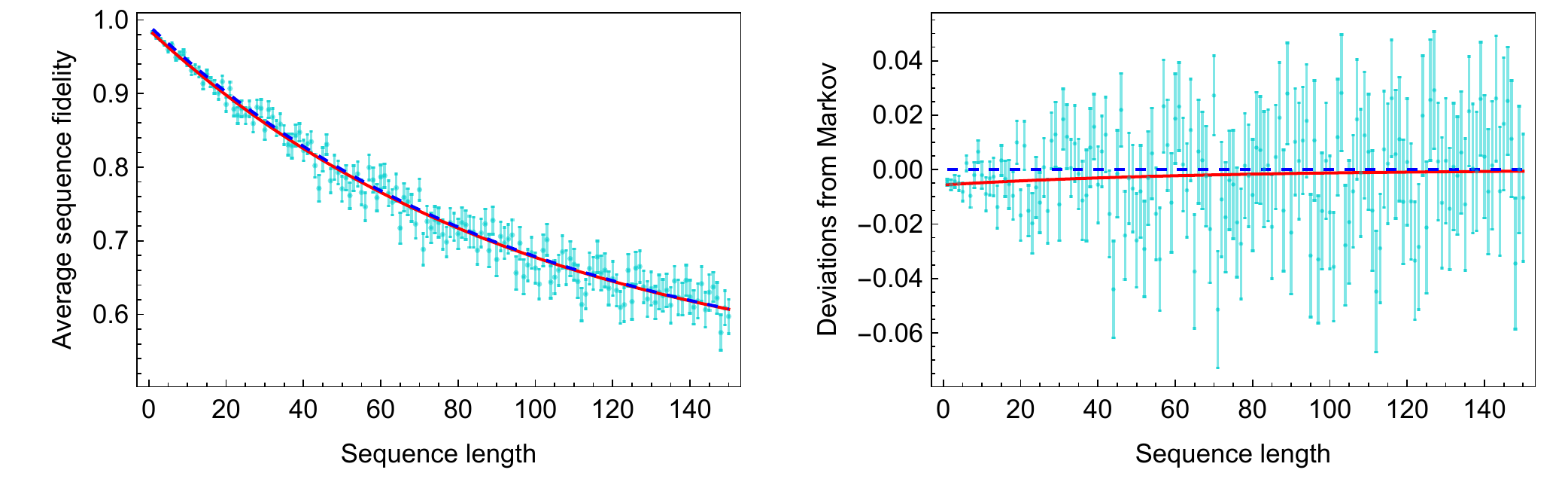}
    \caption{{\textbf{\rb~non-Markovianity blindness on an XX-spin chain.}} Despite being generically non-Markovian, the noise model of Eq.~\eqref{eq: appendix XX spin} displays virtually no deviations from a Markovian noise model when $J=1.2$, $J_y=2.7$.}
    \label{fig: appendix spin XX}
\end{figure}

\textbf{{Absence of non-exponential behavior.}} We notice that for a similar noise model for a couple of qubits,
\begin{align}
    H &= J_x X_1X_2 + J_y Y_1 Y_2 = \begin{pmatrix} 0 & 0 & 0 & J_x - J_y \\
    0 & 0 & J_x + J_y & 0\\
    0 & J_x + J_y & 0 & 0 \\
    J_x - J_y & 0 & 0& 0
    \end{pmatrix},
    \label{eq: appendix XX spin}
\end{align}
essentially no deviation from an exponential is seen. We look again at time-independent noise given by $\lambda=\exp(-iH\delta)$ with small $\delta=0.029475$ and take $\rho=|00\rangle\!\langle00|$, where one of the qubits is identified as system $\syst$ and the other one as the environment $\env$, and take $\mc{M}=|0\rangle\!\langle0|$. We show the corresponding \asf~in Fig.~\ref{fig: appendix spin XX} for the arbitrary choices $J_x=1.2$, $J_y=-2.7$.

Notice that small deviations do occur at very short sequence lengths, although they are practically negligible. While of course, we are not quantifying the non-Markovianity of the model, and also different choices of the couplings might display larger deviations, the point we want to make is that there are going to be models that are blind, or at least myopic, to non-Markovianity when employing \rb, and the circumstances when this occurs are still to be better understood. 

\begin{figure}[ht!]
    \centering
\includegraphics[width=\linewidth]{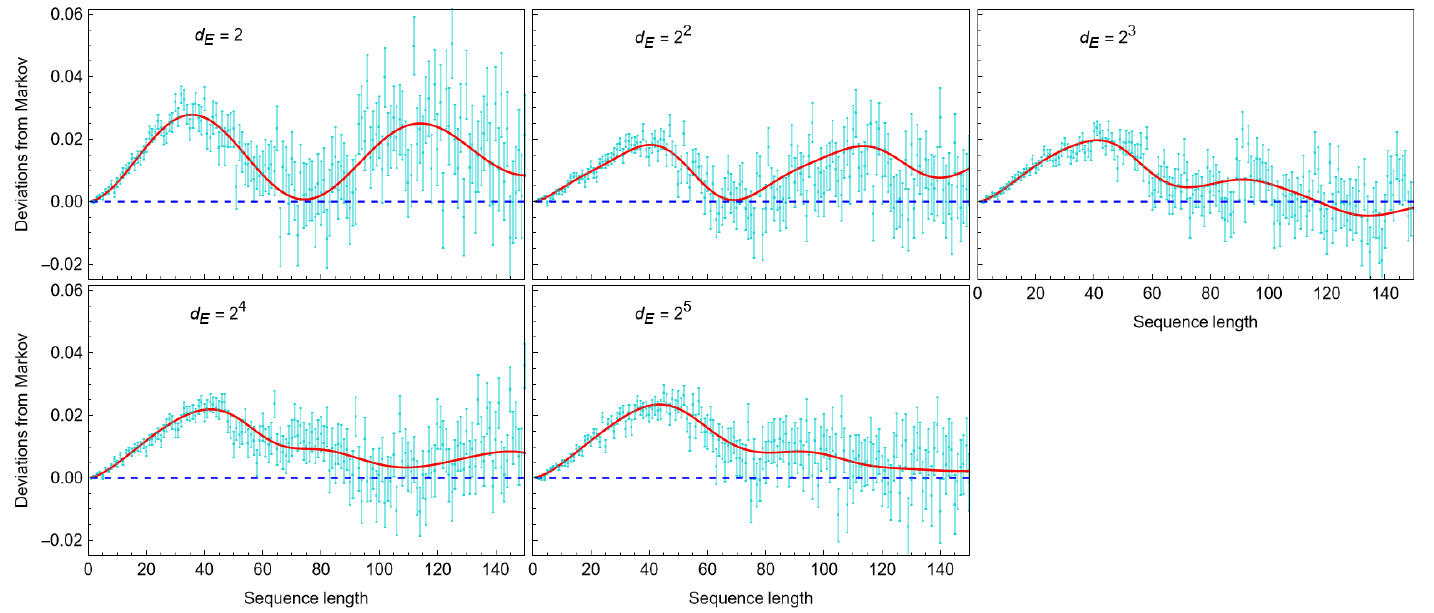}
\caption{\textbf{{Behavior of the \asf~in environment dimension for an Ising spin chain as time-independent noise.}} We take as time-independent noise Eq.~\eqref{eq: appendix Ising} for a single qubit in $\syst$ and a variable number of qubits in $\env$, with site $i=1$ being system $\syst$. For all we pick $J=1.7$, $h_x=0.9$ and $h_y=-1.05$, fixing $\delta=0.029475$ and take $\rho=|0\rangle\!\langle0|^{\otimes3}$ and $\mc{M}=|0\rangle\!\langle0|$, ignoring \spam~errors.}
\label{fig: appendix Ising dE}
\end{figure}

\textbf{{Increasing environment dimension.}} We now look at the effect of increasing the number of qubits in $\env$; noticeably the environment dimension does not show up explicitly in the main \asf~in Eq.~\eqref{eq: average fidelity nonMarkovian}. We now employ similar conditions on the Hamiltonian in Eq.~\eqref{eq: appendix Ising} for a changing value of $N$. In Fig.~\ref{fig: appendix Ising dE} we show the deviations from \rb~non-Markovianity for up to 5 environmental qubits, and notice that the non-exponential deviations get effectively damped, albeit slowly and for longer sequence lengths first. This is expected behavior, but nevertheless it is still a question what is exactly the dependence of the general non-Markovian \asf~in environment dimension.

\textbf{{Finite non-Markovian sequence lengths and non-Markovian deviations.}} Whenever we have finite non-Markovian noise, say over an initial sequence length $\sim\ell$, described by the \compp~maps $\Lambda_1,\Lambda_2,\ldots,\Lambda_\ell$, and an uncorrelated input state, by choosing to fix $\ell-1$ Cliffords after the first one to be identities, by Eq.~\eqref{eq: appendix finite mo}, we get a Markovian decay with $\mc{F}_m \simeq p_{\ell:1}{p}_{\ell+1}\cdots{p}_m\,A+B$, where
\begin{equation}
    p_{\ell:1} := \f{\tr\left[\Lambda_{\ell:1}^\markov\right]-1}{\dimS^2-1},
\end{equation}
with
\begin{equation}
    \Lambda_{\ell:1}^\markov(\cdot) := \tr_\env[\Lambda_\ell\circ\Lambda_{\ell-1}\circ\cdots\circ\Lambda_1(\varepsilon\otimes\cdot\,)],
\end{equation}
with each $\Lambda_n$ in terms of Kraus operators $\lambda_{i_n}$ acting on $\syst\env$ spaces and co-spaces as $\Lambda_n:=\sum_i\lambda_{i_n}\otimes\lambda_{i_n}^\dg$.

That is, the initial block of finite non-Markovian noise looks like a single noise map $\Lambda_{\ell:1}$ if we randomize over a single Clifford within this block, with the remaining ones set to identities.

While this is an idealized scenario, we can use it to estimate sequence lengths at which non-Markovian noise effects are relevant in an \rb~experiment. In the main text we model such a noise process with a noise map at the $n$\textsuperscript{th} step given by
\begin{equation}
    \Lambda_n^{(\ell)} := q_{n-\ell}\Lambda + (1-q_{n-\ell})\Lambda^\markov,\qquad\text{where}\quad~q_n:=\f{1}{1+\exp(n-\ell)}
\end{equation}
where here again $\Lambda=\lambda\otimes\lambda^\dg$ with $\lambda=\exp(-i\delta{H})$ where $H$ is given by Eq.~\eqref{eq: appendix Ising} and $\Lambda^\markov$ acts on $\syst$ as $\Lambda^\markov(\cdot)=\varepsilon\otimes\tr_\env[\Lambda(\cdot)]$. In the main text we also fix the values $J=1.7$, $h_x=0.5$ and $h_y=-1.05$ and set $\delta\approx0.03$, although we now pick a $\delta^\markov=2.5\delta$ for $\Lambda_n^\markov$. This implies that the noise acts jointly over the whole $\syst\env$ throughout the full process, but it acts almost fully as $\Lambda$ for $m<\ell$, whilst it turns almost to act solely on $\syst$ with $\Lambda^\markov$ for $m>\ell$.

In the top figure of Fig.~\ref{fig: plots finite MO} in the main text, we display the \asf{s} for a set of \rb~experiments with $\rho=|00\rangle\!\langle00|$ and $\mc{M}=|0\rangle\!\langle0|$ for different sets of fixed identities at sequence lengths $1,2,\ldots,8$. We fix $\ell=9$ and we describe the way in which an experimenter can estimate this value of $\ell$ from the data of the experiments alone, as well as construct a sensible time-independent Markovian \asf~with which they can quantify the amount of non-Markovian deviations; this is shown in the bottom panel of the same figure.

The procedure is the following given a single \asf, $\mc{F}_m$, displaying a non-exponential decay over a finite sequence length:
\begin{enumerate}
    \item Fix identities at sequence lengths of $\mc{F}_m$ manifestly displaying deviations from an exponential decay and run \rb~experiments for each of them, obtaining corresponding \asf{s} $\mc{F}_{m/\{i,\ldots,j\}}$, where $i,\ldots,j$ are sequence lengths at which identities were fixed.
    \item Identify the section of the original $\mc{F}_m$ manifestly displaying exponential behavior and extract the noise rate $p$ at such section.
    \item Fit an exponential to each $\mc{F}_{m/\{i,\ldots,j\}}$; Eq.~\eqref{eq: appendix finite mo} implies that the curve with an exponential rate $p_{m/\{i^\prime,\ldots,j^\prime\}}$ closest to $p$ will indicate the length at which the noise turns almost Markovian (or where non-Markovian effects become negligible).
    \item Finally, a Markovianized \asf~can be constructed with rate $p_{m/\{i^\prime,\ldots,j^\prime\}}$ and at least two reasonable constraints for the \spam~constants, such as $A+B=1$ and $A\approx{B}$ if the \spam~errors are assumed low and the decay rate is not too high, $p\approx1$.
\end{enumerate}

For the particular example in the main text, step 1 is displayed in the top panel of Fig.~\ref{fig: plots finite MO}, each over 150 samples.

For step 2, we took points $\{m,\mc{F}_m\}$ from $m=12$ to $m=30$, which more manifestly display an exponential decay. These were fitted to an exponential $f_m\approx(0.7847) (0.9325)^m+0.4915$, i.e. we extract $p\simeq0.9325$.

For step 3, we identified the closest decay rate to $p$ occurred for $\mc{F}_{m/\{1,\ldots,8\}}$, with $p_{m/\{1,\ldots,8\}}\simeq0.9278$. This indicates that $\ell\approx9$. Since we fixed $\ell=9$, this procedure is essentially identifying that the non-Markovian effects of the noise on the \asf~become negligible at sequence length $m=9$; notice that at this length $q_0=1/2$, i.e. $\Lambda_9^{(9)}=\f{1}{2}\left(\Lambda+\Lambda^\markov\right)$ so that the noise will still act jointly on $\syst\env$ with at least half probability. In this sense is $\ell$ just approximated numerically. In the bottom panel of Fig.~\ref{fig: plots finite MO}, the dot-dashed line displays the curve given by $\tilde{f}_m\simeq(0.7847) p_{m/\{1,\ldots,8\}}^m+0.4915$, showing the slight offset due to this numerical estimation.

Finally, at step 4 we simply fix $A\approx{B}$ in $\mc{F}^\markov=Ap_{m/\{1,\ldots,8\}}^m+B$ assuming low spam errors; in Fig.~\ref{fig: plots finite MO} we specifically take $A=0.5085$ and $B=0.4915$ with the demand that $B$ converges to the same value as in $f_m$ and $\tilde{f}_m$ for $m\to\infty$. As is the case for \rb, this Markovianized \asf~curve at most informs us about the gate fidelity with respect to the identity of the Markovianized noise through $p_{m/\{1,\ldots,8\}}$.

\begin{thebibliography}{80}%
\makeatletter
\providecommand \@ifxundefined [1]{%
 \@ifx{#1\undefined}
}%
\providecommand \@ifnum [1]{%
 \ifnum #1\expandafter \@firstoftwo
 \else \expandafter \@secondoftwo
 \fi
}%
\providecommand \@ifx [1]{%
 \ifx #1\expandafter \@firstoftwo
 \else \expandafter \@secondoftwo
 \fi
}%
\providecommand \natexlab [1]{#1}%
\providecommand \enquote  [1]{``#1''}%
\providecommand \bibnamefont  [1]{#1}%
\providecommand \bibfnamefont [1]{#1}%
\providecommand \citenamefont [1]{#1}%
\providecommand \href@noop [0]{\@secondoftwo}%
\providecommand \href [0]{\begingroup \@sanitize@url \@href}%
\providecommand \@href[1]{\@@startlink{#1}\@@href}%
\providecommand \@@href[1]{\endgroup#1\@@endlink}%
\providecommand \@sanitize@url [0]{\catcode `\\12\catcode `\$12\catcode
  `\&12\catcode `\#12\catcode `\^12\catcode `\_12\catcode `\%12\relax}%
\providecommand \@@startlink[1]{}%
\providecommand \@@endlink[0]{}%
\providecommand \url  [0]{\begingroup\@sanitize@url \@url }%
\providecommand \@url [1]{\endgroup\@href {#1}{\urlprefix }}%
\providecommand \urlprefix  [0]{URL }%
\providecommand \Eprint [0]{\href }%
\providecommand \doibase [0]{https://doi.org/}%
\providecommand \selectlanguage [0]{\@gobble}%
\providecommand \bibinfo  [0]{\@secondoftwo}%
\providecommand \bibfield  [0]{\@secondoftwo}%
\providecommand \translation [1]{[#1]}%
\providecommand \BibitemOpen [0]{}%
\providecommand \bibitemStop [0]{}%
\providecommand \bibitemNoStop [0]{.\EOS\space}%
\providecommand \EOS [0]{\spacefactor3000\relax}%
\providecommand \BibitemShut  [1]{\csname bibitem#1\endcsname}%
\let\auto@bib@innerbib\@empty
\bibitem [{\citenamefont {Emerson}\ \emph {et~al.}(2005)\citenamefont
  {Emerson}, \citenamefont {Alicki},\ and\ \citenamefont
  {{\.{Z}}yczkowski}}]{Emerson_2005}%
  \BibitemOpen
  \bibfield  {author} {\bibinfo {author} {\bibfnamefont {J.}~\bibnamefont
  {Emerson}}, \bibinfo {author} {\bibfnamefont {R.}~\bibnamefont {Alicki}},\
  and\ \bibinfo {author} {\bibfnamefont {K.}~\bibnamefont {{\.{Z}}yczkowski}},\
  }\bibfield  {title} {\bibinfo {title} {Scalable noise estimation with random
  unitary operators},\ }\href {https://doi.org/10.1088/1464-4266/7/10/021}
  {\bibfield  {journal} {\bibinfo  {journal} {J. Opt. B-Quantum S.O.}\ }\textbf
  {\bibinfo {volume} {7}},\ \bibinfo {pages} {S347} (\bibinfo {year}
  {2005})}\BibitemShut {NoStop}%
\bibitem [{\citenamefont {L\'evi}\ \emph {et~al.}(2007)\citenamefont {L\'evi},
  \citenamefont {L\'opez}, \citenamefont {Emerson},\ and\ \citenamefont
  {Cory}}]{Levi_2007}%
  \BibitemOpen
  \bibfield  {author} {\bibinfo {author} {\bibfnamefont {B.}~\bibnamefont
  {L\'evi}}, \bibinfo {author} {\bibfnamefont {C.~C.}\ \bibnamefont {L\'opez}},
  \bibinfo {author} {\bibfnamefont {J.}~\bibnamefont {Emerson}},\ and\ \bibinfo
  {author} {\bibfnamefont {D.~G.}\ \bibnamefont {Cory}},\ }\bibfield  {title}
  {\bibinfo {title} {Efficient error characterization in quantum information
  processing},\ }\href {https://doi.org/10.1103/PhysRevA.75.022314} {\bibfield
  {journal} {\bibinfo  {journal} {Phys. Rev. A}\ }\textbf {\bibinfo {volume}
  {75}},\ \bibinfo {pages} {022314} (\bibinfo {year} {2007})}\BibitemShut
  {NoStop}%
\bibitem [{\citenamefont {Knill}\ \emph {et~al.}(2008)\citenamefont {Knill},
  \citenamefont {Leibfried}, \citenamefont {Reichle}, \citenamefont {Britton},
  \citenamefont {Blakestad}, \citenamefont {Jost}, \citenamefont {Langer},
  \citenamefont {Ozeri}, \citenamefont {Seidelin},\ and\ \citenamefont
  {Wineland}}]{Knill2008}%
  \BibitemOpen
  \bibfield  {author} {\bibinfo {author} {\bibfnamefont {E.}~\bibnamefont
  {Knill}}, \bibinfo {author} {\bibfnamefont {D.}~\bibnamefont {Leibfried}},
  \bibinfo {author} {\bibfnamefont {R.}~\bibnamefont {Reichle}}, \bibinfo
  {author} {\bibfnamefont {J.}~\bibnamefont {Britton}}, \bibinfo {author}
  {\bibfnamefont {R.~B.}\ \bibnamefont {Blakestad}}, \bibinfo {author}
  {\bibfnamefont {J.~D.}\ \bibnamefont {Jost}}, \bibinfo {author}
  {\bibfnamefont {C.}~\bibnamefont {Langer}}, \bibinfo {author} {\bibfnamefont
  {R.}~\bibnamefont {Ozeri}}, \bibinfo {author} {\bibfnamefont
  {S.}~\bibnamefont {Seidelin}},\ and\ \bibinfo {author} {\bibfnamefont
  {D.~J.}\ \bibnamefont {Wineland}},\ }\bibfield  {title} {\bibinfo {title}
  {Randomized benchmarking of quantum gates},\ }\href
  {https://doi.org/10.1103/PhysRevA.77.012307} {\bibfield  {journal} {\bibinfo
  {journal} {Phys. Rev. A}\ }\textbf {\bibinfo {volume} {77}},\ \bibinfo
  {pages} {012307} (\bibinfo {year} {2008})}\BibitemShut {NoStop}%
\bibitem [{\citenamefont {Magesan}\ \emph {et~al.}(2011)\citenamefont
  {Magesan}, \citenamefont {Gambetta},\ and\ \citenamefont
  {Emerson}}]{PhysRevLett.106.180504}%
  \BibitemOpen
  \bibfield  {author} {\bibinfo {author} {\bibfnamefont {E.}~\bibnamefont
  {Magesan}}, \bibinfo {author} {\bibfnamefont {J.~M.}\ \bibnamefont
  {Gambetta}},\ and\ \bibinfo {author} {\bibfnamefont {J.}~\bibnamefont
  {Emerson}},\ }\bibfield  {title} {\bibinfo {title} {Scalable and robust
  randomized benchmarking of quantum processes},\ }\href
  {https://doi.org/10.1103/PhysRevLett.106.180504} {\bibfield  {journal}
  {\bibinfo  {journal} {Phys. Rev. Lett.}\ }\textbf {\bibinfo {volume} {106}},\
  \bibinfo {pages} {180504} (\bibinfo {year} {2011})}\BibitemShut {NoStop}%
\bibitem [{\citenamefont {Magesan}\ \emph
  {et~al.}(2012{\natexlab{a}})\citenamefont {Magesan}, \citenamefont
  {Gambetta},\ and\ \citenamefont {Emerson}}]{PhysRevA.85.042311}%
  \BibitemOpen
  \bibfield  {author} {\bibinfo {author} {\bibfnamefont {E.}~\bibnamefont
  {Magesan}}, \bibinfo {author} {\bibfnamefont {J.~M.}\ \bibnamefont
  {Gambetta}},\ and\ \bibinfo {author} {\bibfnamefont {J.}~\bibnamefont
  {Emerson}},\ }\bibfield  {title} {\bibinfo {title} {Characterizing quantum
  gates via randomized benchmarking},\ }\href
  {https://doi.org/10.1103/PhysRevA.85.042311} {\bibfield  {journal} {\bibinfo
  {journal} {Phys. Rev. A}\ }\textbf {\bibinfo {volume} {85}},\ \bibinfo
  {pages} {042311} (\bibinfo {year} {2012}{\natexlab{a}})}\BibitemShut
  {NoStop}%
\bibitem [{\citenamefont {Helsen}\ \emph {et~al.}(2020)\citenamefont {Helsen},
  \citenamefont {Roth}, \citenamefont {Onorati}, \citenamefont {Werner},\ and\
  \citenamefont {Eisert}}]{helsen2020general}%
  \BibitemOpen
  \bibfield  {author} {\bibinfo {author} {\bibfnamefont {J.}~\bibnamefont
  {Helsen}}, \bibinfo {author} {\bibfnamefont {I.}~\bibnamefont {Roth}},
  \bibinfo {author} {\bibfnamefont {E.}~\bibnamefont {Onorati}}, \bibinfo
  {author} {\bibfnamefont {A.~H.}\ \bibnamefont {Werner}},\ and\ \bibinfo
  {author} {\bibfnamefont {J.}~\bibnamefont {Eisert}},\ }\href@noop {}
  {\bibinfo {title} {A general framework for randomized benchmarking}}
  (\bibinfo {year} {2020}),\ \Eprint {https://arxiv.org/abs/2010.07974}
  {arXiv:2010.07974 [quant-ph]} \BibitemShut {NoStop}%
\bibitem [{\citenamefont {Chuang}\ and\ \citenamefont
  {Nielsen}(1997)}]{Chuang_1997}%
  \BibitemOpen
  \bibfield  {author} {\bibinfo {author} {\bibfnamefont {I.~L.}\ \bibnamefont
  {Chuang}}\ and\ \bibinfo {author} {\bibfnamefont {M.~A.}\ \bibnamefont
  {Nielsen}},\ }\bibfield  {title} {\bibinfo {title} {Prescription for
  experimental determination of the dynamics of a quantum black box},\ }\href
  {https://doi.org/10.1080/09500349708231894} {\bibfield  {journal} {\bibinfo
  {journal} {J. Mod. Optic}\ }\textbf {\bibinfo {volume} {44}},\ \bibinfo
  {pages} {2455–2467} (\bibinfo {year} {1997})}\BibitemShut {NoStop}%
\bibitem [{\citenamefont {Roth}\ \emph {et~al.}(2018)\citenamefont {Roth},
  \citenamefont {Kueng}, \citenamefont {Kimmel}, \citenamefont {Liu},
  \citenamefont {Gross}, \citenamefont {Eisert},\ and\ \citenamefont
  {Kliesch}}]{PhysRevLett.121.170502}%
  \BibitemOpen
  \bibfield  {author} {\bibinfo {author} {\bibfnamefont {I.}~\bibnamefont
  {Roth}}, \bibinfo {author} {\bibfnamefont {R.}~\bibnamefont {Kueng}},
  \bibinfo {author} {\bibfnamefont {S.}~\bibnamefont {Kimmel}}, \bibinfo
  {author} {\bibfnamefont {Y.-K.}\ \bibnamefont {Liu}}, \bibinfo {author}
  {\bibfnamefont {D.}~\bibnamefont {Gross}}, \bibinfo {author} {\bibfnamefont
  {J.}~\bibnamefont {Eisert}},\ and\ \bibinfo {author} {\bibfnamefont
  {M.}~\bibnamefont {Kliesch}},\ }\bibfield  {title} {\bibinfo {title}
  {Recovering quantum gates from few average gate fidelities},\ }\href
  {https://doi.org/10.1103/PhysRevLett.121.170502} {\bibfield  {journal}
  {\bibinfo  {journal} {Phys. Rev. Lett.}\ }\textbf {\bibinfo {volume} {121}},\
  \bibinfo {pages} {170502} (\bibinfo {year} {2018})}\BibitemShut {NoStop}%
\bibitem [{\citenamefont {Wallman}\ and\ \citenamefont
  {Flammia}(2014)}]{Wallman_2014}%
  \BibitemOpen
  \bibfield  {author} {\bibinfo {author} {\bibfnamefont {J.~J.}\ \bibnamefont
  {Wallman}}\ and\ \bibinfo {author} {\bibfnamefont {S.~T.}\ \bibnamefont
  {Flammia}},\ }\bibfield  {title} {\bibinfo {title} {Randomized benchmarking
  with confidence},\ }\href {https://doi.org/10.1088/1367-2630/16/10/103032}
  {\bibfield  {journal} {\bibinfo  {journal} {New J. Phys.}\ }\textbf {\bibinfo
  {volume} {16}},\ \bibinfo {pages} {103032} (\bibinfo {year}
  {2014})}\BibitemShut {NoStop}%
\bibitem [{\citenamefont {Mohseni}\ \emph {et~al.}(2008)\citenamefont
  {Mohseni}, \citenamefont {Rezakhani},\ and\ \citenamefont
  {Lidar}}]{PhysRevA.77.032322}%
  \BibitemOpen
  \bibfield  {author} {\bibinfo {author} {\bibfnamefont {M.}~\bibnamefont
  {Mohseni}}, \bibinfo {author} {\bibfnamefont {A.~T.}\ \bibnamefont
  {Rezakhani}},\ and\ \bibinfo {author} {\bibfnamefont {D.~A.}\ \bibnamefont
  {Lidar}},\ }\bibfield  {title} {\bibinfo {title} {Quantum-process tomography:
  Resource analysis of different strategies},\ }\href
  {https://doi.org/10.1103/PhysRevA.77.032322} {\bibfield  {journal} {\bibinfo
  {journal} {Phys. Rev. A}\ }\textbf {\bibinfo {volume} {77}},\ \bibinfo
  {pages} {032322} (\bibinfo {year} {2008})}\BibitemShut {NoStop}%
\bibitem [{\citenamefont {Nielsen}\ \emph {et~al.}(2020)\citenamefont
  {Nielsen}, \citenamefont {Gamble}, \citenamefont {Rudinger}, \citenamefont
  {Scholten}, \citenamefont {Young},\ and\ \citenamefont
  {Blume-Kohout}}]{nielsen2020gate}%
  \BibitemOpen
  \bibfield  {author} {\bibinfo {author} {\bibfnamefont {E.}~\bibnamefont
  {Nielsen}}, \bibinfo {author} {\bibfnamefont {J.~K.}\ \bibnamefont {Gamble}},
  \bibinfo {author} {\bibfnamefont {K.}~\bibnamefont {Rudinger}}, \bibinfo
  {author} {\bibfnamefont {T.}~\bibnamefont {Scholten}}, \bibinfo {author}
  {\bibfnamefont {K.}~\bibnamefont {Young}},\ and\ \bibinfo {author}
  {\bibfnamefont {R.}~\bibnamefont {Blume-Kohout}},\ }\href@noop {} {\bibinfo
  {title} {Gate set tomography}} (\bibinfo {year} {2020}),\ \Eprint
  {https://arxiv.org/abs/2009.07301} {arXiv:2009.07301 [quant-ph]} \BibitemShut
  {NoStop}%
\bibitem [{\citenamefont {Gross}\ \emph {et~al.}(2010)\citenamefont {Gross},
  \citenamefont {Liu}, \citenamefont {Flammia}, \citenamefont {Becker},\ and\
  \citenamefont {Eisert}}]{PhysRevLett.105.150401}%
  \BibitemOpen
  \bibfield  {author} {\bibinfo {author} {\bibfnamefont {D.}~\bibnamefont
  {Gross}}, \bibinfo {author} {\bibfnamefont {Y.-K.}\ \bibnamefont {Liu}},
  \bibinfo {author} {\bibfnamefont {S.~T.}\ \bibnamefont {Flammia}}, \bibinfo
  {author} {\bibfnamefont {S.}~\bibnamefont {Becker}},\ and\ \bibinfo {author}
  {\bibfnamefont {J.}~\bibnamefont {Eisert}},\ }\bibfield  {title} {\bibinfo
  {title} {Quantum state tomography via compressed sensing},\ }\href
  {https://doi.org/10.1103/PhysRevLett.105.150401} {\bibfield  {journal}
  {\bibinfo  {journal} {Phys. Rev. Lett.}\ }\textbf {\bibinfo {volume} {105}},\
  \bibinfo {pages} {150401} (\bibinfo {year} {2010})}\BibitemShut {NoStop}%
\bibitem [{\citenamefont {Flammia}\ \emph {et~al.}(2012)\citenamefont
  {Flammia}, \citenamefont {Gross}, \citenamefont {Liu},\ and\ \citenamefont
  {Eisert}}]{Flammia_2012}%
  \BibitemOpen
  \bibfield  {author} {\bibinfo {author} {\bibfnamefont {S.~T.}\ \bibnamefont
  {Flammia}}, \bibinfo {author} {\bibfnamefont {D.}~\bibnamefont {Gross}},
  \bibinfo {author} {\bibfnamefont {Y.-K.}\ \bibnamefont {Liu}},\ and\ \bibinfo
  {author} {\bibfnamefont {J.}~\bibnamefont {Eisert}},\ }\bibfield  {title}
  {\bibinfo {title} {Quantum tomography via compressed sensing: error bounds,
  sample complexity and efficient estimators},\ }\href
  {https://doi.org/10.1088/1367-2630/14/9/095022} {\bibfield  {journal}
  {\bibinfo  {journal} {New J. Phys.}\ }\textbf {\bibinfo {volume} {14}},\
  \bibinfo {pages} {095022} (\bibinfo {year} {2012})}\BibitemShut {NoStop}%
\bibitem [{\citenamefont {Flammia}\ and\ \citenamefont
  {Liu}(2011)}]{PhysRevLett.106.230501}%
  \BibitemOpen
  \bibfield  {author} {\bibinfo {author} {\bibfnamefont {S.~T.}\ \bibnamefont
  {Flammia}}\ and\ \bibinfo {author} {\bibfnamefont {Y.-K.}\ \bibnamefont
  {Liu}},\ }\bibfield  {title} {\bibinfo {title} {Direct fidelity estimation
  from few pauli measurements},\ }\href
  {https://doi.org/10.1103/PhysRevLett.106.230501} {\bibfield  {journal}
  {\bibinfo  {journal} {Phys. Rev. Lett.}\ }\textbf {\bibinfo {volume} {106}},\
  \bibinfo {pages} {230501} (\bibinfo {year} {2011})}\BibitemShut {NoStop}%
\bibitem [{\citenamefont {da~Silva}\ \emph {et~al.}(2011)\citenamefont
  {da~Silva}, \citenamefont {Landon-Cardinal},\ and\ \citenamefont
  {Poulin}}]{PhysRevLett.107.210404}%
  \BibitemOpen
  \bibfield  {author} {\bibinfo {author} {\bibfnamefont {M.~P.}\ \bibnamefont
  {da~Silva}}, \bibinfo {author} {\bibfnamefont {O.}~\bibnamefont
  {Landon-Cardinal}},\ and\ \bibinfo {author} {\bibfnamefont {D.}~\bibnamefont
  {Poulin}},\ }\bibfield  {title} {\bibinfo {title} {Practical characterization
  of quantum devices without tomography},\ }\href
  {https://doi.org/10.1103/PhysRevLett.107.210404} {\bibfield  {journal}
  {\bibinfo  {journal} {Phys. Rev. Lett.}\ }\textbf {\bibinfo {volume} {107}},\
  \bibinfo {pages} {210404} (\bibinfo {year} {2011})}\BibitemShut {NoStop}%
\bibitem [{\citenamefont {Moussa}\ \emph {et~al.}(2012)\citenamefont {Moussa},
  \citenamefont {da~Silva}, \citenamefont {Ryan},\ and\ \citenamefont
  {Laflamme}}]{PhysRevLett.109.070504}%
  \BibitemOpen
  \bibfield  {author} {\bibinfo {author} {\bibfnamefont {O.}~\bibnamefont
  {Moussa}}, \bibinfo {author} {\bibfnamefont {M.~P.}\ \bibnamefont
  {da~Silva}}, \bibinfo {author} {\bibfnamefont {C.~A.}\ \bibnamefont {Ryan}},\
  and\ \bibinfo {author} {\bibfnamefont {R.}~\bibnamefont {Laflamme}},\
  }\bibfield  {title} {\bibinfo {title} {Practical experimental certification
  of computational quantum gates using a twirling procedure},\ }\href
  {https://doi.org/10.1103/PhysRevLett.109.070504} {\bibfield  {journal}
  {\bibinfo  {journal} {Phys. Rev. Lett.}\ }\textbf {\bibinfo {volume} {109}},\
  \bibinfo {pages} {070504} (\bibinfo {year} {2012})}\BibitemShut {NoStop}%
\bibitem [{Note1()}]{Note1}%
  \BibitemOpen
  \bibinfo {note} {In Ref.~\cite {PhysRevLett.119.130502}, it was pointed out
  that this relation between the \protect \textsf {ASF}~and the gate fidelity
  of the noise is not unique due to gauge invariance. An in-depth discussion
  can be seen in Ref.~\cite {Wallman_2018, merkel2018randomized} with an
  overview and generalisation in \cite {helsen2020general}.}\BibitemShut
  {Stop}%
\bibitem [{\citenamefont {Proctor}\ \emph {et~al.}(2017)\citenamefont
  {Proctor}, \citenamefont {Rudinger}, \citenamefont {Young}, \citenamefont
  {Sarovar},\ and\ \citenamefont {Blume-Kohout}}]{PhysRevLett.119.130502}%
  \BibitemOpen
  \bibfield  {author} {\bibinfo {author} {\bibfnamefont {T.}~\bibnamefont
  {Proctor}}, \bibinfo {author} {\bibfnamefont {K.}~\bibnamefont {Rudinger}},
  \bibinfo {author} {\bibfnamefont {K.}~\bibnamefont {Young}}, \bibinfo
  {author} {\bibfnamefont {M.}~\bibnamefont {Sarovar}},\ and\ \bibinfo {author}
  {\bibfnamefont {R.}~\bibnamefont {Blume-Kohout}},\ }\bibfield  {title}
  {\bibinfo {title} {What randomized benchmarking actually measures},\ }\href
  {https://doi.org/10.1103/PhysRevLett.119.130502} {\bibfield  {journal}
  {\bibinfo  {journal} {Phys. Rev. Lett.}\ }\textbf {\bibinfo {volume} {119}},\
  \bibinfo {pages} {130502} (\bibinfo {year} {2017})}\BibitemShut {NoStop}%
\bibitem [{\citenamefont {Wallman}(2018)}]{Wallman_2018}%
  \BibitemOpen
  \bibfield  {author} {\bibinfo {author} {\bibfnamefont {J.~J.}\ \bibnamefont
  {Wallman}},\ }\bibfield  {title} {\bibinfo {title} {Randomized benchmarking
  with gate-dependent noise},\ }\href
  {https://doi.org/10.22331/q-2018-01-29-47} {\bibfield  {journal} {\bibinfo
  {journal} {Quantum}\ }\textbf {\bibinfo {volume} {2}},\ \bibinfo {pages} {47}
  (\bibinfo {year} {2018})}\BibitemShut {NoStop}%
\bibitem [{\citenamefont {Helsen}\ \emph {et~al.}(2019)\citenamefont {Helsen},
  \citenamefont {Xue}, \citenamefont {Vandersypen},\ and\ \citenamefont
  {Wehner}}]{Helsen2019}%
  \BibitemOpen
  \bibfield  {author} {\bibinfo {author} {\bibfnamefont {J.}~\bibnamefont
  {Helsen}}, \bibinfo {author} {\bibfnamefont {X.}~\bibnamefont {Xue}},
  \bibinfo {author} {\bibfnamefont {L.~M.~K.}\ \bibnamefont {Vandersypen}},\
  and\ \bibinfo {author} {\bibfnamefont {S.}~\bibnamefont {Wehner}},\
  }\bibfield  {title} {\bibinfo {title} {A new class of efficient randomized
  benchmarking protocols},\ }\href {https://doi.org/10.1038/s41534-019-0182-7}
  {\bibfield  {journal} {\bibinfo  {journal} {npj Quantum Information}\
  }\textbf {\bibinfo {volume} {5}},\ \bibinfo {pages} {71} (\bibinfo {year}
  {2019})}\BibitemShut {NoStop}%
\bibitem [{\citenamefont {Carignan-Dugas}\ \emph {et~al.}(2015)\citenamefont
  {Carignan-Dugas}, \citenamefont {Wallman},\ and\ \citenamefont
  {Emerson}}]{PhysRevA.92.060302}%
  \BibitemOpen
  \bibfield  {author} {\bibinfo {author} {\bibfnamefont {A.}~\bibnamefont
  {Carignan-Dugas}}, \bibinfo {author} {\bibfnamefont {J.~J.}\ \bibnamefont
  {Wallman}},\ and\ \bibinfo {author} {\bibfnamefont {J.}~\bibnamefont
  {Emerson}},\ }\bibfield  {title} {\bibinfo {title} {Characterizing universal
  gate sets via dihedral benchmarking},\ }\href
  {https://doi.org/10.1103/PhysRevA.92.060302} {\bibfield  {journal} {\bibinfo
  {journal} {Phys. Rev. A}\ }\textbf {\bibinfo {volume} {92}},\ \bibinfo
  {pages} {060302} (\bibinfo {year} {2015})}\BibitemShut {NoStop}%
\bibitem [{\citenamefont {Hashagen}\ \emph {et~al.}(2018)\citenamefont
  {Hashagen}, \citenamefont {Flammia}, \citenamefont {Gross},\ and\
  \citenamefont {Wallman}}]{Hashagen_2018}%
  \BibitemOpen
  \bibfield  {author} {\bibinfo {author} {\bibfnamefont {A.~K.}\ \bibnamefont
  {Hashagen}}, \bibinfo {author} {\bibfnamefont {S.~T.}\ \bibnamefont
  {Flammia}}, \bibinfo {author} {\bibfnamefont {D.}~\bibnamefont {Gross}},\
  and\ \bibinfo {author} {\bibfnamefont {J.~J.}\ \bibnamefont {Wallman}},\
  }\bibfield  {title} {\bibinfo {title} {Real randomized benchmarking},\ }\href
  {https://doi.org/10.22331/q-2018-08-22-85} {\bibfield  {journal} {\bibinfo
  {journal} {Quantum}\ }\textbf {\bibinfo {volume} {2}},\ \bibinfo {pages} {85}
  (\bibinfo {year} {2018})}\BibitemShut {NoStop}%
\bibitem [{\citenamefont {Magesan}\ \emph
  {et~al.}(2012{\natexlab{b}})\citenamefont {Magesan}, \citenamefont
  {Gambetta}, \citenamefont {Johnson}, \citenamefont {Ryan}, \citenamefont
  {Chow}, \citenamefont {Merkel}, \citenamefont {da~Silva}, \citenamefont
  {Keefe}, \citenamefont {Rothwell}, \citenamefont {Ohki}, \citenamefont
  {Ketchen},\ and\ \citenamefont {Steffen}}]{PhysRevLett.109.080505}%
  \BibitemOpen
  \bibfield  {author} {\bibinfo {author} {\bibfnamefont {E.}~\bibnamefont
  {Magesan}}, \bibinfo {author} {\bibfnamefont {J.~M.}\ \bibnamefont
  {Gambetta}}, \bibinfo {author} {\bibfnamefont {B.~R.}\ \bibnamefont
  {Johnson}}, \bibinfo {author} {\bibfnamefont {C.~A.}\ \bibnamefont {Ryan}},
  \bibinfo {author} {\bibfnamefont {J.~M.}\ \bibnamefont {Chow}}, \bibinfo
  {author} {\bibfnamefont {S.~T.}\ \bibnamefont {Merkel}}, \bibinfo {author}
  {\bibfnamefont {M.~P.}\ \bibnamefont {da~Silva}}, \bibinfo {author}
  {\bibfnamefont {G.~A.}\ \bibnamefont {Keefe}}, \bibinfo {author}
  {\bibfnamefont {M.~B.}\ \bibnamefont {Rothwell}}, \bibinfo {author}
  {\bibfnamefont {T.~A.}\ \bibnamefont {Ohki}}, \bibinfo {author}
  {\bibfnamefont {M.~B.}\ \bibnamefont {Ketchen}},\ and\ \bibinfo {author}
  {\bibfnamefont {M.}~\bibnamefont {Steffen}},\ }\bibfield  {title} {\bibinfo
  {title} {Efficient measurement of quantum gate error by interleaved
  randomized benchmarking},\ }\href
  {https://doi.org/10.1103/PhysRevLett.109.080505} {\bibfield  {journal}
  {\bibinfo  {journal} {Phys. Rev. Lett.}\ }\textbf {\bibinfo {volume} {109}},\
  \bibinfo {pages} {080505} (\bibinfo {year} {2012}{\natexlab{b}})}\BibitemShut
  {NoStop}%
\bibitem [{\citenamefont {Wallman}\ \emph {et~al.}(2015)\citenamefont
  {Wallman}, \citenamefont {Granade}, \citenamefont {Harper},\ and\
  \citenamefont {Flammia}}]{Wallman_2015}%
  \BibitemOpen
  \bibfield  {author} {\bibinfo {author} {\bibfnamefont {J.}~\bibnamefont
  {Wallman}}, \bibinfo {author} {\bibfnamefont {C.}~\bibnamefont {Granade}},
  \bibinfo {author} {\bibfnamefont {R.}~\bibnamefont {Harper}},\ and\ \bibinfo
  {author} {\bibfnamefont {S.~T.}\ \bibnamefont {Flammia}},\ }\bibfield
  {title} {\bibinfo {title} {Estimating the coherence of noise},\ }\href
  {https://doi.org/10.1088/1367-2630/17/11/113020} {\bibfield  {journal}
  {\bibinfo  {journal} {New J. Phys.}\ }\textbf {\bibinfo {volume} {17}},\
  \bibinfo {pages} {113020} (\bibinfo {year} {2015})}\BibitemShut {NoStop}%
\bibitem [{\citenamefont {Ryan}\ \emph {et~al.}(2009)\citenamefont {Ryan},
  \citenamefont {Laforest},\ and\ \citenamefont {Laflamme}}]{Ryan_2009}%
  \BibitemOpen
  \bibfield  {author} {\bibinfo {author} {\bibfnamefont {C.~A.}\ \bibnamefont
  {Ryan}}, \bibinfo {author} {\bibfnamefont {M.}~\bibnamefont {Laforest}},\
  and\ \bibinfo {author} {\bibfnamefont {R.}~\bibnamefont {Laflamme}},\
  }\bibfield  {title} {\bibinfo {title} {Randomized benchmarking of single- and
  multi-qubit control in liquid-state {NMR} quantum information processing},\
  }\href {https://doi.org/10.1088/1367-2630/11/1/013034} {\bibfield  {journal}
  {\bibinfo  {journal} {New Journal of Physics}\ }\textbf {\bibinfo {volume}
  {11}},\ \bibinfo {pages} {013034} (\bibinfo {year} {2009})}\BibitemShut
  {NoStop}%
\bibitem [{\citenamefont {Park}\ \emph {et~al.}(2016)\citenamefont {Park},
  \citenamefont {Feng}, \citenamefont {Rahimi}, \citenamefont {Baugh},\ and\
  \citenamefont {Laflamme}}]{Park2016}%
  \BibitemOpen
  \bibfield  {author} {\bibinfo {author} {\bibfnamefont {D.~K.}\ \bibnamefont
  {Park}}, \bibinfo {author} {\bibfnamefont {G.}~\bibnamefont {Feng}}, \bibinfo
  {author} {\bibfnamefont {R.}~\bibnamefont {Rahimi}}, \bibinfo {author}
  {\bibfnamefont {J.}~\bibnamefont {Baugh}},\ and\ \bibinfo {author}
  {\bibfnamefont {R.}~\bibnamefont {Laflamme}},\ }\bibfield  {title} {\bibinfo
  {title} {Randomized benchmarking of quantum gates implemented by electron
  spin resonance},\ }\href
  {https://doi.org/https://doi.org/10.1016/j.jmr.2016.04.010} {\bibfield
  {journal} {\bibinfo  {journal} {J. Magn. Reson.}\ }\textbf {\bibinfo {volume}
  {267}},\ \bibinfo {pages} {68} (\bibinfo {year} {2016})}\BibitemShut
  {NoStop}%
\bibitem [{\citenamefont {Epstein}\ \emph {et~al.}(2014)\citenamefont
  {Epstein}, \citenamefont {Cross}, \citenamefont {Magesan},\ and\
  \citenamefont {Gambetta}}]{PhysRevA.89.062321}%
  \BibitemOpen
  \bibfield  {author} {\bibinfo {author} {\bibfnamefont {J.~M.}\ \bibnamefont
  {Epstein}}, \bibinfo {author} {\bibfnamefont {A.~W.}\ \bibnamefont {Cross}},
  \bibinfo {author} {\bibfnamefont {E.}~\bibnamefont {Magesan}},\ and\ \bibinfo
  {author} {\bibfnamefont {J.~M.}\ \bibnamefont {Gambetta}},\ }\bibfield
  {title} {\bibinfo {title} {Investigating the limits of randomized
  benchmarking protocols},\ }\href {https://doi.org/10.1103/PhysRevA.89.062321}
  {\bibfield  {journal} {\bibinfo  {journal} {Phys. Rev. A}\ }\textbf {\bibinfo
  {volume} {89}},\ \bibinfo {pages} {062321} (\bibinfo {year}
  {2014})}\BibitemShut {NoStop}%
\bibitem [{\citenamefont {Fogarty}\ \emph {et~al.}(2015)\citenamefont
  {Fogarty}, \citenamefont {Veldhorst}, \citenamefont {Harper}, \citenamefont
  {Yang}, \citenamefont {Bartlett}, \citenamefont {Flammia},\ and\
  \citenamefont {Dzurak}}]{PhysRevA.92.022326}%
  \BibitemOpen
  \bibfield  {author} {\bibinfo {author} {\bibfnamefont {M.~A.}\ \bibnamefont
  {Fogarty}}, \bibinfo {author} {\bibfnamefont {M.}~\bibnamefont {Veldhorst}},
  \bibinfo {author} {\bibfnamefont {R.}~\bibnamefont {Harper}}, \bibinfo
  {author} {\bibfnamefont {C.~H.}\ \bibnamefont {Yang}}, \bibinfo {author}
  {\bibfnamefont {S.~D.}\ \bibnamefont {Bartlett}}, \bibinfo {author}
  {\bibfnamefont {S.~T.}\ \bibnamefont {Flammia}},\ and\ \bibinfo {author}
  {\bibfnamefont {A.~S.}\ \bibnamefont {Dzurak}},\ }\bibfield  {title}
  {\bibinfo {title} {Nonexponential fidelity decay in randomized benchmarking
  with low-frequency noise},\ }\href
  {https://doi.org/10.1103/PhysRevA.92.022326} {\bibfield  {journal} {\bibinfo
  {journal} {Phys. Rev. A}\ }\textbf {\bibinfo {volume} {92}},\ \bibinfo
  {pages} {022326} (\bibinfo {year} {2015})}\BibitemShut {NoStop}%
\bibitem [{\citenamefont {Mavadia}\ \emph {et~al.}(2018)\citenamefont
  {Mavadia}, \citenamefont {Edmunds}, \citenamefont {Hempel}, \citenamefont
  {Ball}, \citenamefont {Roy}, \citenamefont {Stace},\ and\ \citenamefont
  {Biercuk}}]{Mavadia_2018}%
  \BibitemOpen
  \bibfield  {author} {\bibinfo {author} {\bibfnamefont {S.}~\bibnamefont
  {Mavadia}}, \bibinfo {author} {\bibfnamefont {C.~L.}\ \bibnamefont
  {Edmunds}}, \bibinfo {author} {\bibfnamefont {C.}~\bibnamefont {Hempel}},
  \bibinfo {author} {\bibfnamefont {H.}~\bibnamefont {Ball}}, \bibinfo {author}
  {\bibfnamefont {F.}~\bibnamefont {Roy}}, \bibinfo {author} {\bibfnamefont
  {T.~M.}\ \bibnamefont {Stace}},\ and\ \bibinfo {author} {\bibfnamefont
  {M.~J.}\ \bibnamefont {Biercuk}},\ }\bibfield  {title} {\bibinfo {title}
  {Experimental quantum verification in the presence of temporally correlated
  noise},\ }\href {https://doi.org/10.1038/s41534-017-0052-0} {\bibfield
  {journal} {\bibinfo  {journal} {npj Quantum Inf.}\ }\textbf {\bibinfo
  {volume} {4}},\ \bibinfo {pages} {7} (\bibinfo {year} {2018})}\BibitemShut
  {NoStop}%
\bibitem [{\citenamefont {Ball}\ \emph {et~al.}(2016)\citenamefont {Ball},
  \citenamefont {Stace}, \citenamefont {Flammia},\ and\ \citenamefont
  {Biercuk}}]{PhysRevA.93.022303}%
  \BibitemOpen
  \bibfield  {author} {\bibinfo {author} {\bibfnamefont {H.}~\bibnamefont
  {Ball}}, \bibinfo {author} {\bibfnamefont {T.~M.}\ \bibnamefont {Stace}},
  \bibinfo {author} {\bibfnamefont {S.~T.}\ \bibnamefont {Flammia}},\ and\
  \bibinfo {author} {\bibfnamefont {M.~J.}\ \bibnamefont {Biercuk}},\
  }\bibfield  {title} {\bibinfo {title} {Effect of noise correlations on
  randomized benchmarking},\ }\href
  {https://doi.org/10.1103/PhysRevA.93.022303} {\bibfield  {journal} {\bibinfo
  {journal} {Phys. Rev. A}\ }\textbf {\bibinfo {volume} {93}},\ \bibinfo
  {pages} {022303} (\bibinfo {year} {2016})}\BibitemShut {NoStop}%
\bibitem [{\citenamefont {Fong}\ and\ \citenamefont
  {Merkel}(2017)}]{fong2017randomized}%
  \BibitemOpen
  \bibfield  {author} {\bibinfo {author} {\bibfnamefont {B.~H.}\ \bibnamefont
  {Fong}}\ and\ \bibinfo {author} {\bibfnamefont {S.~T.}\ \bibnamefont
  {Merkel}},\ }\href@noop {} {\bibinfo {title} {Randomized benchmarking,
  correlated noise, and ising models}} (\bibinfo {year} {2017}),\ \Eprint
  {https://arxiv.org/abs/1703.09747} {arXiv:1703.09747 [quant-ph]} \BibitemShut
  {NoStop}%
\bibitem [{\citenamefont {Qi}\ and\ \citenamefont
  {Ng}(2021)}]{PhysRevA.103.022607}%
  \BibitemOpen
  \bibfield  {author} {\bibinfo {author} {\bibfnamefont {J.}~\bibnamefont
  {Qi}}\ and\ \bibinfo {author} {\bibfnamefont {H.~K.}\ \bibnamefont {Ng}},\
  }\bibfield  {title} {\bibinfo {title} {Randomized benchmarking in the
  presence of time-correlated dephasing noise},\ }\href
  {https://doi.org/10.1103/PhysRevA.103.022607} {\bibfield  {journal} {\bibinfo
   {journal} {Phys. Rev. A}\ }\textbf {\bibinfo {volume} {103}},\ \bibinfo
  {pages} {022607} (\bibinfo {year} {2021})}\BibitemShut {NoStop}%
\bibitem [{\citenamefont {Gambetta}\ \emph {et~al.}(2012)\citenamefont
  {Gambetta}, \citenamefont {C\'orcoles}, \citenamefont {Merkel}, \citenamefont
  {Johnson}, \citenamefont {Smolin}, \citenamefont {Chow}, \citenamefont
  {Ryan}, \citenamefont {Rigetti}, \citenamefont {Poletto}, \citenamefont
  {Ohki}, \citenamefont {Ketchen},\ and\ \citenamefont
  {Steffen}}]{PhysRevLett.109.240504}%
  \BibitemOpen
  \bibfield  {author} {\bibinfo {author} {\bibfnamefont {J.~M.}\ \bibnamefont
  {Gambetta}}, \bibinfo {author} {\bibfnamefont {A.~D.}\ \bibnamefont
  {C\'orcoles}}, \bibinfo {author} {\bibfnamefont {S.~T.}\ \bibnamefont
  {Merkel}}, \bibinfo {author} {\bibfnamefont {B.~R.}\ \bibnamefont {Johnson}},
  \bibinfo {author} {\bibfnamefont {J.~A.}\ \bibnamefont {Smolin}}, \bibinfo
  {author} {\bibfnamefont {J.~M.}\ \bibnamefont {Chow}}, \bibinfo {author}
  {\bibfnamefont {C.~A.}\ \bibnamefont {Ryan}}, \bibinfo {author}
  {\bibfnamefont {C.}~\bibnamefont {Rigetti}}, \bibinfo {author} {\bibfnamefont
  {S.}~\bibnamefont {Poletto}}, \bibinfo {author} {\bibfnamefont {T.~A.}\
  \bibnamefont {Ohki}}, \bibinfo {author} {\bibfnamefont {M.~B.}\ \bibnamefont
  {Ketchen}},\ and\ \bibinfo {author} {\bibfnamefont {M.}~\bibnamefont
  {Steffen}},\ }\bibfield  {title} {\bibinfo {title} {Characterization of
  addressability by simultaneous randomized benchmarking},\ }\href
  {https://doi.org/10.1103/PhysRevLett.109.240504} {\bibfield  {journal}
  {\bibinfo  {journal} {Phys. Rev. Lett.}\ }\textbf {\bibinfo {volume} {109}},\
  \bibinfo {pages} {240504} (\bibinfo {year} {2012})}\BibitemShut {NoStop}%
\bibitem [{\citenamefont {Milz}\ and\ \citenamefont
  {Modi}(2021)}]{milz2020quantum}%
  \BibitemOpen
  \bibfield  {author} {\bibinfo {author} {\bibfnamefont {S.}~\bibnamefont
  {Milz}}\ and\ \bibinfo {author} {\bibfnamefont {K.}~\bibnamefont {Modi}},\
  }\bibfield  {title} {\bibinfo {title} {Quantum stochastic processes and
  quantum non-markovian phenomena},\ }\href
  {https://doi.org/10.1103/PRXQuantum.2.030201} {\bibfield  {journal} {\bibinfo
   {journal} {PRX Quantum}\ }\textbf {\bibinfo {volume} {2}},\ \bibinfo {pages}
  {030201} (\bibinfo {year} {2021})}\BibitemShut {NoStop}%
\bibitem [{\citenamefont {Chiribella}\ \emph {et~al.}(2008)\citenamefont
  {Chiribella}, \citenamefont {D'Ariano},\ and\ \citenamefont
  {Perinotti}}]{PhysRevLett.101.060401}%
  \BibitemOpen
  \bibfield  {author} {\bibinfo {author} {\bibfnamefont {G.}~\bibnamefont
  {Chiribella}}, \bibinfo {author} {\bibfnamefont {G.~M.}\ \bibnamefont
  {D'Ariano}},\ and\ \bibinfo {author} {\bibfnamefont {P.}~\bibnamefont
  {Perinotti}},\ }\bibfield  {title} {\bibinfo {title} {Quantum circuit
  architecture},\ }\href {https://doi.org/10.1103/PhysRevLett.101.060401}
  {\bibfield  {journal} {\bibinfo  {journal} {Phys. Rev. Lett.}\ }\textbf
  {\bibinfo {volume} {101}},\ \bibinfo {pages} {060401} (\bibinfo {year}
  {2008})}\BibitemShut {NoStop}%
\bibitem [{\citenamefont {Chiribella}\ \emph {et~al.}(2009)\citenamefont
  {Chiribella}, \citenamefont {D'Ariano},\ and\ \citenamefont
  {Perinotti}}]{PhysRevA.80.022339}%
  \BibitemOpen
  \bibfield  {author} {\bibinfo {author} {\bibfnamefont {G.}~\bibnamefont
  {Chiribella}}, \bibinfo {author} {\bibfnamefont {G.~M.}\ \bibnamefont
  {D'Ariano}},\ and\ \bibinfo {author} {\bibfnamefont {P.}~\bibnamefont
  {Perinotti}},\ }\bibfield  {title} {\bibinfo {title} {Theoretical framework
  for quantum networks},\ }\href {https://doi.org/10.1103/PhysRevA.80.022339}
  {\bibfield  {journal} {\bibinfo  {journal} {Phys. Rev. A}\ }\textbf {\bibinfo
  {volume} {80}},\ \bibinfo {pages} {022339} (\bibinfo {year}
  {2009})}\BibitemShut {NoStop}%
\bibitem [{\citenamefont {Costa}\ and\ \citenamefont
  {Shrapnel}(2016)}]{Costa_2016}%
  \BibitemOpen
  \bibfield  {author} {\bibinfo {author} {\bibfnamefont {F.}~\bibnamefont
  {Costa}}\ and\ \bibinfo {author} {\bibfnamefont {S.}~\bibnamefont
  {Shrapnel}},\ }\bibfield  {title} {\bibinfo {title} {Quantum causal
  modelling},\ }\href {https://doi.org/10.1088/1367-2630/18/6/063032}
  {\bibfield  {journal} {\bibinfo  {journal} {New J. Phys.}\ }\textbf {\bibinfo
  {volume} {18}},\ \bibinfo {pages} {063032} (\bibinfo {year}
  {2016})}\BibitemShut {NoStop}%
\bibitem [{\citenamefont {Pollock}\ \emph
  {et~al.}(2018{\natexlab{a}})\citenamefont {Pollock}, \citenamefont
  {Rodr\'{\i}guez-Rosario}, \citenamefont {Frauenheim}, \citenamefont
  {Paternostro},\ and\ \citenamefont {Modi}}]{PhysRevLett.120.040405}%
  \BibitemOpen
  \bibfield  {author} {\bibinfo {author} {\bibfnamefont {F.~A.}\ \bibnamefont
  {Pollock}}, \bibinfo {author} {\bibfnamefont {C.}~\bibnamefont
  {Rodr\'{\i}guez-Rosario}}, \bibinfo {author} {\bibfnamefont {T.}~\bibnamefont
  {Frauenheim}}, \bibinfo {author} {\bibfnamefont {M.}~\bibnamefont
  {Paternostro}},\ and\ \bibinfo {author} {\bibfnamefont {K.}~\bibnamefont
  {Modi}},\ }\bibfield  {title} {\bibinfo {title} {Operational markov condition
  for quantum processes},\ }\href
  {https://doi.org/10.1103/PhysRevLett.120.040405} {\bibfield  {journal}
  {\bibinfo  {journal} {Phys. Rev. Lett.}\ }\textbf {\bibinfo {volume} {120}},\
  \bibinfo {pages} {040405} (\bibinfo {year} {2018}{\natexlab{a}})}\BibitemShut
  {NoStop}%
\bibitem [{\citenamefont {Pollock}\ \emph
  {et~al.}(2018{\natexlab{b}})\citenamefont {Pollock}, \citenamefont
  {Rodr\'{\i}guez-Rosario}, \citenamefont {Frauenheim}, \citenamefont
  {Paternostro},\ and\ \citenamefont {Modi}}]{PhysRevA.97.012127}%
  \BibitemOpen
  \bibfield  {author} {\bibinfo {author} {\bibfnamefont {F.~A.}\ \bibnamefont
  {Pollock}}, \bibinfo {author} {\bibfnamefont {C.}~\bibnamefont
  {Rodr\'{\i}guez-Rosario}}, \bibinfo {author} {\bibfnamefont {T.}~\bibnamefont
  {Frauenheim}}, \bibinfo {author} {\bibfnamefont {M.}~\bibnamefont
  {Paternostro}},\ and\ \bibinfo {author} {\bibfnamefont {K.}~\bibnamefont
  {Modi}},\ }\bibfield  {title} {\bibinfo {title} {Non-markovian quantum
  processes: Complete framework and efficient characterization},\ }\href
  {https://doi.org/10.1103/PhysRevA.97.012127} {\bibfield  {journal} {\bibinfo
  {journal} {Phys. Rev. A}\ }\textbf {\bibinfo {volume} {97}},\ \bibinfo
  {pages} {012127} (\bibinfo {year} {2018}{\natexlab{b}})}\BibitemShut
  {NoStop}%
\bibitem [{\citenamefont {Gottesman}(1997)}]{Gottesman1997}%
  \BibitemOpen
  \bibfield  {author} {\bibinfo {author} {\bibfnamefont {D.}~\bibnamefont
  {Gottesman}},\ }\href@noop {} {\bibinfo {title} {Stabilizer codes and quantum
  error correction}} (\bibinfo {year} {1997}),\ \Eprint
  {https://arxiv.org/abs/quant-ph/9705052} {arXiv:quant-ph/9705052 [quant-ph]}
  \BibitemShut {NoStop}%
\bibitem [{\citenamefont {Gottesman}(1998)}]{Gottesman1998}%
  \BibitemOpen
  \bibfield  {author} {\bibinfo {author} {\bibfnamefont {D.}~\bibnamefont
  {Gottesman}},\ }\href@noop {} {\bibinfo {title} {The heisenberg
  representation of quantum computers}} (\bibinfo {year} {1998}),\ \Eprint
  {https://arxiv.org/abs/quant-ph/9807006} {arXiv:quant-ph/9807006 [quant-ph]}
  \BibitemShut {NoStop}%
\bibitem [{\citenamefont {Aaronson}\ and\ \citenamefont
  {Gottesman}(2004)}]{PhysRevA.70.052328}%
  \BibitemOpen
  \bibfield  {author} {\bibinfo {author} {\bibfnamefont {S.}~\bibnamefont
  {Aaronson}}\ and\ \bibinfo {author} {\bibfnamefont {D.}~\bibnamefont
  {Gottesman}},\ }\bibfield  {title} {\bibinfo {title} {Improved simulation of
  stabilizer circuits},\ }\href {https://doi.org/10.1103/PhysRevA.70.052328}
  {\bibfield  {journal} {\bibinfo  {journal} {Phys. Rev. A}\ }\textbf {\bibinfo
  {volume} {70}},\ \bibinfo {pages} {052328} (\bibinfo {year}
  {2004})}\BibitemShut {NoStop}%
\bibitem [{\citenamefont {Dankert}\ \emph {et~al.}(2009)\citenamefont
  {Dankert}, \citenamefont {Cleve}, \citenamefont {Emerson},\ and\
  \citenamefont {Livine}}]{PhysRevA.80.012304}%
  \BibitemOpen
  \bibfield  {author} {\bibinfo {author} {\bibfnamefont {C.}~\bibnamefont
  {Dankert}}, \bibinfo {author} {\bibfnamefont {R.}~\bibnamefont {Cleve}},
  \bibinfo {author} {\bibfnamefont {J.}~\bibnamefont {Emerson}},\ and\ \bibinfo
  {author} {\bibfnamefont {E.}~\bibnamefont {Livine}},\ }\bibfield  {title}
  {\bibinfo {title} {Exact and approximate unitary 2-designs and their
  application to fidelity estimation},\ }\href
  {https://doi.org/10.1103/PhysRevA.80.012304} {\bibfield  {journal} {\bibinfo
  {journal} {Phys. Rev. A}\ }\textbf {\bibinfo {volume} {80}},\ \bibinfo
  {pages} {012304} (\bibinfo {year} {2009})}\BibitemShut {NoStop}%
\bibitem [{\citenamefont {Nakata}\ \emph {et~al.}(2021)\citenamefont {Nakata},
  \citenamefont {Zhao}, \citenamefont {Okuda}, \citenamefont {Bannai},
  \citenamefont {Suzuki}, \citenamefont {Tamiya}, \citenamefont {Heya},
  \citenamefont {Yan}, \citenamefont {Zuo}, \citenamefont {Tamate},
  \citenamefont {Tabuchi},\ and\ \citenamefont {Nakamura}}]{nakata2021quantum}%
  \BibitemOpen
  \bibfield  {author} {\bibinfo {author} {\bibfnamefont {Y.}~\bibnamefont
  {Nakata}}, \bibinfo {author} {\bibfnamefont {D.}~\bibnamefont {Zhao}},
  \bibinfo {author} {\bibfnamefont {T.}~\bibnamefont {Okuda}}, \bibinfo
  {author} {\bibfnamefont {E.}~\bibnamefont {Bannai}}, \bibinfo {author}
  {\bibfnamefont {Y.}~\bibnamefont {Suzuki}}, \bibinfo {author} {\bibfnamefont
  {S.}~\bibnamefont {Tamiya}}, \bibinfo {author} {\bibfnamefont
  {K.}~\bibnamefont {Heya}}, \bibinfo {author} {\bibfnamefont {Z.}~\bibnamefont
  {Yan}}, \bibinfo {author} {\bibfnamefont {K.}~\bibnamefont {Zuo}}, \bibinfo
  {author} {\bibfnamefont {S.}~\bibnamefont {Tamate}}, \bibinfo {author}
  {\bibfnamefont {Y.}~\bibnamefont {Tabuchi}},\ and\ \bibinfo {author}
  {\bibfnamefont {Y.}~\bibnamefont {Nakamura}},\ }\href@noop {} {\bibinfo
  {title} {Quantum circuits for exact unitary $t$-designs and applications to
  higher-order randomized benchmarking}} (\bibinfo {year} {2021}),\ \Eprint
  {https://arxiv.org/abs/2102.12617} {arXiv:2102.12617 [quant-ph]} \BibitemShut
  {NoStop}%
\bibitem [{\citenamefont {Taranto}(2020)}]{Taranto_2020}%
  \BibitemOpen
  \bibfield  {author} {\bibinfo {author} {\bibfnamefont {P.}~\bibnamefont
  {Taranto}},\ }\bibfield  {title} {\bibinfo {title} {Memory effects in quantum
  processes},\ }\href {https://doi.org/10.1142/s0219749919410028} {\bibfield
  {journal} {\bibinfo  {journal} {Int. J. Quantum Inf.}\ }\textbf {\bibinfo
  {volume} {18}},\ \bibinfo {pages} {1941002} (\bibinfo {year}
  {2020})}\BibitemShut {NoStop}%
\bibitem [{\citenamefont {Milz}\ \emph {et~al.}(2017)\citenamefont {Milz},
  \citenamefont {Pollock},\ and\ \citenamefont {Modi}}]{Milz_2017}%
  \BibitemOpen
  \bibfield  {author} {\bibinfo {author} {\bibfnamefont {S.}~\bibnamefont
  {Milz}}, \bibinfo {author} {\bibfnamefont {F.~A.}\ \bibnamefont {Pollock}},\
  and\ \bibinfo {author} {\bibfnamefont {K.}~\bibnamefont {Modi}},\ }\bibfield
  {title} {\bibinfo {title} {An introduction to operational quantum dynamics},\
  }\href {https://doi.org/10.1142/s1230161217400169} {\bibfield  {journal}
  {\bibinfo  {journal} {Open Syst. Inf. Dyn.}\ }\textbf {\bibinfo {volume}
  {24}},\ \bibinfo {pages} {1740016} (\bibinfo {year} {2017})}\BibitemShut
  {NoStop}%
\bibitem [{\citenamefont {Watrous}(2018)}]{watrous2018theory}%
  \BibitemOpen
  \bibfield  {author} {\bibinfo {author} {\bibfnamefont {J.}~\bibnamefont
  {Watrous}},\ }\href {https://books.google.com.au/books?id=GRNSDwAAQBAJ}
  {\emph {\bibinfo {title} {The Theory of Quantum Information}}}\ (\bibinfo
  {publisher} {Cambridge University Press},\ \bibinfo {year}
  {2018})\BibitemShut {NoStop}%
\bibitem [{Note2()}]{Note2}%
  \BibitemOpen
  \bibinfo {note} {Strictly speaking, here we are defining $\protect \mathfrak
  {C}_m=\protect \mathds 1\otimes \protect \mathfrak {A}_m$, where $\protect
  \mathfrak {A}_m$ is the Choi state of the sequence of gates $\protect
  \mathcal {G}_i$, and $\protect \mathcal {G}_i$ can be defined to act on
  either auxiliary space $\protect \mathsf {A}_i$ or $\protect \mathsf {B}_i$,
  the choice only depends on what auxiliary space the swap $\protect \mathscr
  {S}_i$ on the definition of $\Upsilon _m$ swaps with, so that the contraction
  $\Upsilon _m\protect \mathfrak {C}_m^\protect \mathrm {T}$ contracts the
  correct spaces.}\BibitemShut {Stop}%
\bibitem [{\citenamefont {Milz}\ \emph {et~al.}(2020)\citenamefont {Milz},
  \citenamefont {Sakuldee}, \citenamefont {Pollock},\ and\ \citenamefont
  {Modi}}]{Milz2020kolmogorovextension}%
  \BibitemOpen
  \bibfield  {author} {\bibinfo {author} {\bibfnamefont {S.}~\bibnamefont
  {Milz}}, \bibinfo {author} {\bibfnamefont {F.}~\bibnamefont {Sakuldee}},
  \bibinfo {author} {\bibfnamefont {F.~A.}\ \bibnamefont {Pollock}},\ and\
  \bibinfo {author} {\bibfnamefont {K.}~\bibnamefont {Modi}},\ }\bibfield
  {title} {\bibinfo {title} {Kolmogorov extension theorem for (quantum) causal
  modelling and general probabilistic theories},\ }\href
  {https://doi.org/10.22331/q-2020-04-20-255} {\bibfield  {journal} {\bibinfo
  {journal} {{Quantum}}\ }\textbf {\bibinfo {volume} {4}},\ \bibinfo {pages}
  {255} (\bibinfo {year} {2020})}\BibitemShut {NoStop}%
\bibitem [{\citenamefont {Nurdin}\ and\ \citenamefont
  {Gough}(2021)}]{nurdin2021heisenberg}%
  \BibitemOpen
  \bibfield  {author} {\bibinfo {author} {\bibfnamefont {H.~I.}\ \bibnamefont
  {Nurdin}}\ and\ \bibinfo {author} {\bibfnamefont {J.~E.}\ \bibnamefont
  {Gough}},\ }\href@noop {} {\bibinfo {title} {From the {H}eisenberg to the
  {S}chr\"{o}dinger picture: Quantum stochastic processes and process tensors}}
  (\bibinfo {year} {2021}),\ \Eprint {https://arxiv.org/abs/2109.09256}
  {arXiv:2109.09256 [quant-ph]} \BibitemShut {NoStop}%
\bibitem [{\citenamefont {Milz}\ \emph {et~al.}(2019)\citenamefont {Milz},
  \citenamefont {Kim}, \citenamefont {Pollock},\ and\ \citenamefont
  {Modi}}]{PhysRevLett.123.040401}%
  \BibitemOpen
  \bibfield  {author} {\bibinfo {author} {\bibfnamefont {S.}~\bibnamefont
  {Milz}}, \bibinfo {author} {\bibfnamefont {M.~S.}\ \bibnamefont {Kim}},
  \bibinfo {author} {\bibfnamefont {F.~A.}\ \bibnamefont {Pollock}},\ and\
  \bibinfo {author} {\bibfnamefont {K.}~\bibnamefont {Modi}},\ }\bibfield
  {title} {\bibinfo {title} {Completely positive divisibility does not mean
  markovianity},\ }\href {https://doi.org/10.1103/PhysRevLett.123.040401}
  {\bibfield  {journal} {\bibinfo  {journal} {Phys. Rev. Lett.}\ }\textbf
  {\bibinfo {volume} {123}},\ \bibinfo {pages} {040401} (\bibinfo {year}
  {2019})}\BibitemShut {NoStop}%
\bibitem [{\citenamefont {Figueroa-Romero}\ \emph {et~al.}(2019)\citenamefont
  {Figueroa-Romero}, \citenamefont {Modi},\ and\ \citenamefont
  {Pollock}}]{2019almostmarkovian}%
  \BibitemOpen
  \bibfield  {author} {\bibinfo {author} {\bibfnamefont {P.}~\bibnamefont
  {Figueroa-Romero}}, \bibinfo {author} {\bibfnamefont {K.}~\bibnamefont
  {Modi}},\ and\ \bibinfo {author} {\bibfnamefont {F.~A.}\ \bibnamefont
  {Pollock}},\ }\bibfield  {title} {\bibinfo {title} {Almost {M}arkovian
  processes from closed dynamics},\ }\href
  {https://doi.org/10.22331/q-2019-04-30-136} {\bibfield  {journal} {\bibinfo
  {journal} {{Quantum}}\ }\textbf {\bibinfo {volume} {3}},\ \bibinfo {pages}
  {136} (\bibinfo {year} {2019})}\BibitemShut {NoStop}%
\bibitem [{\citenamefont {Figueroa-Romero}\ \emph {et~al.}(2021)\citenamefont
  {Figueroa-Romero}, \citenamefont {Pollock},\ and\ \citenamefont
  {Modi}}]{2020markovianization}%
  \BibitemOpen
  \bibfield  {author} {\bibinfo {author} {\bibfnamefont {P.}~\bibnamefont
  {Figueroa-Romero}}, \bibinfo {author} {\bibfnamefont {F.~A.}\ \bibnamefont
  {Pollock}},\ and\ \bibinfo {author} {\bibfnamefont {K.}~\bibnamefont
  {Modi}},\ }\bibfield  {title} {\bibinfo {title} {Markovianization with
  approximate unitary designs},\ }\href
  {https://doi.org/10.1038/s42005-021-00629-w} {\bibfield  {journal} {\bibinfo
  {journal} {Commun. Phys.}\ }\textbf {\bibinfo {volume} {4}},\ \bibinfo
  {pages} {127} (\bibinfo {year} {2021})}\BibitemShut {NoStop}%
\bibitem [{Note3()}]{Note3}%
  \BibitemOpen
  \bibinfo {note} {The action of the map $\protect \$_{\Lambda _n}$ can
  alternatively be written as $\protect \$_{\Lambda }(\cdot )=\DOTSB \sum@
  \slimits@ _\mu \protect \tr _\protect \mathsf {S}(\lambda _\mu )(\cdot
  )\protect \tr _\protect \mathsf {S}(\lambda _\mu ^\dagger )$ with $\lambda
  _\mu $ the Kraus operators of $\Lambda $.}\BibitemShut {Stop}%
\bibitem [{\citenamefont {Modi}(2012)}]{modiscirep}%
  \BibitemOpen
  \bibfield  {author} {\bibinfo {author} {\bibfnamefont {K.}~\bibnamefont
  {Modi}},\ }\bibfield  {title} {\bibinfo {title} {Operational approach to open
  dynamics and quantifying initial correlations},\ }\href
  {http://www.nature.com/articles/srep00581} {\bibfield  {journal} {\bibinfo
  {journal} {Sci. Rep.}\ }\textbf {\bibinfo {volume} {2}},\ \bibinfo {pages}
  {581} (\bibinfo {year} {2012})}\BibitemShut {NoStop}%
\bibitem [{\citenamefont {Ringbauer}\ \emph {et~al.}(2015)\citenamefont
  {Ringbauer}, \citenamefont {Wood}, \citenamefont {Modi}, \citenamefont
  {Gilchrist}, \citenamefont {White},\ and\ \citenamefont
  {Fedrizzi}}]{PhysRevLett.114.090402}%
  \BibitemOpen
  \bibfield  {author} {\bibinfo {author} {\bibfnamefont {M.}~\bibnamefont
  {Ringbauer}}, \bibinfo {author} {\bibfnamefont {C.~J.}\ \bibnamefont {Wood}},
  \bibinfo {author} {\bibfnamefont {K.}~\bibnamefont {Modi}}, \bibinfo {author}
  {\bibfnamefont {A.}~\bibnamefont {Gilchrist}}, \bibinfo {author}
  {\bibfnamefont {A.~G.}\ \bibnamefont {White}},\ and\ \bibinfo {author}
  {\bibfnamefont {A.}~\bibnamefont {Fedrizzi}},\ }\bibfield  {title} {\bibinfo
  {title} {Characterizing quantum dynamics with initial system-environment
  correlations},\ }\href {https://doi.org/10.1103/PhysRevLett.114.090402}
  {\bibfield  {journal} {\bibinfo  {journal} {Phys. Rev. Lett.}\ }\textbf
  {\bibinfo {volume} {114}},\ \bibinfo {pages} {090402} (\bibinfo {year}
  {2015})}\BibitemShut {NoStop}%
\bibitem [{\citenamefont {Rudinger}\ \emph {et~al.}(2019)\citenamefont
  {Rudinger}, \citenamefont {Proctor}, \citenamefont {Langharst}, \citenamefont
  {Sarovar}, \citenamefont {Young},\ and\ \citenamefont
  {Blume-Kohout}}]{PhysRevX.9.021045}%
  \BibitemOpen
  \bibfield  {author} {\bibinfo {author} {\bibfnamefont {K.}~\bibnamefont
  {Rudinger}}, \bibinfo {author} {\bibfnamefont {T.}~\bibnamefont {Proctor}},
  \bibinfo {author} {\bibfnamefont {D.}~\bibnamefont {Langharst}}, \bibinfo
  {author} {\bibfnamefont {M.}~\bibnamefont {Sarovar}}, \bibinfo {author}
  {\bibfnamefont {K.}~\bibnamefont {Young}},\ and\ \bibinfo {author}
  {\bibfnamefont {R.}~\bibnamefont {Blume-Kohout}},\ }\bibfield  {title}
  {\bibinfo {title} {Probing context-dependent errors in quantum processors},\
  }\href {https://doi.org/10.1103/PhysRevX.9.021045} {\bibfield  {journal}
  {\bibinfo  {journal} {Phys. Rev. X}\ }\textbf {\bibinfo {volume} {9}},\
  \bibinfo {pages} {021045} (\bibinfo {year} {2019})}\BibitemShut {NoStop}%
\bibitem [{\citenamefont {Giarmatzi}\ and\ \citenamefont
  {Costa}(2021)}]{Giarmatzi_2021}%
  \BibitemOpen
  \bibfield  {author} {\bibinfo {author} {\bibfnamefont {C.}~\bibnamefont
  {Giarmatzi}}\ and\ \bibinfo {author} {\bibfnamefont {F.}~\bibnamefont
  {Costa}},\ }\bibfield  {title} {\bibinfo {title} {Witnessing quantum memory
  in non-markovian processes},\ }\href
  {https://doi.org/10.22331/q-2021-04-26-440} {\bibfield  {journal} {\bibinfo
  {journal} {Quantum}\ }\textbf {\bibinfo {volume} {5}},\ \bibinfo {pages}
  {440} (\bibinfo {year} {2021})}\BibitemShut {NoStop}%
\bibitem [{\citenamefont {Milz}\ \emph {et~al.}(2021)\citenamefont {Milz},
  \citenamefont {Spee}, \citenamefont {Xu}, \citenamefont {Pollock},
  \citenamefont {Modi},\ and\ \citenamefont {Gühne}}]{milz2020genuine}%
  \BibitemOpen
  \bibfield  {author} {\bibinfo {author} {\bibfnamefont {S.}~\bibnamefont
  {Milz}}, \bibinfo {author} {\bibfnamefont {C.}~\bibnamefont {Spee}}, \bibinfo
  {author} {\bibfnamefont {Z.-P.}\ \bibnamefont {Xu}}, \bibinfo {author}
  {\bibfnamefont {F.~A.}\ \bibnamefont {Pollock}}, \bibinfo {author}
  {\bibfnamefont {K.}~\bibnamefont {Modi}},\ and\ \bibinfo {author}
  {\bibfnamefont {O.}~\bibnamefont {Gühne}},\ }\bibfield  {title} {\bibinfo
  {title} {{Genuine Multipartite Entanglement in Time}},\ }\href
  {https://doi.org/10.21468/SciPostPhys.10.6.141} {\bibfield  {journal}
  {\bibinfo  {journal} {SciPost Phys.}\ }\textbf {\bibinfo {volume} {10}},\
  \bibinfo {pages} {141} (\bibinfo {year} {2021})}\BibitemShut {NoStop}%
\bibitem [{\citenamefont {Arenz}\ \emph {et~al.}(2015)\citenamefont {Arenz},
  \citenamefont {Hillier}, \citenamefont {Fraas},\ and\ \citenamefont
  {Burgarth}}]{PhysRevA.92.022102}%
  \BibitemOpen
  \bibfield  {author} {\bibinfo {author} {\bibfnamefont {C.}~\bibnamefont
  {Arenz}}, \bibinfo {author} {\bibfnamefont {R.}~\bibnamefont {Hillier}},
  \bibinfo {author} {\bibfnamefont {M.}~\bibnamefont {Fraas}},\ and\ \bibinfo
  {author} {\bibfnamefont {D.}~\bibnamefont {Burgarth}},\ }\bibfield  {title}
  {\bibinfo {title} {Distinguishing decoherence from alternative quantum
  theories by dynamical decoupling},\ }\href
  {https://doi.org/10.1103/PhysRevA.92.022102} {\bibfield  {journal} {\bibinfo
  {journal} {Phys. Rev. A}\ }\textbf {\bibinfo {volume} {92}},\ \bibinfo
  {pages} {022102} (\bibinfo {year} {2015})}\BibitemShut {NoStop}%
\bibitem [{\citenamefont {Arenz}\ \emph {et~al.}(2018)\citenamefont {Arenz},
  \citenamefont {Burgarth}, \citenamefont {Facchi},\ and\ \citenamefont
  {Hillier}}]{Arenz_2018}%
  \BibitemOpen
  \bibfield  {author} {\bibinfo {author} {\bibfnamefont {C.}~\bibnamefont
  {Arenz}}, \bibinfo {author} {\bibfnamefont {D.}~\bibnamefont {Burgarth}},
  \bibinfo {author} {\bibfnamefont {P.}~\bibnamefont {Facchi}},\ and\ \bibinfo
  {author} {\bibfnamefont {R.}~\bibnamefont {Hillier}},\ }\bibfield  {title}
  {\bibinfo {title} {Dynamical decoupling of unbounded hamiltonians},\ }\href
  {https://doi.org/10.1063/1.5016495} {\bibfield  {journal} {\bibinfo
  {journal} {J. Math. Phys.}\ }\textbf {\bibinfo {volume} {59}},\ \bibinfo
  {pages} {032203} (\bibinfo {year} {2018})}\BibitemShut {NoStop}%
\bibitem [{\citenamefont {Taranto}\ \emph
  {et~al.}(2019{\natexlab{a}})\citenamefont {Taranto}, \citenamefont
  {Pollock},\ and\ \citenamefont {Modi}}]{taranto2019memory}%
  \BibitemOpen
  \bibfield  {author} {\bibinfo {author} {\bibfnamefont {P.}~\bibnamefont
  {Taranto}}, \bibinfo {author} {\bibfnamefont {F.~A.}\ \bibnamefont
  {Pollock}},\ and\ \bibinfo {author} {\bibfnamefont {K.}~\bibnamefont
  {Modi}},\ }\href@noop {} {\bibinfo {title} {Memory strength and
  recoverability of non-markovian quantum stochastic processes}} (\bibinfo
  {year} {2019}{\natexlab{a}}),\ \Eprint {https://arxiv.org/abs/1907.12583}
  {arXiv:1907.12583 [quant-ph]} \BibitemShut {NoStop}%
\bibitem [{Note4()}]{Note4}%
  \BibitemOpen
  \bibinfo {note} {This observation can be seen to follow e.g. because we may
  upper-bound the total \protect \textsf {RB}~non-Markovianity of a sequence
  length $m$ experiment, $\protect \mathcal {N}_q^{\protect \mathcal {F}_m}$ in
  Eq.~\protect \textup {\hbox {\mathsurround \z@ \protect \normalfont
  (\ignorespaces \ref {eq: rb non-Markovianity}\unskip \@@italiccorr )}}, as
  $\protect \mathcal {N}_q^{\protect \mathcal {F}_m}\leq \DOTSB \sum@ \slimits@
  \protect \mathcal {N}_q$, where $\protect \mathcal {N}_q$ is general
  non-Markovianity for each intermediate step up to $m$ as in Eq.~\protect
  \textup {\hbox {\mathsurround \z@ \protect \normalfont (\ignorespaces \ref
  {eq: general nonMarkovianity measure}\unskip \@@italiccorr )}} with $D$ being
  a Schatten $q$-norm, $D=\|X\|_q:=(\protect \tr |X|^q)^{1/q}$.}\BibitemShut
  {Stop}%
\bibitem [{\citenamefont {Taranto}\ \emph
  {et~al.}(2019{\natexlab{b}})\citenamefont {Taranto}, \citenamefont {Pollock},
  \citenamefont {Milz}, \citenamefont {Tomamichel},\ and\ \citenamefont
  {Modi}}]{PhysRevLett.122.140401}%
  \BibitemOpen
  \bibfield  {author} {\bibinfo {author} {\bibfnamefont {P.}~\bibnamefont
  {Taranto}}, \bibinfo {author} {\bibfnamefont {F.~A.}\ \bibnamefont
  {Pollock}}, \bibinfo {author} {\bibfnamefont {S.}~\bibnamefont {Milz}},
  \bibinfo {author} {\bibfnamefont {M.}~\bibnamefont {Tomamichel}},\ and\
  \bibinfo {author} {\bibfnamefont {K.}~\bibnamefont {Modi}},\ }\bibfield
  {title} {\bibinfo {title} {Quantum markov order},\ }\href
  {https://doi.org/10.1103/PhysRevLett.122.140401} {\bibfield  {journal}
  {\bibinfo  {journal} {Phys. Rev. Lett.}\ }\textbf {\bibinfo {volume} {122}},\
  \bibinfo {pages} {140401} (\bibinfo {year} {2019}{\natexlab{b}})}\BibitemShut
  {NoStop}%
\bibitem [{\citenamefont {Taranto}\ \emph
  {et~al.}(2019{\natexlab{c}})\citenamefont {Taranto}, \citenamefont {Milz},
  \citenamefont {Pollock},\ and\ \citenamefont {Modi}}]{PhysRevA.99.042108}%
  \BibitemOpen
  \bibfield  {author} {\bibinfo {author} {\bibfnamefont {P.}~\bibnamefont
  {Taranto}}, \bibinfo {author} {\bibfnamefont {S.}~\bibnamefont {Milz}},
  \bibinfo {author} {\bibfnamefont {F.~A.}\ \bibnamefont {Pollock}},\ and\
  \bibinfo {author} {\bibfnamefont {K.}~\bibnamefont {Modi}},\ }\bibfield
  {title} {\bibinfo {title} {Structure of quantum stochastic processes with
  finite markov order},\ }\href {https://doi.org/10.1103/PhysRevA.99.042108}
  {\bibfield  {journal} {\bibinfo  {journal} {Phys. Rev. A}\ }\textbf {\bibinfo
  {volume} {99}},\ \bibinfo {pages} {042108} (\bibinfo {year}
  {2019}{\natexlab{c}})}\BibitemShut {NoStop}%
\bibitem [{\citenamefont {White}\ \emph {et~al.}(2021)\citenamefont {White},
  \citenamefont {Pollock}, \citenamefont {Hollenberg}, \citenamefont {Modi},\
  and\ \citenamefont {Hill}}]{white2021nonmarkovian}%
  \BibitemOpen
  \bibfield  {author} {\bibinfo {author} {\bibfnamefont {G.~A.~L.}\
  \bibnamefont {White}}, \bibinfo {author} {\bibfnamefont {F.~A.}\ \bibnamefont
  {Pollock}}, \bibinfo {author} {\bibfnamefont {L.~C.~L.}\ \bibnamefont
  {Hollenberg}}, \bibinfo {author} {\bibfnamefont {K.}~\bibnamefont {Modi}},\
  and\ \bibinfo {author} {\bibfnamefont {C.~D.}\ \bibnamefont {Hill}},\
  }\href@noop {} {\bibinfo {title} {Non-markovian quantum process tomography}}
  (\bibinfo {year} {2021}),\ \Eprint {https://arxiv.org/abs/2106.11722}
  {arXiv:2106.11722 [quant-ph]} \BibitemShut {NoStop}%
\bibitem [{\citenamefont {Heinz}\ and\ \citenamefont
  {Burkard}(2021)}]{heinz2021crosstalk}%
  \BibitemOpen
  \bibfield  {author} {\bibinfo {author} {\bibfnamefont {I.}~\bibnamefont
  {Heinz}}\ and\ \bibinfo {author} {\bibfnamefont {G.}~\bibnamefont
  {Burkard}},\ }\href@noop {} {\bibinfo {title} {Crosstalk analysis for
  single-qubit and two-qubit gates in spin qubit arrays}} (\bibinfo {year}
  {2021}),\ \Eprint {https://arxiv.org/abs/2105.10221} {arXiv:2105.10221
  [cond-mat.mes-hall]} \BibitemShut {NoStop}%
\bibitem [{\citenamefont {Tamascelli}\ \emph {et~al.}(2018)\citenamefont
  {Tamascelli}, \citenamefont {Smirne}, \citenamefont {Huelga},\ and\
  \citenamefont {Plenio}}]{PhysRevLett.120.030402}%
  \BibitemOpen
  \bibfield  {author} {\bibinfo {author} {\bibfnamefont {D.}~\bibnamefont
  {Tamascelli}}, \bibinfo {author} {\bibfnamefont {A.}~\bibnamefont {Smirne}},
  \bibinfo {author} {\bibfnamefont {S.~F.}\ \bibnamefont {Huelga}},\ and\
  \bibinfo {author} {\bibfnamefont {M.~B.}\ \bibnamefont {Plenio}},\ }\bibfield
   {title} {\bibinfo {title} {Nonperturbative treatment of non-markovian
  dynamics of open quantum systems},\ }\href
  {https://doi.org/10.1103/PhysRevLett.120.030402} {\bibfield  {journal}
  {\bibinfo  {journal} {Phys. Rev. Lett.}\ }\textbf {\bibinfo {volume} {120}},\
  \bibinfo {pages} {030402} (\bibinfo {year} {2018})}\BibitemShut {NoStop}%
\bibitem [{\citenamefont {Luchnikov}\ \emph {et~al.}(2019)\citenamefont
  {Luchnikov}, \citenamefont {Vintskevich}, \citenamefont {Ouerdane},\ and\
  \citenamefont {Filippov}}]{PhysRevLett.122.160401}%
  \BibitemOpen
  \bibfield  {author} {\bibinfo {author} {\bibfnamefont {I.~A.}\ \bibnamefont
  {Luchnikov}}, \bibinfo {author} {\bibfnamefont {S.~V.}\ \bibnamefont
  {Vintskevich}}, \bibinfo {author} {\bibfnamefont {H.}~\bibnamefont
  {Ouerdane}},\ and\ \bibinfo {author} {\bibfnamefont {S.~N.}\ \bibnamefont
  {Filippov}},\ }\bibfield  {title} {\bibinfo {title} {Simulation complexity of
  open quantum dynamics: Connection with tensor networks},\ }\href
  {https://doi.org/10.1103/PhysRevLett.122.160401} {\bibfield  {journal}
  {\bibinfo  {journal} {Phys. Rev. Lett.}\ }\textbf {\bibinfo {volume} {122}},\
  \bibinfo {pages} {160401} (\bibinfo {year} {2019})}\BibitemShut {NoStop}%
\bibitem [{\citenamefont {Dirkse}\ \emph {et~al.}(2019)\citenamefont {Dirkse},
  \citenamefont {Helsen},\ and\ \citenamefont {Wehner}}]{PhysRevA.99.012315}%
  \BibitemOpen
  \bibfield  {author} {\bibinfo {author} {\bibfnamefont {B.}~\bibnamefont
  {Dirkse}}, \bibinfo {author} {\bibfnamefont {J.}~\bibnamefont {Helsen}},\
  and\ \bibinfo {author} {\bibfnamefont {S.}~\bibnamefont {Wehner}},\
  }\bibfield  {title} {\bibinfo {title} {Efficient unitarity randomized
  benchmarking of few-qubit clifford gates},\ }\href
  {https://doi.org/10.1103/PhysRevA.99.012315} {\bibfield  {journal} {\bibinfo
  {journal} {Phys. Rev. A}\ }\textbf {\bibinfo {volume} {99}},\ \bibinfo
  {pages} {012315} (\bibinfo {year} {2019})}\BibitemShut {NoStop}%
\bibitem [{\citenamefont {Girling}\ \emph {et~al.}(2021)\citenamefont
  {Girling}, \citenamefont {Cirstoiu},\ and\ \citenamefont
  {Jennings}}]{girling2021estimation}%
  \BibitemOpen
  \bibfield  {author} {\bibinfo {author} {\bibfnamefont {M.}~\bibnamefont
  {Girling}}, \bibinfo {author} {\bibfnamefont {C.}~\bibnamefont {Cirstoiu}},\
  and\ \bibinfo {author} {\bibfnamefont {D.}~\bibnamefont {Jennings}},\
  }\href@noop {} {\bibinfo {title} {Estimation of correlations and
  non-separability in quantum channels via unitarity benchmarking}} (\bibinfo
  {year} {2021}),\ \Eprint {https://arxiv.org/abs/2104.04352} {arXiv:2104.04352
  [quant-ph]} \BibitemShut {NoStop}%
\bibitem [{\citenamefont {Brown}\ and\ \citenamefont
  {Brown}(2019)}]{PhysRevA.100.032325}%
  \BibitemOpen
  \bibfield  {author} {\bibinfo {author} {\bibfnamefont {N.~C.}\ \bibnamefont
  {Brown}}\ and\ \bibinfo {author} {\bibfnamefont {K.~R.}\ \bibnamefont
  {Brown}},\ }\bibfield  {title} {\bibinfo {title} {Leakage mitigation for
  quantum error correction using a mixed qubit scheme},\ }\href
  {https://doi.org/10.1103/PhysRevA.100.032325} {\bibfield  {journal} {\bibinfo
   {journal} {Phys. Rev. A}\ }\textbf {\bibinfo {volume} {100}},\ \bibinfo
  {pages} {032325} (\bibinfo {year} {2019})}\BibitemShut {NoStop}%
\bibitem [{\citenamefont {Strikis}\ \emph {et~al.}(2020)\citenamefont
  {Strikis}, \citenamefont {Qin}, \citenamefont {Chen}, \citenamefont
  {Benjamin},\ and\ \citenamefont {Li}}]{arm2020learningbased}%
  \BibitemOpen
  \bibfield  {author} {\bibinfo {author} {\bibfnamefont {A.}~\bibnamefont
  {Strikis}}, \bibinfo {author} {\bibfnamefont {D.}~\bibnamefont {Qin}},
  \bibinfo {author} {\bibfnamefont {Y.}~\bibnamefont {Chen}}, \bibinfo {author}
  {\bibfnamefont {S.~C.}\ \bibnamefont {Benjamin}},\ and\ \bibinfo {author}
  {\bibfnamefont {Y.}~\bibnamefont {Li}},\ }\href@noop {} {\bibinfo {title}
  {Learning-based quantum error mitigation}} (\bibinfo {year} {2020}),\ \Eprint
  {https://arxiv.org/abs/2005.07601} {arXiv:2005.07601 [quant-ph]} \BibitemShut
  {NoStop}%
\bibitem [{\citenamefont {Parrado-Rodr{\'{i}}guez}\ \emph
  {et~al.}(2021)\citenamefont {Parrado-Rodr{\'{i}}guez}, \citenamefont
  {Ryan-Anderson}, \citenamefont {Bermudez},\ and\ \citenamefont
  {M{\"{u}}ller}}]{ParradoRodriguez2021crosstalk}%
  \BibitemOpen
  \bibfield  {author} {\bibinfo {author} {\bibfnamefont {P.}~\bibnamefont
  {Parrado-Rodr{\'{i}}guez}}, \bibinfo {author} {\bibfnamefont
  {C.}~\bibnamefont {Ryan-Anderson}}, \bibinfo {author} {\bibfnamefont
  {A.}~\bibnamefont {Bermudez}},\ and\ \bibinfo {author} {\bibfnamefont
  {M.}~\bibnamefont {M{\"{u}}ller}},\ }\bibfield  {title} {\bibinfo {title}
  {Crosstalk {S}uppression for {F}ault-tolerant {Q}uantum {E}rror {C}orrection
  with {T}rapped {I}ons},\ }\href {https://doi.org/10.22331/q-2021-06-29-487}
  {\bibfield  {journal} {\bibinfo  {journal} {{Quantum}}\ }\textbf {\bibinfo
  {volume} {5}},\ \bibinfo {pages} {487} (\bibinfo {year} {2021})}\BibitemShut
  {NoStop}%
\bibitem [{\citenamefont {Berk}\ \emph {et~al.}(2021)\citenamefont {Berk},
  \citenamefont {Milz}, \citenamefont {Pollock},\ and\ \citenamefont
  {Modi}}]{berk2021extracting}%
  \BibitemOpen
  \bibfield  {author} {\bibinfo {author} {\bibfnamefont {G.~D.}\ \bibnamefont
  {Berk}}, \bibinfo {author} {\bibfnamefont {S.}~\bibnamefont {Milz}}, \bibinfo
  {author} {\bibfnamefont {F.~A.}\ \bibnamefont {Pollock}},\ and\ \bibinfo
  {author} {\bibfnamefont {K.}~\bibnamefont {Modi}},\ }\href@noop {} {\bibinfo
  {title} {Extracting quantum dynamical resources: Consumption of
  non-markovianity for noise reduction}} (\bibinfo {year} {2021}),\ \Eprint
  {https://arxiv.org/abs/2110.02613} {arXiv:2110.02613 [quant-ph]} \BibitemShut
  {NoStop}%
\bibitem [{\citenamefont {Merkel}\ \emph {et~al.}(2018)\citenamefont {Merkel},
  \citenamefont {Pritchett},\ and\ \citenamefont
  {Fong}}]{merkel2018randomized}%
  \BibitemOpen
  \bibfield  {author} {\bibinfo {author} {\bibfnamefont {S.~T.}\ \bibnamefont
  {Merkel}}, \bibinfo {author} {\bibfnamefont {E.~J.}\ \bibnamefont
  {Pritchett}},\ and\ \bibinfo {author} {\bibfnamefont {B.~H.}\ \bibnamefont
  {Fong}},\ }\href@noop {} {\bibinfo {title} {Randomized benchmarking as
  convolution: Fourier analysis of gate dependent errors}} (\bibinfo {year}
  {2018}),\ \Eprint {https://arxiv.org/abs/1804.05951} {arXiv:1804.05951
  [quant-ph]} \BibitemShut {NoStop}%
\bibitem [{\citenamefont {Bengtsson}\ and\ \citenamefont
  {{\.Z}yczkowski}(2017)}]{bengtsson2017geometry}%
  \BibitemOpen
  \bibfield  {author} {\bibinfo {author} {\bibfnamefont {I.}~\bibnamefont
  {Bengtsson}}\ and\ \bibinfo {author} {\bibfnamefont {K.}~\bibnamefont
  {{\.Z}yczkowski}},\ }\href
  {https://books.google.com.tw/books?id=sYswDwAAQBAJ} {\emph {\bibinfo {title}
  {Geometry of Quantum States: An Introduction to Quantum Entanglement}}}\
  (\bibinfo  {publisher} {Cambridge University Press},\ \bibinfo {year}
  {2017})\BibitemShut {NoStop}%
\bibitem [{\citenamefont
  {Figueroa-Romero}(2021)}]{figueroaromero2021equilibration}%
  \BibitemOpen
  \bibfield  {author} {\bibinfo {author} {\bibfnamefont {P.}~\bibnamefont
  {Figueroa-Romero}},\ }\href@noop {} {\bibinfo {title} {Equilibration and
  typicality in quantum processes}} (\bibinfo {year} {2021}),\ \Eprint
  {https://arxiv.org/abs/2102.02289} {arXiv:2102.02289 [quant-ph]} \BibitemShut
  {NoStop}%
\bibitem [{\citenamefont {Collins}(2003)}]{Collins_2003}%
  \BibitemOpen
  \bibfield  {author} {\bibinfo {author} {\bibfnamefont {B.}~\bibnamefont
  {Collins}},\ }\bibfield  {title} {\bibinfo {title} {Moments and cumulants of
  polynomial random variables on unitary groups, the {I}tzykson-{Z}uber
  integral, and free probability},\ }\href
  {https://doi.org/10.1155/s107379280320917x} {\bibfield  {journal} {\bibinfo
  {journal} {Int. Math. Res. Notices}\ }\textbf {\bibinfo {volume} {2003}},\
  \bibinfo {pages} {953} (\bibinfo {year} {2003})}\BibitemShut {NoStop}%
\bibitem [{\citenamefont {Collins}\ and\ \citenamefont
  {Śniady}(2006)}]{Collins_2006}%
  \BibitemOpen
  \bibfield  {author} {\bibinfo {author} {\bibfnamefont {B.}~\bibnamefont
  {Collins}}\ and\ \bibinfo {author} {\bibfnamefont {P.}~\bibnamefont
  {Śniady}},\ }\bibfield  {title} {\bibinfo {title} {Integration with respect
  to the haar measure on unitary, orthogonal and symplectic group},\ }\href
  {https://doi.org/10.1007/s00220-006-1554-3} {\bibfield  {journal} {\bibinfo
  {journal} {Commun. Math. Phys.}\ }\textbf {\bibinfo {volume} {264}},\
  \bibinfo {pages} {773–795} (\bibinfo {year} {2006})}\BibitemShut {NoStop}%
\end{thebibliography}
\end{document}